\documentclass[a4paper,11pt]{article}
\pdfoutput=1 

\usepackage{jheppub} 

\usepackage[T1]{fontenc} 
\usepackage{threeparttable}
\usepackage[figuresright]{rotating}
\usepackage{pdflscape} 
\usepackage{bm}

\usepackage[%
  colorlinks=true,
  urlcolor=blue,
  linkcolor=blue,
  citecolor=blue
]{hyperref}

\usepackage{slashed,color,amsmath,amssymb,mathrsfs}
\usepackage{graphicx}
\usepackage{subfigure}
\usepackage{booktabs}
\usepackage{hyperref}
\usepackage{longtable}
\usepackage{array}
\usepackage{enumerate}
\usepackage{indentfirst}
\usepackage{verbatim}
\usepackage{makecell}
\usepackage{multirow}
\allowdisplaybreaks[4]
\newcommand{\mt}{M}

\title{\boldmath One-loop analysis of the interactions between doubly charmed baryons and Nambu-Goldstone bosons}

\author[a]{Ze-Rui~Liang,}
\author[a]{Peng-Cheng~Qiu,}
\author[a,b,1]{De-Liang~Yao\note{Corresponding author.}}

\affiliation[a]{School of Physics and Electronics, Hunan University,\\ 410082 Changsha, China}
\affiliation[b]{Hunan Provincial Key Laboratory of High-Energy Scale Physics and Applications, \\ Hunan University, 410082 Changsha, China}

\emailAdd{zeruiliang@hnu.edu.cn}
\emailAdd{qiupengcheng@hnu.edu.cn}
\emailAdd{yaodeliang@hnu.edu.cn}

\abstract{
The interactions between the spin-$1/2$ doubly charmed baryons and Nambu-Goldstone bosons are analyzed within a manifestly relativistic baryon chiral perturbation theory up to next-to-next-to leading order, by using the so-called extended-on-mass-shell scheme.  We utilize heavy diquark-antiquark symmetry to estimate the low-energy constants in the chiral effective Lagrangians. The $S$- and $P$-wave scattering lengths are predicted. We find that those diagrams, vanishing exactly in the heavy-quark limit, do contribute slightly to the $S$-wave scattering lengths in reality.  The influence of the spin-$3/2$ doubly charmed baryons, as heavy-quark spin partners of the spin-$1/2$ ones, on the scattering lengths is discussed as well. Finally, $S$-wave phase shifts for elastic scattering processes are presented in the energy region near threshold. Our results in this work will not only be very useful for performing chiral extrapolations of future lattice QCD data, but also provide us chiral inputs for the investigation of the spectroscopy of doubly heavy baryons.
}

\arxivnumber{2303.03370}

\begin{document} 
\maketitle
\flushbottom

\section{Introduction}\label{sec:1}

Over the past two decades, hadrons containing heavy quarks have been extensively explored, see e.g. Refs.~\cite{Klempt:2009pi,Chen:2016spr,Guo:2017jvc,Chen:2021ftn,Chen:2022asf,Meng:2022ozq} for recent reviews. This is not only because of the high experimental interest of completing the hadron spectroscopy, but also due to the theoretical importance in understanding the low-energy dynamics of quantum chromodynamics (QCD). Amongst them, the doubly heavy baryons, predicted by conventional quark model~\cite{Gell-Mann:1964ewy}, have attracted quite a few attentions. They offer a unique platform for investigating the non-perturbative dynamics of light quarks in the environment of two heavy quarks and, consequently, can be used to test the correctness of theoretical frameworks such as the QCD-inspired quark model, the non-relativistic factorization theory~\cite{Bodwin:1994jh} and so on.

Experimental attempts have been made to hunt for doubly heavy-flavored baryons since 2002~\cite{SELEX:2002wqn,SELEX:2004lln}, but their existence has been under controversy till 2017. In 2017, the LHCb collaboration announced the first successful observation of such states~\cite{LHCb:2017iph}. It was reported that a doubly charmed baryon $\Xi_{cc}^{++}$ with high significance is observed via the decay mode $\Xi_{cc}^{++}\to \Lambda_c^+K^-\pi^+\pi^+$. The existence of the $\Xi_{cc}^{++}$ state was later confirmed through another decay mode $\Xi_{cc}^{++}\to \Xi_c^+\pi^+$~\cite{LHCb:2018pcs}. It should be noted that searching for doubly charmed baryons in the above two decay modes was previously suggested by the theoretical work~\cite{Yu:2017zst}. Determinations of the properties of the $\Xi_{cc}^{++}$ baryon are conducted subsequently, e.g., measurements of its life time~\cite{LHCb:2018zpl}, mass~\cite{LHCb:2019epo} and production cross section~\cite{LHCb:2019qed}. On the other hand, the quark content of $\Xi_{cc}^{++}$ is assigned to be $[ccu]$ in the quark model. Therefore, its SU(3) partners, $\Xi_{cc}^+$ and $\Omega_{cc}^+$ with contents $[ccd]$ and $[ccs]$ respectively, are expected to exist and are hoped to be observed at LHCb. Thereby, searches for the two states are carried out and are still ongoing. Yet, no evidence of observation has been found~\cite{LHCb:2019gqy,LHCb:2021eaf,LHCb:2021rkb}. 

In fact, the search for $\Xi_{cc}^+$ was initially performed by the SELEX collaboration twenty years ago~\cite{SELEX:2002wqn}. However, the SELEX results of $\Xi_{cc}^+$ reported in Refs.~\cite{SELEX:2002wqn,SELEX:2004lln} are not confirmed by any other collaborations: FOCUS at Tevatron (proton-antiproton) collider~\cite{Ratti:2003ez}, BaBar and Belle at electron-positron colliders~\cite{BaBar:2006bab,Belle:2006edu}, and LHCb at proton-proton collider~\cite{LHCb:2019gqy,LHCb:2021eaf}. Even worse, the experimental value of the $\Xi_{cc}^+$ mass by SELEX is inconsistent with theoretical determinations obtained by relativistic quark model~\cite{Ebert:2002ig}, effective potential models~\cite{Karliner:2014gca}, heavy quark effective theory~\cite{Korner:1994nh}, lattice QCD~\cite{Lewis:2001iz,Liu:2009jc,Brown:2014ena}, etc.. See also e.g. Ref.~\cite{Yao:2020bxx} for a brief review. The aforementioned facts lead to a long-standing puzzle in questioning the existence of doubly charmed baryons. The puzzle has been solved partly by the observation of $\Xi_{cc}^{++}$ at LHCb~\cite{LHCb:2017iph,LHCb:2018pcs}. And hopefully, it will be addressed completely in the foreseeable future, as the LHC record condition keeps improved so as to overcome the drawbacks caused by the short life times of $\Xi_{cc}^+$ and $\Omega_{cc}^+$~\cite{Cerri:2018ypt}.

The discovery of the $\Xi_{cc}^{++}$ has stimulated a multitude of works on the theoretical side. Various approaches have been employed to decipher the underlying information of these doubly heavy baryons. Static properties, decays and productions of the doubly charmed baryons have been widely studied via quark model~\cite{Lu:2017meb,Wang:2017mqp,Ke:2019lcf,Ke:2022gxm}, extended chromomagnetic model~\cite{Weng:2018mmf}, SU(3) symmetry method~\cite{Wang:2017azm,Geng:2017mxn}, operator product expansion~\cite{Berezhnoy:2018bde}, QCD sum rules~\cite{Chen:2017sbg,Wang:2017qvg,Wang:2018lhz,Shi:2019hbf}, lattice QCD~\cite{Mathur:2018rwu,Bahtiyar:2020uuj} and so on. 

The doubly heavy baryons are composed of two heavy quarks ($Q$) and one light quark ($q$), where both heavy quark symmetry as $m_Q\to \infty$ and chiral symmetry as $m_q\to 0$ will manifest. Therefore, the doubly heavy baryons provide a novel platform to study heavy quark symmetry and chiral symmetry of light quarks simultaneously. 
In particular, the two heavy quarks in the baryons usually act as a form of compact heavy diquark. In the heavy quark limit (HQL), the heavy diquark belongs to the color $\bar{3}_c$ representation and serves as a static color source for the light quarks. The same color dynamics arises in the mesons containing a single heavy antiquark. The correspondence between a heavy diquark and a single heavy antiquark is known as heavy diquark-antiquark (HDA) symmetry. In consequence, the doubly heavy baryons can be related to the mesons with a heavy antiquark component through HDA symmetry~\cite{Savage:1990di}. In practice, effective field theories (EFTs) are efficient and powerful in the sense that they can easily implement all such kind of symmetries through the approach of constructing pertinent effective Lagrangians.

Chiral perturbation theory (ChPT)~\cite{Weinberg:1978kz,Gasser:1983yg,Gasser:1984gg} is the low-energy EFT of QCD, which is a powerful tool to explore hadron physics in the non-perturbative regime. It has achieved great triumphs in the pure mesonic sector. Meanwhile, various extensions have been developed in the past years~\cite{Bernard:1996gq,Bernard:2007zu,book:789407}. Note that, ChPT for heavy hadrons is comprehensively reviewed in Ref.~\cite{Meng:2022ozq}. The version extended to the single-baryon sector~\cite{Gasser:1987rb} is known as baryon chiral perturbation theory (BChPT). However, when baryon fields are incorporated as explicit degrees of freedom in the theory, the notable power counting breaking (PCB) problem arises due to the non-zero mass of the baryons in the chiral limit. Many approaches have been proposed in order to remedy this issue. The most popular ones are the heavy-baryon (HB) formalism~\cite{Jenkins:1990jv}, the infrared regularization (IR) prescription~\cite{Becher:1999he} and the extended-on-mass-shell (EOMS) renormalization scheme~\cite{Fuchs:2003qc}. Thereinto, the EOMS scheme not only restores the correct power counting but also respects the original analytic structure. Furthermore, it has been demonstrated that better convergence properties can be established compared to other approaches; see Refs.~\cite{Bernard:2007zu,Geng:2013xn} for reviews.

Within the framework of BChPT, the properties such as masses and magnetic moments of the doubly charmed baryons have been investigated in Refs.~\cite{Brodsky:2011zs,Sun:2014aya,Sun:2016wzh,Li:2017pxa,Li:2017cfz, HillerBlin:2018gjw,Yao:2018ifh,Liu:2018euh}. 
For the scattering of pseudoscalar Nambu-Goldstone bosons (NGB) off doubly charmed baryons, Ref.~\cite{Guo:2017vcf} conducts a lead-order (LO) BChPT calculation. Unitarization of the LO chiral amplitude is carried out so as to search for possible exotic doubly charmed states. The low-lying spectrum of the double-charm baryons with negative parity was further studied in Ref.~\cite{Yan:2018zdt} by using a chiral potential of next-to-leading order (NLO). In Ref.~\cite{Meng:2018zbl}, the scattering lengths of NGB and doubly charmed baryons are calculated up to next-to-next-to-leading order (NNLO) in the HB formalism. For a relativistic chiral description of the doubly charmed baryons, complete and minimal set of chiral effective Lagrangians have already been constructed up to $O(p^4)$ in Ref.~\cite{Qiu:2020omj}, with $p$ collectively denoting the chiral expansion parameters. However, a systematical one-loop analysis of the interactions between the doubly charmed baryons and NGBs in manifestly relativistic BChPT at one-loop order is still lacking. 

In present work, we have calculated the scattering amplitudes for the interactions of NGBs and doubly charmed baryons in the covariant BChPT up to NNLO, i.e. the leading one-loop order. The ultraviolet (UV) divergences stemming from the loops are removed by utilizing the modified minimal subtraction scheme, namely, the $\overline{\rm MS}-1$ scheme. Furthermore, we have explicitly checked that the PCB terms can be absorbed exactly via a finite shifts of the low energy constants (LECs) according to the essence of EOMS scheme. In this way, we obtain the EOMS-renormalized chiral amplitude, which is renormalization-scale independent and can be used to derive physical observable of interest. Moreover, the obtained one-loop amplitudes possess proper analytic structure and correct power counting. And hence they are very suited to perform chiral extrapolation, when relevant lattice QCD data are available in future. In addition, our one-loop amplitudes can be applied to evaluate finite-volume corrections of lattice QCD results by merely substituting all the involved one-loop integrals with their finite counterparts that are uniformly formulated in Ref.~\cite{Liang:2022tcj}.

For the lack of available data from experiments or lattice QCD at present, we have to estimate the unknown LECs in chiral effective Lagrangians by imposing the HDA symmetry mentioned above. Specifically, the unknown LECs in our case can be related to the ones involved in the $D\phi$ scatterings, where $D$ and $\phi$ stand for charmed $D$ mesons and NGBs, respectively. Fortunately, the values $D\phi$ LECs have been well determined by performing fits to lattice QCD data of the $S$-wave $D\phi$ scatting lengths in Refs.~\cite{Liu:2012zya,Altenbuchinger:2013vwa,Yao:2015qia,Du:2017ttu}. In fact, two of the $\mathcal{O}(p^2)$ LECs, $b_1$ and $b_2$, have been determined by fitting to lattice QCD data on the doubly charmed baryon masses in Ref.~\cite{Yao:2018ifh}. We find that the values of $b_1$ and $b_2$ obtained in Ref.~\cite{Yao:2018ifh} agree well with those by HDA symmetry, indicating the feasibility of the use of HDA symmetry in this work. The remaining LECs, that can not be constrained by HDA symmetry, are set to zero in line with the ansatz of naturalness of LECs. 

Given that the LECs are pinned down, we predict the $S$- and $P$-wave scattering lengths for the elastic channels with definite strangeness $S$ and isospin $I$. For $S$-wave scattering lengths, our relativistic results turn out to be qualitatively consistent with the ones obtained in the HB approach~\cite{Meng:2018zbl}. Sizable relativistic recoil corrections exist, for instance, in the channel of $\Xi_{cc}\bar{K}\to\Xi_{cc}\bar{K}$ with $(S,I)=(-1,0)$. In Ref.~\cite{Meng:2018zbl}, the contributions from the diagrams, which should vanish exactly only in the HQL, are ignored. We have explicitly verified that those HQL-vanishing diagrams indeed contribute marginally to the $S$-wave scattering lengths in the real world with a finite heavy quark mass. In HQL, since the spin-$3/2$ and spin-$1/2$ doubly charmed baryons degenerate into a heavy quark spin (HQS) multiplet, they should be treated on equal footing. Therefore, we assess the influence of the spin-$3/2$ states on the scattering lengths by incorporating them as explicit degrees of freedom into the effective Lagrangians. As expected, they mainly affect the $P$-wave scattering lengths with quantum numbers $J^P=\frac{3}{2}^+$. Their contributions to the $S$-wave scattering lengths are negligible. As byproducts, we also discuss the so-called resonance contribution to the LECs by integrating out the spin-$3/2$ baryons. 

For future reference, $S$-wave phase shifts are calculated for the channels of elastic scatterings in the energy regions close to the lowest thresholds under consideration. Future lattice QCD simulations of the low-energy interactions between doubly charmed baryons and NGBs are necessary to explore the spectrum of the doubly heavy baryons. Our $S$-wave phase shifts can be associated directly with the energy levels at non-zero momenta via the famous L{\"u}scher formula~\cite{Luscher:1986pf} and its generalizations~\cite{Rummukainen:1995vs,Kim:2005gf,Gockeler:2012yj,Briceno:2014oea}. 

The layout of this manuscript is described as follows. Formal aspects for the scattering amplitude are introduced in section~\ref{sec:2}, where Lorentz decomposition, the strangeness-isospin structure and partial wave projection are briefly illustrated. Details on the calculation of the chiral amplitudes are exhibited in section~\ref{sec:ampcalc}. Chiral effective lagrangians relevant to our calculation up to NNLO are displayed in subsection~\ref{sec:3.2}. Tree-level and loop amplitudes are given in subsections~\ref{sec:3.3} and~\ref{sec:3.4}, respectively. Section~\ref{sec:4} complies the procedure of renormalization of the one-loop amplitudes within the EOMS scheme. In section~\ref{sec:5.1}, the values of the LECs are estimated. The $S$-wave and $P$-wave scattering lengths are calculated in subsection~\ref{sec:5.2}. Subsection~\ref{sec:5.3} discusses the impact of the spin-$3/2$ doubly charmed baryons. The $S$-wave phase shifts for the channels of elastic scatterings are presented in subsection~\ref{sec:5.4}. Finally, section~\ref{sec:6} comprises our summary and outlook. The $\beta$-functions concerning the UV- and EOMS-subtractions are relegated to appendices~\ref{app:UV-divergent} and~\ref{app:EOMS-PCBterms}, respectively. Appendix~\ref{app.wave} contains the definition of loop integrals and explicit expressions of loop corrections for the masses and wave function renormalization constants. Application of HDA symmetry to the estimation of the LEC values is detailed in appendix~\ref{app.hda}.


\section{Formal aspects of scattering amplitude}\label{sec:2}

\subsection{Lorentz structure of the amplitude}\label{sec:2.1}
The scattering process of $\psi_{1}(p) \phi_{1}(q) \rightarrow \psi_{2}(p^\prime) \phi_{2}(q^\prime)$, with momenta indicated in parentheses, is described by the Lorentz-invariant amplitude, which can be decomposed as
\begin{align}\label{eq:ABform}
\mathcal{T}_{\psi_1\phi_1\to\psi_2\phi_2}(s,t)=\bar{u}(p^\prime,\sigma^\prime)\left\{A(s,t)+\frac{1}{2}(\slashed{q}+\slashed{q}^\prime)B(s,t)\right\}u(p,\sigma)\ .
\end{align}
Here, $\psi_{1,2}\in\{\Xi_{cc}^{++},\Xi_{cc}^{+},\Omega_{cc}^{+}\}$ represent the incoming and outgoing doubly charmed baryons respectively, while $\phi_{1,2}\in\{\pi^\pm,\pi^0,K^\pm,K^0,\bar{K^0},\eta\}$ are the incoming and outgoing Goldstone bosons in order. The symbols $\sigma$ and $\sigma^\prime$ denote the spins of the corresponding baryons. The Mandelstam variables are defined by
\begin{align}\label{Mandvar}
 s = \left(p + q\right)^2, \quad t = \left(p - p^\prime\right)^2, \quad u = \left(p - q^\prime\right)^2,
\end{align}
which satisfy the constraint $s+t+u=m_{\psi_1}^2+m_{\phi_1}^2+m_{\psi_2}^2+m_{\phi_2}^2\equiv \Sigma$. 
The Lorentz decomposition of the scattering amplitude is not unique, an alternative form is given by
\begin{align}\label{eq:DBform}
\mathcal{T}_{\psi_1\phi_1\to\psi_2\phi_2}(s,t)=\bar{u}(p^\prime,\sigma^\prime)\left\{D(s,t)+\frac{i\sigma^{\mu\nu}q^\prime_\mu q_\nu}{m_{\psi_1}+m_{\psi_2}} B(s,t)\right\}u(p,\sigma)\ ,
\end{align}
with $\sigma^{\mu\nu}=\frac{i}{2}[\gamma^\mu,\gamma^\nu]$. The new function $D(s,t)$ is related to $A$ and $B$ via 
\begin{align}
D(s,t)=A(s,t)+\nu B(s,t)\ ,\quad \nu=\frac{s-u}{2(m_{\psi_1}+m_{\psi_2})}\ .
\end{align}
Similar to the situation in pion-nucleon scattering, the decomposition in terms of $D$ and $B$ is more suited to perform chiral expansion~\cite{Becher:2001hv}, while the other is more practically convenient for the extraction of the structure functions $A$ and $B$.

\subsection{Amplitudes for given strangeness and isospin}\label{sec:2.2}

The doubly charmed baryons and the pseudoscalar NGBs fill the representations of the SU(3) flavor group of the light $(u,d,s)$ quarks. Namely, they belong to the SU(3) triplets and octets, respectively. In consequence, there exist a multitude of physical scattering amplitudes corresponding to the various charge states showing up in the multiplets. These scattering amplitudes can be classified, thanks to the conservation of strangeness ($S$) and isospin ($I$). As a result, the general meson-baryon scattering processes can categorized into $7$ independent channels of interactions: $4$ single-channel processes and $3$ coupled-channel ones. In practice, the amplitudes with definite $(S,I)$ quantum numbers can be expressed in terms of the physical amplitudes. Explicit relations are listed in the following.
\begin{enumerate}[(i)]
\item For the single channels with $(S,I)=\left(-2, \frac{1}{2}\right)$, $\left(1, 1\right) $, $\left(1, 0\right) $, $\left(0, \frac{3}{2}\right) $, one has
\begin{align}
  \mathcal{T}^{(-2, \frac{1}{2})}_{\Omega_{cc} \bar{K} \rightarrow \Omega_{cc} \bar{K}}\left(s, t, u\right)&=\mathcal{T}_{\Omega^{+}_{cc} K^{-} \rightarrow \Omega^{+}_{cc} K^{-}}\left(s, t, u\right), \\
  \mathcal{T}^{(1, 1)}_{\Xi_{cc} K \rightarrow \Xi_{cc} K}\left(s, t, u\right) &= \mathcal{T}_{\Xi^{++}_{cc} K^{+} \rightarrow \Xi^{++}_{cc} K^{+}}\left(s, t, u\right), \\
  \mathcal{T}^{(1, 0)}_{\Xi_{cc} K \rightarrow \Xi_{cc} K}\left(s, t, u\right) &= 2\mathcal{T}_{\Xi^{++}_{cc} K^{0} \rightarrow \Xi^{++}_{cc} K^{0}}\left(s, t, u\right)-\mathcal{T}_{\Xi^{++}_{cc} K^{+} \rightarrow \Xi^{++}_{cc} K^{+}}\left(s, t, u\right), \\
  \mathcal{T}^{(0, \frac{3}{2})}_{\Xi_{cc} \pi \rightarrow \Xi_{cc} \pi}\left(s, t, u\right) &= \mathcal{T}_{\Xi^{++}_{cc} \pi^{+} \rightarrow \Xi^{++}_{cc} \pi^{+}}\left(s, t, u\right). 
\end{align}
\item For the two-coupled channel with $(S,I)=\left(-1, 1\right)$, the relations are given by 
\begin{align}
  \mathcal{T}^{(-1, 1)}_{\Omega_{cc} \pi \rightarrow \Omega_{cc} \pi}\left(s, t, u\right) &= \mathcal{T}_{\Omega^{+}_{cc} \pi^{0} \rightarrow \Omega^{+}_{cc} \pi^{0}}\left(s, t, u\right), \\
  \mathcal{T}^{(-1, 1)}_{\Xi_{cc} \bar{K}\rightarrow \Xi_{cc} \bar{K}}\left(s, t, u\right) &= \mathcal{T}_{\Xi^{++}_{cc} K^{0} \rightarrow \Xi^{++}_{cc} K^{0}}\left(u, t, s\right), \\
  \mathcal{T}^{(-1, 1)}_{\Omega_{cc} \pi \rightarrow \Xi_{cc} \bar{K}}\left(s, t, u\right) &=\sqrt{2} \mathcal{T}_{\Xi^{++}_{cc} K^- \rightarrow \Omega^{+}_{cc} \pi^{0} }\left(s, t, u\right).
\end{align}
\item For the two-coupled channel with $(S,I)=\left(-1, 0\right)$, the relations can be written as
\begin{align}
  \mathcal{T}^{(-1, 0)}_{\Xi_{cc} \bar{K}\rightarrow \Xi_{cc} \bar{K}}\left(s, t, u\right) &= 2\mathcal{T}_{\Xi^{++}_{cc} K^{+} \rightarrow \Xi^{++}_{cc} K^{+}}\left(u, t, s\right) - \mathcal{T}_{\Xi^{++}_{cc} K^{0} \rightarrow \Xi^{++}_{cc} K^{0}} \left(u, t, s\right), \\
  \mathcal{T}^{(-1, 0)}_{\Omega_{cc} \eta \rightarrow \Omega_{cc} \eta}\left(s, t, u\right) &= \mathcal{T}_{\Omega^{+}_{cc} \eta \rightarrow \Omega^{+}_{cc} \eta}\left(s, t, u\right), \\
  \mathcal{T}^{(-1, 0)}_{\Xi_{cc} \bar{K}  \rightarrow \Omega_{cc} \eta}\left(s, t, u\right) &= \sqrt{2}\mathcal{T}_{\Xi^{+}_{cc} \bar{K}^{0}  \rightarrow \Omega^{+}_{cc} \eta}\left(s, t, u\right). 
\end{align}
\item For the three-coupled channel with $(S,I)=\left(0, \frac{1}{2}\right)$, the relations read
\begin{align}
  \mathcal{T}^{(0, \frac{1}{2})}_{\Xi_{cc} \pi \rightarrow \Xi_{cc} \pi}\left(s, t, u\right) &= \frac{3}{2}\mathcal{T}_{\Xi^{++}_{cc} \pi^{+} \rightarrow \Xi^{++}_{cc} \pi^{+}}\left(u, t, s\right) - \frac{1}{2}\mathcal{T}_{\Xi^{++}_{cc} \pi^{+} \rightarrow \Xi^{++}_{cc} \pi^{+}}\left(s, t, u\right), \\
  \mathcal{T}^{(0, \frac{1}{2})}_{\Xi_{cc} \eta \rightarrow \Xi_{cc} \eta}\left(s, t, u\right) &= \mathcal{T}_{\Xi^{++}_{cc} \eta \rightarrow \Xi^{++}_{cc} \eta }\left(s, t, u\right), \\
  \mathcal{T}^{(0, \frac{1}{2})}_{\Omega_{cc} K  \rightarrow \Omega_{cc} K }\left(s, t, u\right) &= \mathcal{T}_{\Omega^{+}_{cc} K^{-} \rightarrow \Omega^{+}_{cc} K^{-}}\left(u, t, s\right), \\
  \mathcal{T}^{(0, \frac{1}{2})}_{\Xi_{cc} \pi \rightarrow \Xi_{cc} \eta }\left(s, t, u\right) &= \sqrt{3}\mathcal{T}_{\Xi^{++}_{cc} \pi^{0} \rightarrow \Xi^{++}_{cc} \eta }\left(s, t, u\right), \\
  \mathcal{T}^{(0, \frac{1}{2})}_{\Xi_{cc} \pi \rightarrow  \Omega_{cc} K}\left(s, t, u\right) &= \sqrt{3}\mathcal{T}_{\Xi^{++}_{cc}K^-\rightarrow \Omega^{+}_{cc}\pi^{0}}\left(u, t, s\right), \\
  \mathcal{T}^{(0, \frac{1}{2})}_{\Xi_{cc} \eta  \rightarrow  \Omega_{cc} K}\left(s, t, u\right) &= \mathcal{T}_{\Xi^{+}_{cc} \bar{K}^{0}  \rightarrow \Omega^{+}_{cc} \eta}\left(u, t, s\right).
\end{align}
\end{enumerate}
In above, there are in total $16$ scattering amplitudes with definite strangeness and isospin. Nevertheless, they can be expressed in terms of $10$ physical amplitudes, provided that crossing symmetry is implemented.   

\subsection{Partial wave projection}\label{sec:2.3}
The partial wave expansion of the scattering amplitude takes the form~\cite{Chew:1957zz,goldberger2004collision} 
\begin{align}
\mathcal{T}^{(S,I)}_{\psi_1\phi_1\to\psi_2\phi_2}(s,t) 
&=\sqrt{2m_{\psi_1}}\sqrt{2m_{\psi_2}}\chi_{\psi_2}^\dagger\sum_{\ell=0}^{\infty}\bigg\{
\big[(\ell+1)\,\mathcal{T}_{\ell+}^{(S,I)}(s)+\ell\, \mathcal{T}_{\ell-}^{(S,I)}(s)\big]P_{\ell}(z)\notag\\
&-\big[\mathcal{T}_{\ell+}^{(S,I)}(s)-\mathcal{T}_{\ell-}^{(S,I)}(s)\big]i{\bm\sigma}\cdot\left(\hat{\bf q}^\prime\times \hat{\bf q}\right)P_\ell^\prime(z)
\bigg\}\chi_{\psi_1}\ ,\label{eq:PWE}
\end{align}
where ${\bm\sigma}=(\sigma_1,\sigma_2,\sigma_3)$ is a vector with components being Pauli matrices, and $\chi$ ($\chi^\dagger$) is the spinor of the incoming (outgoing) baryon. Furthermore, $\hat{\bf q}={\bf q}/|{\bf q}|$ and $\hat{\bf q}^\prime={\bf q}^\prime/|{\bf q}^\prime|$ are the directions of the incoming and outgoing Goldstone mesons, respectively. $P_{\ell}(z)$ is the Legendre polynomial and $P_{\ell}^\prime(z)$ is its derivative. $z=\cos\theta$ with $\theta$ the scattering angle. Here $\ell$ is the orbital momentum, and the notation $\ell\pm$ indicates the total angular momentum is $J=\ell\pm\frac{1}{2}$. 

In Eq.~\eqref{eq:PWE}, the amplitude $\mathcal{T}_{\ell\pm}^{(S,I)}$ is the so-called dimensionless partial-wave amplitude, which possesses definite quantum numbers of strangeness $S$, isospin $I$ and total angular momentum $J$. It is popular to redefine a new partial wave amplitude $f_{\ell\pm}^{(S,I)}$ via
\begin{align}
    \mathcal{T}_{\ell\pm}^{(S,I)}(s)&=\frac{8\pi\sqrt{s}}{\sqrt{2m_{\psi_1}}\sqrt{2m_{\psi_2}}}f_{\ell\pm}^{(S,I)}(s)\ ,
\end{align}
which is more preferable in investigations of the analytic properties~\cite{hohler1983pion}. The inversion of Eq.~\eqref{eq:PWE} yields 
\begin{align}
  f_{\ell\pm }^{(S,I)}(s) &= \frac{\sqrt{E_{\psi_{1}}+m_{\psi_{1}}}\sqrt{E_{\psi_{2}}+m_{\psi_{2}}}}{16\pi\sqrt{s} }\left\{A_{\ell}^{(S,I)}(s)+ \frac{E_{\phi_{1}}+E_{\phi_{2}}}{2}B_{\ell}^{(S,I)}(s) + \left[ \frac{\left\lvert \bf{q } \right\rvert^2 }{2(E_{\psi_{1}}+m_{\psi_{1}})} \right. \right. \notag \\
  & \left. + \frac{\left\lvert \bf{q }^{\prime} \right\rvert^2 }{2(E_{\psi_{2}}+m_{\psi_{2}})}  \right] B_{\ell}^{(S,I)}(s) +\left\lvert \bf{q } \right\rvert \left\lvert \bf{q }^{\prime} \right\rvert \left[ \frac{B_{\ell\pm 1}^{(S,I)}(s)}{2(E_{\psi_{1}}+m_{\psi_{1}})} + \frac{B_{\ell\pm 1}^{(S,I)}(s)}{2(E_{\psi_{2}}+m_{\psi_{2}})}\right.  \notag \\
  & \left. \left.- \frac{A_{\ell\pm 1}^{(S,I)}(s)-\frac{E_{\phi_{1}}+E_{\phi_{2}}}{2} B_{\ell\pm 1}^{(S,I)}(s)}{(E_{\psi_{1}}+m_{\psi_{1}}) (E_{\psi_{2}}+m_{\psi_{2}})} \right]\right\}\ ,\label{gen-fl}
\end{align}
where $A_\ell^{(S,I)}$ and $B_\ell^{(S,I)}$ are given by
\begin{align}
  & A_{\ell}^{(S,I)}(s)=\int_{-1}^{1}  \,{\rm d}\cos\theta~P_{\ell}(\cos\theta)A^{(S,I)}(s,t)\mid _{t=m_{\psi_1}^2+m_{\psi_2}^2-2E_{\psi_{1}}E_{\psi_{2}} + 2\left\lvert \bf{q } \right\rvert \left\lvert \bf{q }^{\prime} \right\rvert \cos\theta},  \notag \\
  & B_{\ell}^{(S,I)}(s)=\int_{-1}^{1}  \,{\rm d}\cos\theta~P_{\ell}(\cos\theta)B^{(S,I)}(s,t)\mid _{t=m_{\psi_1}^2+m_{\psi_2}^2-2E_{\psi_{1}}E_{\psi_{2}} + 2\left\lvert \bf{q } \right\rvert \left\lvert \bf{q }^{\prime} \right\rvert \cos\theta}. 
\end{align}
In the above equations, $E_{\psi_{1}}$ ($E_{\psi_{2}}$) is the energy of the incoming (outgoing) doubly charmed baryon, $E_{\phi_{1}}$ ($E_{\phi_{2}}$) denotes the energy of the incoming (outgoing) meson, and $\bf{q}$ ($\bf{q }^{\prime}$) stands for the three momentum of the incoming (outgoing) meson. In the center-of-mass (CM) frame, one has
\begin{align}
  & E_{\psi_{i}}=\frac{s+m_{\psi_i}^2-m_{\phi_i}^2}{2 \sqrt{s}}\ , \quad
  E_{\phi_{i}}=\frac{s-m_{\psi_i}^2+m_{\phi_i}^2}{2 \sqrt{s}}\ ,\quad (i=1,2)  \notag \\
  & \left\lvert \bf{q } \right\rvert^2 =\frac{1}{4s}\lambda(s,m_{\psi_1}^2,m_{\phi_1}^2)\ ,
  \quad 
  \left\lvert \bf{q }^{\prime} \right\rvert^2 =\frac{1}{4s} \lambda(s,m_{\psi_2}^2,m_{\phi_2}^2)\ ,
\end{align}
with $\lambda(a,b,c)=a^2+b^2+c^2-2ab-2ac-2bc$ being the K\"all\'{e}n function.

For an elastic scattering process $\psi\phi\to \psi\phi$, the relevant partial wave amplitude $f_{\ell\pm }^{(S, I)}(s)$ in Eq.~\eqref{gen-fl} can be simplified to~\cite{Frazer:1960zz,Hoferichter:2015hva}
\begin{align}\label{fl}
  f_{\ell \pm}^{(S, I)}(s) &= \frac{1}{16\pi\sqrt{s} } \left\{(E + m_{\psi}) \left[A_{\ell}^{(S, I)}(s)+ \left(\sqrt{s} - m_{\psi}\right)B_{\ell}^{(S, I)}(s)\right] \right. \notag \\
  & \left. + (E - m_{\psi}) \left[-A_{\ell\pm 1}^{(S, I)}(s) + \left(\sqrt{s} + m_{\psi}\right)B_{\ell\pm 1}^{(S, I)}(s)\right]\right\} ,
\end{align}
with
\begin{align}\label{iofabl}
  & A_{\ell}^{(S, I)}(s) = \int_{-1}^{1} \,{\rm d}\cos\theta~P_{\ell}(\cos\theta)A^{(S, I)}(s, t) \mid_{t = -2{\bf q}^2(1 - \cos\theta)}, \notag \\
  & B_{\ell}^{(S, I)}(s) = \int_{-1}^{1} \,{\rm d}\cos\theta~P_{\ell}(\cos\theta)B^{(S, I)}(s, t) \mid_{t = -2{\bf q}^2(1 - \cos\theta)} .
\end{align}
Here $E=({s + m_{\psi}^2 - m_{\phi}^2})/({2 \sqrt{s}})$ is the CM energy of doubly charmed baryon and $|{\bf q}|=\lambda^{1/2}(s,m_{\psi}^2,m_{\phi}^2)/(2\sqrt{s})$ refers to the modulus of the CM momentum.

\subsection{Scattering length}\label{sec:2.4}
Scattering length is one of the most important quantities characterizing the properties of strong interaction. In what follows, we derive the formulae for the calculation of the scattering lengths in the elastic scattering channels.

In the vicinity of threshold, the amplitudes $f_{\ell \pm}^{(S, I)}(s)$ in Eq.~\eqref{fl} can be expanded in terms of the three momentum squared ${\bf q}^{2}$,
\begin{align}\label{eoffl}
  f_{\ell \pm}^{(S, I)}(s) = a_{\ell \pm}^{(S, I)} {\bf q}^{2\ell} + b_{\ell \pm}^{(S, I)} {\bf q}^{2\ell + 2} + c_{\ell \pm}^{(S, I)} {\bf q}^{2\ell + 4} + \mathcal{O}({\bf q}^{2\ell + 6}).
\end{align}
The coefficients on the right hand side of the above equation are referred to as threshold parameters. Specifically, the coefficients of the first three terms, $a_{\ell \pm}^{(S, I)}$, $b_{\ell \pm}^{(S, I)}$ and $c_{\ell \pm}^{(S, I)}$, are called scattering lengths\footnote{The scattering lengths for the $P$ waves are also called scattering volumes.}, effective ranges and shape parameters, in order. 

In view of Eq.~\eqref{fl}, the threshold expansion of $f_{\ell \pm}^{(S, I)}(s)$ can be obtained by expanding $A_{\ell}^{(S, I)}(s)$ and $B_{\ell}^{(S, I)}(s)$ in the same way as Eq.~\eqref{eoffl}, i.e.,
\begin{align}\label{eofabl}
  A_{\ell}^{(S, I)}(s) &= A_{\ell}^{(S, I, \ell)} {\bf q}^{2\ell} + A_{\ell}^{(S, I, \ell + 1)} {\bf q}^{2\ell + 2} + A_{\ell}^{(S, I, \ell + 2)} {\bf q}^{2\ell + 4} + \mathcal{O}({\bf q}^{2\ell + 6})\ , \notag \\
  B_{\ell}^{(S, I)}(s) &= B_{\ell}^{(S, I, \ell)} {\bf q}^{2\ell} + B_{\ell}^{(S, I, \ell + 1)} {\bf q}^{2\ell + 2} + B_{\ell}^{(S, I, \ell + 2)} {\bf q}^{2\ell + 4} + \mathcal{O}({\bf q}^{2\ell + 6})\ .
\end{align}
Therefore, the $S$- and $P$-wave scattering lengths of our interest can be written as
\begin{align}\label{scattlenv1}
   a_{0+}^{(S, I)} &= \frac{m_{\psi}}{8\pi\left(m_{\psi} + m_{\phi}\right)} \left(A_{0}^{(S, I, 0)} + m_{\phi} B_{0}^{(S, I, 0)}\right)\ , \notag \\
   a_{1+}^{(S, I)} &= \frac{m_{\psi}}{8\pi\left(m_{\psi} + m_{\phi}\right) }\left(A_{1}^{(S, I, 1)} + m_{\phi} B_{1}^{(S, I, 1)}\right)\ , \notag \\
   a_{1-}^{(S, I)} &= a_{1+}^{(S, I)} + \frac{1}{32\pi m_{\psi}\left(m_{\psi} + m_{\phi}\right)} \left(-A_{0}^{(S, I, 0)} + \left(2m_{\psi} + m_{\phi}\right) B_{0}^{(S, I, 0)}\right)\ .
\end{align}

On the other hand, for small $t$, $A^{(S,I)}(s,t)$ and $B^{(S,I)}(s,t)$ can be expressed as
\begin{align}\label{eofabst}
  A^{(S, I)}(s, t) &= A^{(S, I)}(s, 0) + t\left[\partial_t A^{(S, I)}(s, t)\right]_{t = 0} + \frac{t^2}{2}\left[\partial_t^2 A^{(S, I)}(s, t)\right]_{t = 0} + \cdots, \notag \\
  B^{(S, I)}(s, t) &= B^{(S, I)}(s, 0) + t\left[\partial_t B^{(S, I)}(s, t)\right]_{t = 0} + \frac{t^2}{2}\left[\partial_t^2 B^{(S, I)}(s, t)\right]_{t = 0} + \cdots.      
\end{align}
Substituting Eq.~\eqref{eofabst} in to Eq.~\eqref{iofabl},  one can obtain the coefficients in Eq.~\eqref{eofabl}. For $S$ and $P$ waves, we get
\begin{align}\label{absi}
  A_{0}^{(S, I, 0)} &= 2\left[A^{(S, I)}(s, 0)\right]_{{\bf q}^2 = 0}, \quad
  B_{0}^{(S, I, 0)} = 2\left[B^{(S, I)}(s, 0)\right]_{{\bf q}^2 = 0}, \notag\\
  A_{1}^{(S, I, 1)} &= \frac{4}{3}\left[\partial_{t}A^{(S, I)}(s, t)\right]_{t = 0,{\bf q}^2 = 0},\quad
  B_{1}^{(S, I, 1)} = \frac{4}{3}\left[\partial_{t}B^{(S, I)}(s, t)\right]_{t = 0, {\bf q}^2 = 0}.
\end{align}

Eventually, with the help of Eq.~\eqref{absi} and Eq.~\eqref{scattlenv1}, the $S$- and $P$-wave scattering lengths in terms of $A$ and $B$ amplitudes are expressed as\footnote{Our notation of the $S$-wave scattering length is the same as the one used in Ref.~\cite{Meng:2018zbl}, but different from the definition in Ref.~\cite{Guo:2017vcf} by a factor of $1/2$, i.e., $a_{0+}=a_{0+}^{\rm Guo}/2$.}
\begin{align}
  a_{0+}^{(S, I)} &= \frac{m_{\psi}}{4\pi\left(m_{\psi} + m_{\phi}\right)} \bigg\{\left[A^{(S, I)}(s, 0)\right]_{{\bf q}^2 = 0} + m_{\phi} \left[B^{(S, I)}(s, 0)\right]_{{\bf q}^2 = 0}\bigg\}, \notag \\
  a_{1+}^{(S, I)} &= \frac{m_{\psi}}{6\pi\left(m_{\psi} + m_{\phi}\right)} \bigg\{\left[\partial_{t}A^{(S, I)}(s, t)\right]_{t = 0, {\bf q}^2 = 0} + m_{\phi} \left[\partial_{t}B^{(S, I)}(s, t)\right]_{t = 0, {\bf q}^2 = 0}\bigg\}, \notag \\
  a_{1-}^{(S, I)} &= a_{1+}^{(S, I)} - \frac{1}{16\pi m_{\psi}\left(m_{\psi} + m_{\phi}\right)}\bigg\{\left[A^{(S, I)}(s, 0)\right]_{{\bf q}^2 = 0} - \left(2m_{\psi} + m_{\phi}\right) \left[B^{(S, I)}(s, 0)\right]_{{\bf q}^2 = 0}\bigg\}.
  \label{eq:sca.len.ana}
\end{align}
Given that the expressions of the $A$ and $B$ functions are known, the above formulae can be readily applied to obtain the scattering lengths analytically.

\section{Calculation of the amplitude in BChPT\label{sec:ampcalc}}
\subsection{Power counting}\label{sec:3.1}
The amplitudes for the processes of on-shell scatterings are multivariate functions of masses and Mandelstam variables. Since the baryon masses are non-zero in the chiral limit, the chiral expansion of the amplitudes in the vicinity of threshold can be organized in powers of the following quantities,
\begin{align}
  \frac{s - m_{\psi_1/\psi_2}^2}{\Lambda_\chi^2} \sim \frac{u - m_{\psi_1/\psi_2}^2}{\Lambda_\chi^2} \sim \frac{m_{\phi_1/\phi_2}}{\Lambda_\chi} \ll 1, \quad \frac{t}{\Lambda_\chi^2} \ll 1,
\end{align}
with $\Lambda_\chi$ denoting the chiral symmetry breaking scale. Accordingly, the power counting rules for those parameters are set as
\begin{align}
  & m_{\psi_1/\psi_2} \sim \mathcal{O}(p^0), \quad m_{\phi_1/\phi_2} \sim \mathcal{O}(p^1), \quad s - m_{\psi_1/\psi_2}^2 \sim \mathcal{O}(p^1), \notag \\
  & u - m_{\psi_1/\psi_2}^2 \sim \mathcal{O}(p^1), \quad t \sim \mathcal{O}(p^2),\label{eq:3.2}
\end{align}
where $p$ is a collective symbol representing the small parameters.

An important feature in BChPT is that each Feynman diagram under consideration is characterized by a chiral dimension $D$. Namely, the importance of the diagram is regarded to be the order of $(p/\Lambda_\chi )^{D}$. In our case of one-baryon sector, the chiral dimension $D$ for a given diagram can be determined by the naive power counting rule,
\begin{align}\label{chiraldim}
  D = 4L + \sum_{n} nV_n - 2I_{\phi} - I_{\psi}\ ,
\end{align}
with $L$ the number of loops, $V_n$ the number of the $n^{th}$-order vertices, $I_{\phi}$ the number of internal pion lines, and $I_{\psi}$ the number of internal doubly charmed baryon lines. 

However, there exist pieces in the loop amplitudes, originating from the diagrams with internal baryon lines, which violates the above power counting rule~\eqref{chiraldim}. These pieces are known as PCB terms, as mentioned in the Introduction. The emergence of PCB terms is due to the fact that the masses of doubly charmed baryons do not vanish in the chiral limit, as pointed out by Ref.~\cite{Gasser:1987rb}. We will address this issue by using the EOMS scheme in section~\ref{sec:4}.

\subsection{Chiral effective Lagrangian}\label{sec:3.2}
The chiral effective Lagrangian, which is relevant to our calculation of the meson-baryon scattering amplitude up to $\mathcal{O}(p^3)$, takes the form
\begin{eqnarray}
\mathscr{L}_{\rm eff} = \sum_{i=1}^{2} \mathscr{L}_{\phi\phi}^{(2i)} + \sum_{j=1}^{3} \mathscr{L}_{\psi \phi}^{(j)},
\end{eqnarray}
with the superscripts \lq$2i$' and \lq $j$' in the brackets representing the chiral dimensions. For the purely mesonic sector, we need the following terms~\cite{Gasser:1984gg}
\begin{align}
  \mathscr{L}_{\phi\phi}^{(2)} &= \frac{F^2}{4} \left\langle \partial_{\mu} U \partial^\mu U^\dagger \right\rangle + \frac{F^2}{4} \left\langle \chi U^\dagger + U \chi^\dagger  \right\rangle \ , \\
  \mathscr{L}_{\phi\phi}^{(4)} &= L_4 \left\langle \partial_\mu U \partial^\mu U^\dagger \right\rangle\left\langle \chi U^\dagger + U \chi^\dagger \right\rangle +L_5 \left\langle \partial_\mu U \partial^\mu U^\dagger \left(\chi U^\dagger + U \chi^\dagger \right) \right\rangle + L_6 \left\langle \chi U^\dagger +  U \chi^\dagger \right\rangle ^2  \notag \\
  &+ L_8 \left\langle U \chi^\dagger U \chi^\dagger + \chi U^\dagger \chi U^\dagger \right\rangle  +\cdots \ ,
\end{align}
where $\left\langle \cdots \right\rangle$ stands for the trace in the flavor space, $F$ is the pion decay constant in the SU(3) chiral limit~\cite{Oller:2006yh,book:789407}, and $L_{k}$ ($k=4, 5, 6, 8$) denote the mesonic  LECs. 
The NGB fields are collected in $U$, which reads
\begin{gather}
U = \exp\left(i\sqrt{2}\phi/F\right)\ ,\quad
  \phi=
\begin{pmatrix} 
  \frac{1}{\sqrt{2}}\pi^0 + \frac{1}{\sqrt{6}}\eta & \pi^+ & K^+ \\
  \pi^- & -\frac{1}{\sqrt{2}}\pi^0 + \frac{1}{\sqrt{6}}\eta & K^0 \\
  K^- & \bar{K}^0 & -\frac{2}{\sqrt{6}}\eta
\end{pmatrix}.
\end{gather}
Furthermore, in the isospin limit $m_u=m_d=\hat{m}$, the chiral operator $\chi$ can be written as 
\begin{eqnarray}\label{eq:GBmass}
\chi = 2B_{0}s = {\rm diag}\left(m_\pi^2,m_\pi^2,2m_K^2-m_\pi^2\right),
\end{eqnarray}
where $B_0 = -\left\langle 0 | \bar{q}q|0 \right\rangle /3F^2$, and $\left\langle 0 | \bar{q}q|0 \right\rangle$ is the quark condensate in the chiral limit~\cite{Gasser:1983yg}. 

For the baryonic sector, we proceed with the SU(3) triplet $\psi$ in which the doubly charmed baryons are contained. The physical states $\Xi_{cc}^{++}$, $\Xi_{cc}^{+}$ and $\Omega_{cc}^{+}$ form the baryon triplet $\psi$, which reads
\begin{gather}
\psi=
\begin{pmatrix} 
\Xi_{cc}^{++} \\
\Xi_{cc}^{+}   \\
\Omega_{cc}^{+} \\
\end{pmatrix}.
\end{gather}
The full set of the chiral operators up to and including $\mathcal{O}(p^4)$, describing the interactions between Goldstone bosons and doubly charmed baryons, is constructed in Ref.~\cite{Qiu:2020omj}. In our current case, the required terms are given as follows:
\begin{align}  
\label{op1baryon} \mathscr{L}_{\psi \phi}^{(1)} &= \bar{\psi}\left(i\slashed{D} - m\right)\psi + \frac{g}{2}\bar{\psi}\slashed{u}\gamma_{5}\psi \ , \\
\label{op2baryon} \mathscr{L}_{\psi \phi}^{(2)} &= b_{1}\bar{\psi} \left \langle \chi_{+} \right \rangle \psi + b_{2}\bar{\psi} \widetilde{\chi}_{+} \psi + b_{3}\bar{\psi} u^{2} \psi + b_{4}\bar{\psi} \left \langle u^{2}\right \rangle \psi + \frac{b_{5}}{m^2}\bar{\psi} \left(\{u^{\mu},u^{\nu}\}D_{\mu\nu} + H.c.\right) \psi \notag \\
&+ \frac{b_{6}}{m^2} \bar{\psi} \left(\left \langle u^{\mu}u^{\nu} \right \rangle D_{\mu\nu} + H.c.\right) \psi + i b_{7}\bar{\psi} \left[u^{\mu},u^{\nu} \right]\sigma_{\mu\nu} \psi \ ,  \\
\label{op3baryon} \mathscr{L}_{\psi \phi}^{(3)} &= ic_{11}\bar{\psi} \left[u_{\mu},h^{\mu \nu} \right]\gamma_{\nu} \psi + \frac{c_{12}}{m^2}\bar{\psi} \left(i\left[u^{\mu},h^{\nu\rho}\right]\gamma_{\mu} D_{\nu\rho} + H.c.\right) \psi + \frac{c_{13}}{m}\bar{\psi} \left( i\left\{u^{\mu},h^{\nu\rho}\right\}\sigma_{\mu\nu}D_{\rho} \right.  \notag \\
& \left. + H.c.\right)\psi + \frac{c_{14}}{m}\bar{\psi} \left(i\sigma_{\mu\nu} \left \langle u^{\mu}h^{\nu\rho} \right \rangle D_{\rho} + H.c.\right) \psi +  c_{15}\bar{\psi} \left\{u^{\mu},\widetilde{\chi}_{+}\right\}\gamma_{5}\gamma_{\mu} \psi  \notag \\
&+ c_{16}\bar{\psi} u^{\mu}\gamma_{5}\gamma_{\mu} \left \langle  \chi_{+} \right \rangle \psi + c_{17}\bar{\psi}\gamma_{5}\gamma_{\mu} \left \langle u^{\mu}\widetilde{\chi}_{+}\right \rangle \psi + ic_{18}\bar{\psi} \gamma_{5}\gamma_{\mu}\left[D^{\mu},\widetilde{\chi}_{-}\right] \psi \notag \\
&+ ic_{19}\bar{\psi} \gamma_{5}\gamma_{\mu}\left \langle \left[D^{\mu},\chi_{-} \right]\right \rangle \psi + c_{20}\bar{\psi}\left[\widetilde{\chi}_{-},u^{\mu}\right]\gamma_{\mu} \psi \ ,
\end{align}
where 
\begin{align}
  D_{\mu} &= \partial_{\mu} + \Gamma_{\mu}, & \Gamma_{\mu} &= \frac{1}{2}\left[u^{\dagger}\partial_{\mu}u + u\partial_{\mu}u^{\dagger}\right], \notag \\
  u_{\mu} &= i\left[u^{\dagger}\partial_{\mu}u - u\partial_{\mu}u^{\dagger}\right], & u &= U^{\frac{1}{2}}, \notag\\
  \chi_{\pm } &= u^\dagger \chi u^\dagger \pm u \chi^\dagger u, & \widetilde{\chi}_{\pm } &= \chi_{\pm} - \frac{1}{3} \left\langle \chi_{\pm} \right\rangle, \notag \\
  h_{\mu\nu} &= D_{\mu}u_{\nu} + D_{\nu}u_{\mu} ,   & D_{\mu\nu} &=\{ D_{\mu} , D_{\nu} \} \ .\label{eq:blocks}
\end{align}
Here, $m$ is the baryon mass in the chiral limit and $g$ denotes the bare axial-vector coupling constant. The baryonic LECs $b_{j}$ ($j=1,\cdots,7$) and $c_{k}$ ($k=11,\cdots,20$) are unknown parameters in units of GeV$^{-1}$ and GeV$^{-2}$, respectively. $H.c.$ stands for terms obtained by hermitian conjugation.


\subsection{Tree amplitudes\label{sec:3.3}}
\begin{figure}[!htb] 
\centering
\includegraphics[width=0.75\textwidth]{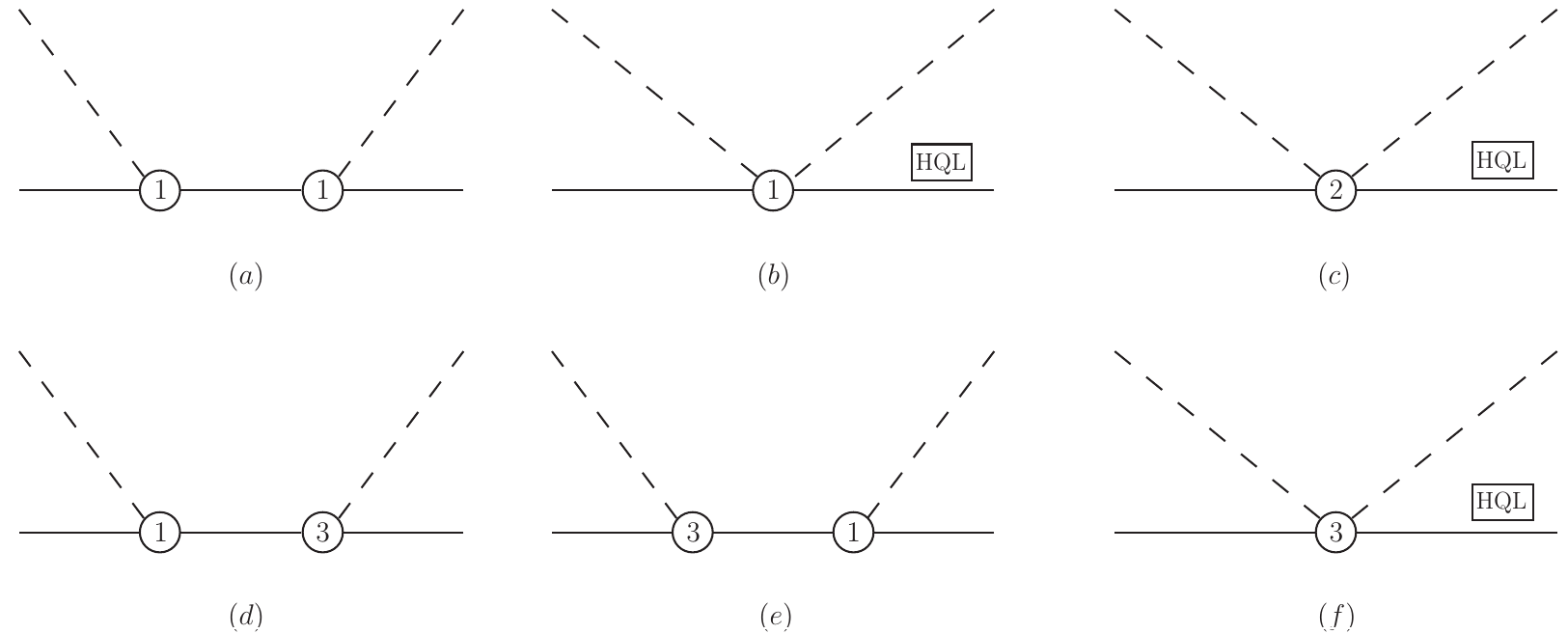}
\caption{Tree diagrams for $\psi$$\phi$ scattering up to $\mathcal{O}(p^3)$. The solid and dashed lines stand for doubly charmed baryons and NGBs, respectively. The relevant crossed diagrams are not shown explicitly. Diagrams surviving in the HQL are indicated by boxed \lq\lq HQL".}
\label{treediagram}
\end{figure}

According to the aforementioned power counting rule, the tree-level Feynman diagrams contributing to the meson-baryon scattering amplitude up to $\mathcal{O}(p^3)$ are shown in Figure~\ref{treediagram}. Note that the corresponding crossed diagrams are not displayed in Figure~\ref{treediagram}.

The LO tree-level amplitude takes the following form
\begin{align}
 A_{\rm tree}^{(1)}&=\frac{g^2}{8F^2}\left[\mathcal{C}_S^{(1)}\mathcal{F}(s) +\mathcal{C}_U^{(1)}\mathcal{F}(u)\right]\ ,\notag\\
 B_{\rm tree}^{(1)}&= \frac{\mathcal{C}_{\rm WT}^{(1)}}{4F^2}-
 \frac{g^2}{4F^2}\left[\mathcal{C}_S^{(1)}
\mathcal{G}(s)-
 \mathcal{C}_U^{(1)}\mathcal{G}(u)
 \right] \ ,\label{o1tofT}
\end{align}
with the functions $\mathcal{F}$ and $\mathcal{G}$ defined by
\begin{align}
\mathcal{F}(s)&=
\frac{(m_{\psi_1}+m_{\psi_2})(s-m_{\psi_1}m_{\psi_2})+m(2s-m_{\psi_1}^2-m_{\psi_2}^2)}{s-m^2}\ ,\notag\\
\mathcal{G}(s)&= \frac{s+m_{\psi_1}m_{\psi_2}+m(m_{\psi_1}+m_{\psi_2})}{s-m^2}  \ ,
\end{align}
where $m$ is the mass of the intermediate doubly charmed baryons in the chiral limit. All the involved coefficients are given in Table~\ref{tab:op1coeff}. The Weinberg-Tomazawa (WT) term, accompanied by the coefficient $\mathcal{C}_{\rm WT}^{(1)}$, stems from diagram ($b$) in Figure~\ref{treediagram}. The $s$-channel diagram ($a$) gives the terms proportional to $C_S^{(1)}$, while its crossed partner yields the pieces with coefficient $C_U^{(1)}$.

\begin{table*}[htbp]
  \centering
  \caption{\label{tab:op1coeff} Coefficients appearing in the tree amplitudes of $\mathcal{O}(p^1)$. The exchanged doubly charmed baryons are indicated in the square brackets. The scattering processes are categorized by strangeness $S$ and isospin $I$.}
  {
  \begin{tabular}{c c | c c c}
  \hline
  \hline
   $(S, I)$ & Processes & $\mathcal{C}^{(1)}_{\rm WT}$ & $\mathcal{C}^{(1)}_S$ & $\mathcal{C}^{(1)}_U$ \\
  \hline
  $(-2, \frac{1}{2})$ & $\Omega_{cc}\bar{K}\rightarrow \Omega_{cc}\bar{K}$ &$-2$  & $0$  & $2$ $[\Xi_{cc}]$ \\
  \hline
  $(1, 1)$ & $\Xi_{cc}K\rightarrow \Xi_{cc}K$ 
  & $-2$ & $0$  & $2$ $[\Omega_{cc}]$ \\
  \hline
  $(1, 0)$ & $\Xi_{cc}K\rightarrow \Xi_{cc}K$ 
  & $2$ & $0$  & $-2$ $[\Omega_{cc}]$ \\
  \hline
  $(0, \frac{3}{2})$ & $\Xi_{cc}\pi \rightarrow \Xi_{cc}\pi$ & $-2$  & $0$  & $2$ $[\Xi_{cc}]$ \\
  \hline
  $(-1, 0)$ & $\Xi_{cc}\bar{K} \rightarrow \Xi_{cc}\bar{K}$ & $4$  & $4$ $[\Omega_{cc}]$ & $0$  \\
  $       $ & $\Omega_{cc} \eta \rightarrow \Omega_{cc} \eta$ & $0$  & $\frac{4}{3}$ $[\Omega_{cc}]$ & $\frac{4}{3}$ $[\Omega_{cc}]$ \\
  $       $ & $\Xi_{cc}\bar{K}\rightarrow \Omega_{cc}\eta$  & $-2\sqrt{3}$ 
  & $-\frac{4}{\sqrt{3}}$ $[\Omega_{cc}]$  & $\frac{2}{\sqrt{3}}$ $[\Xi_{cc}]$ \\
  \hline
  $(-1, 1)$ & $\Omega_{cc} \pi \rightarrow \Omega_{cc} \pi$  & $0$ & $0$ & $0$  \\
  $       $ & $\Xi_{cc}\bar{K} \rightarrow \Xi_{cc}\bar{K}$  & $0$ & $0$ & $0$  \\
  $       $ & $\Omega_{cc} \pi \rightarrow \Xi_{cc}\bar{K}$  & $-2$ & $0$   & $2$ $[\Xi_{cc}]$ \\
  \hline
  $(0, \frac{1}{2})$ & $\Xi_{cc}\pi\rightarrow \Xi_{cc}\pi$  & $4$  & $3$ $[\Xi_{cc}]$   & $-1$ $[\Xi_{cc}]$ \\
  $      $ & $\Xi_{cc}\eta\rightarrow \Xi_{cc}\eta$ & $0$  & $\frac{1}{3}$ $[\Xi_{cc}]$  & $\frac{1}{3}$  $[\Xi_{cc}]$ \\
  $      $ & $\Omega_{cc}K\rightarrow \Omega_{cc}K$  & $2$ & $2$ $[\Xi_{cc}]$ & $0$  \\
  $      $ & $\Xi_{cc}\pi\rightarrow \Xi_{cc}\eta$  & $0$ & $1$ $[\Xi_{cc}]$ & $1$
  $[\Xi_{cc}]$ \\
  $      $ & $\Xi_{cc}\pi\rightarrow \Omega_{cc}K$  & $\sqrt{6}$  & $\sqrt{6}$ $[\Xi_{cc}]$ & $0$ \\
  $      $ & $\Xi_{cc}\eta\rightarrow \Omega_{cc}K$  & $\sqrt{6}$  
  & $\frac{\sqrt{6}}{3}$ $[\Xi_{cc}]$   & $-\frac{2\sqrt{6}}{3}$ $[\Omega_{cc}]$\\
  \hline
  \hline
  \end{tabular}
  }
\end{table*}

The $\mathcal{O}(p^2)$ meson-baryon scattering amplitude reads
\begin{align}
A_{\rm tree}^{(2)}&=\frac{\mathcal{C}_1^{(2)}}{6F^2}
+\mathcal{C}_2^{(2)}\frac{m_{\phi_1}^2+m_{\phi_2}^2-t}{2F^2}
+
\frac{\mathcal{C}_3^{(2)}}{2 m^2F^2}\mathcal{H}(s,t)
+\mathcal{C}_4^{(2)}\frac{u-s}{F^2}  \ ,\notag\\
B_{\rm tree}^{(2)}&=\mathcal{C}_4^{(2)}\frac{2(m_{\psi_1}+m_{\psi_2})}{F^2} \ ,\label{o2tofT}
\end{align}
with
\begin{align}
    \mathcal{H}(s,t)=2su-(s+u)\Sigma+m_{\psi_1}^4+m_{\psi_2}^4+2m_{\phi_1}^2m_{\phi_2}^2+(m_{\psi_1}^2+m_{\psi_2}^2)(m_{\phi_1}^2+m_{\phi_2}^2) \ .
\end{align}
The coefficients $\mathcal{C}^{(2)}_{i}$ ($i=1,\cdots,4$) are compiled in Table~\ref{tab:op2coeff}. All of these coefficients are obtained from diagram ($c$) of Figure~\ref{treediagram}.

\begin{table*}[htbp]
  \centering
  \renewcommand{\tabcolsep}{0.3pc}
  \caption{\label{tab:op2coeff} Coefficients in the tree amplitudes of $\mathcal{O}(p^2)$.}
  {\small
  \begin{tabular}{c c | c c c c}
  \hline
  \hline
   $(S, I)$ & Processes & $\mathcal{C}^{(2)}_{1}$ & $\mathcal{C}^{(2)}_{2}$ &$\mathcal{C}^{(2)}_{3}$ & $\mathcal{C}^{(2)}_{4}$  \\
  \hline
  $(-2, \frac{1}{2})$ & $\Omega_{cc}\bar{K}\rightarrow \Omega_{cc}\bar{K}$ & $-4(6b_1+b_2)m_K^2$  & $2(b_3+2b_4)$  & $4(b_5+b_6)$ & $-2b_7$   \\
  \hline
  $(1, 1)$ & $\Xi_{cc}K\rightarrow \Xi_{cc}K$ & $-4(6b_1+b_2)m_K^2$  & $2(b_3+2b_4)$  & $4(b_5+b_6)$ & $-2b_7$ \\
  \hline
  $(1, 0)$ & $\Xi_{cc}K\rightarrow \Xi_{cc}K$ & $-4(6b_1-5b_2)m_K^2$  & $-2(b_3-2b_4)$  & $-4(b_5-b_6)$  & $2b_7$ \\
  \hline
  $(0, \frac{3}{2})$ & $\Xi_{cc}\pi \rightarrow \Xi_{cc}\pi$ & $-4(6b_1+b_2)m_{\pi}^2$  & $2(b_3+2b_4)$  & $4(b_5+b_6)$   & $-2b_7$ \\
  \hline
  $(-1, 0)$ & $\Xi_{cc}\bar{K} \rightarrow \Xi_{cc}\bar{K}$  & $-8(3b_1+2b_2)m_K^2$  & $4(b_3+b_4)$ & $4(2b_5+b_6)$  & $4b_7$ \\
  $       $ & $\Omega_{cc} \eta \rightarrow \Omega_{cc} \eta$ 
  & $-\frac{32}{3}(3b_1+2b_2)m_K^2+(8b_1+\frac{40}{3}b_2)m_{\pi}^2$  
  & $\frac{4}{3}(2b_3+3b_4)$  & $\frac{4}{3}(4b_5+3b_6)$  & $0$  \\
  $       $ & $\Xi_{cc}\bar{K}\rightarrow \Omega_{cc}\eta$ &$2\sqrt{3}b_2(5m_K^2-3m_{\pi}^2)$  & $-\frac{2}{\sqrt{3}}b_3$ & $-\frac{4}{\sqrt{3}}b_5$  & $-2\sqrt{3}b_7$ \\
  \hline
  $(-1, 1)$ & $\Omega_{cc} \pi \rightarrow \Omega_{cc} \pi$ 
  & $-8(3b_1-b_2)m_{\pi}^2$  & $4b_4$  & $4b_6$ &$0$  \\
  $       $ & $\Xi_{cc}\bar{K} \rightarrow \Xi_{cc}\bar{K}$ 
  & $-8(3b_1-b_2)m_K^2$ & $4b_4$   &$4b_6$  &$0$  \\
  $       $ & $\Omega_{cc} \pi \rightarrow \Xi_{cc}\bar{K}$ 
  & $-6b_2(m_K^2+m_{\pi}^2)$ & $2b_3$  & $4b_5$  &$-2b_7$  \\
  \hline
  $(0, \frac{1}{2})$ & $\Xi_{cc}\pi\rightarrow \Xi_{cc}\pi$ 
  & $-4(6b_1+b_2)m_{\pi}^2$  & $2(b_3+2b_4)$ & $4(b_5+b_6)$  & $4b_7$ \\
  $      $ & $\Xi_{cc}\eta\rightarrow \Xi_{cc}\eta$ 
  & $-\frac{32}{3}(3b_1-b_2)m_K^2+(8b_1-\frac{20}{3}b_2)m_{\pi}^2$   & $\frac{2}{3}(b_3+6b_4)$ & $\frac{4}{3}(b_5+3b_6)$ & $0$ \\
  $      $ & $\Omega_{cc}K\rightarrow \Omega_{cc}K$ 
  & $-4(6b_1+b_2)m_K^2$ & $2(b_3+2b_4)$ & $4(b_5+b_6)$  & $2b_7$  \\
  $      $ & $\Xi_{cc}\pi\rightarrow \Xi_{cc}\eta$ 
  & $-12b_2m_{\pi}^2$  & $2b_3$  & $4b_5$  & $0$  \\
  $      $ & $\Xi_{cc}\pi\rightarrow \Omega_{cc}K$ 
  & $-3\sqrt{6}b_2(m_{K}^2+m_{\pi}^2)$ & $\sqrt{6}b_3$ & $2\sqrt{6}b_5$  &$\sqrt{6}b_7$ \\
  $      $ & $\Xi_{cc}\eta\rightarrow \Omega_{cc}K$ 
   & $\sqrt{6}b_2(5m_{K}^2-3m_{\pi}^2)$ & $-\frac{\sqrt{6}}{3}b_3$ & $-\frac{2\sqrt{6}}{3}b_5$  &$\sqrt{6}b_7$ \\
  \hline
  \hline
  \end{tabular}
  }
\end{table*}

The tree-level amplitude at $\mathcal{O}(p^3)$ can be written as
\begin{align}
A_{\rm tree}^{(3)}
&=\mathcal{C}_2^{(3)}
\frac{(m_{\psi_2}-m_{\psi_1})}{m^2F^2}\bigg[(m_{\phi_1}^2-m_{\phi_2}^2)t+(m_{\psi_2}^2-m_{\psi_1}^2)(s-u)\bigg]\notag\\
&+\mathcal{C}_3^{(3)}\frac{(s-u)^2}{mF^2}
+(\mathcal{C}_4^{(3)}+\mathcal{C}_5^{(3)})\frac{m_{\psi_2}-m_{\psi_1}}{2F^2}\notag \\
&+\frac{g}{4F^2}\bigg[(\mathcal{C}_6^{(3)}+\mathcal{C}_8^{(3)})\mathcal{F}(s)+(\mathcal{C}_7^{(3)}+\mathcal{C}_9^{(3)})\mathcal{F}(u)\bigg] \ , \notag\\
B_{\rm tree}^{(3)}
&=\mathcal{C}_1^{(3)}\frac{2(m_{\phi_1}^2+m_{\phi_2}^2-t)}{F^2}
-\frac{2\,\mathcal{C}_2^{(3)}}{m^2F^2}\bigg[\mathcal{H}(s,t)+(s-u)^2+(m_{\phi_1}^2-m_{\phi_2}^2)^2\bigg] 
\notag\\
&-\mathcal{C}_3^{(3)}
\frac{2(s-u)(m_{\psi_1}+m_{\psi_2})}{mF^2}
-\frac{2(\mathcal{C}_4^{(3)}-\mathcal{C}_5^{(3)})}{F^2}
\notag \\
&-\frac{g}{2F^2}\bigg[(\mathcal{C}_6^{(3)}+\mathcal{C}_8^{(3)})\mathcal{G}(s)-(\mathcal{C}_7^{(3)}+\mathcal{C}_9^{(3)})\mathcal{G}(u)\bigg] \ .\label{o3tofT}
\end{align}
The coefficients are shown in Tables~\ref{tab:op3coeff1},~\ref{tab:op3coeff2} and~\ref{tab:op3coeff3}. The $\mathcal{C}^{(3)}_{i}$ ($i=1,\cdots,5$) are obtained from diagram ($f$), which corresponds to the contact contribution of $\mathcal{O}(p^3)$. The $s$-channel exchange diagrams, ($d$) and ($e$), generate the terms with coefficients $\mathcal{C}^{(3)}_{6}$ and $\mathcal{C}^{(3)}_{8}$, and their crossed diagrams are responsible for the contributions with coefficients $\mathcal{C}^{(3)}_{7}$ and $\mathcal{C}^{(3)}_{9}$.

\begin{table*}[htbp]
\centering
\renewcommand{\tabcolsep}{0.3pc}
  \caption{\label{tab:op3coeff1} Coefficients in the tree amplitudes of $\mathcal{O}(p^3)$.}
 {\footnotesize
  \begin{tabular}{c c | c c c c c}
  \hline
  \hline
   $(S, I)$ & Processes 
   &$\mathcal{C}^{(3)}_{1}$ 
   &$\mathcal{C}^{(3)}_{2}$ &$\mathcal{C}^{(3)}_{3}$ 
   &$\mathcal{C}^{(3)}_{4}$ 
   &$\mathcal{C}^{(3)}_{5}$
  \\
  \hline
  $(-2, \frac{1}{2})$ & $\Omega_{cc}\bar{K}\rightarrow \Omega_{cc}\bar{K}$ 
  & $-2c_{11}$  & $-2c_{12}$ & $2(c_{13}+c_{14})$    & $2c_{20}m_K^2$  & $-2c_{20}m_K^2$ \\
  \hline
  $(1, 1)$ & $\Xi_{cc}K\rightarrow \Xi_{cc}K$ 
  & $-2c_{11}$  & $-2c_{12}$  & $2(c_{13}+c_{14})$  & $2c_{20}m_K^2$  & $-2c_{20}m_K^2$ \\
  \hline
  $(1, 0)$ & $\Xi_{cc}K\rightarrow \Xi_{cc}K$ 
  & $2c_{11}$   & $2c_{12}$   & $-2(c_{13}-c_{14})$  & $-2c_{20}m_K^2$  & $2c_{20}m_K^2$ \\
  \hline
  $(0, \frac{3}{2})$ & $\Xi_{cc}\pi \rightarrow \Xi_{cc}\pi$ 
  & $-2c_{11}$  & $-2c_{12}$  & $2(c_{13}+c_{14})$  & $2c_{20}m_{\pi}^2$   &$-2c_{20}m_{\pi}^2$ \\
  \hline
  $(-1, 0)$ & $\Xi_{cc}\bar{K} \rightarrow \Xi_{cc}\bar{K}$  
  & $4c_{11}$   & $4c_{12}$   & $2(2c_{13}+c_{14})$  & $-4c_{20}m_K^2$  & $4c_{20}m_K^2$ \\
  $       $ & $\Omega_{cc} \eta \rightarrow \Omega_{cc} \eta$ 
  & $0$  & $0$  & $\frac{2}{3}(4c_{13}+3c_{14})$ & $0$  &$0$ \\
  $       $ & $\Xi_{cc}\bar{K}\rightarrow \Omega_{cc}\eta$ 
  & $-2\sqrt{3}c_{11}$ & $-2\sqrt{3}c_{12}$ & $-\frac{2\sqrt{3}}{3}c_{13}$  & $\frac{1}{\sqrt{3}}c_{20}(5m_K^2+m_{\pi}^2)$
  & $-\sqrt{3}c_{20}(3m_K^2-m_{\pi}^2)$
  \\
  \hline
  $(-1, 1)$ & $\Omega_{cc} \pi \rightarrow \Omega_{cc} \pi$ 
  & $0$  & $0$  & $2c_{14}$ & $0$   & $0$ \\
  $       $ & $\Xi_{cc}\bar{K} \rightarrow \Xi_{cc}\bar{K}$ 
  & $0$  & $0$  & $2c_{14}$  & $0$  & $0$ \\
  $       $ & $\Omega_{cc} \pi \rightarrow \Xi_{cc}\bar{K}$ 
  & $-2c_{11}$  & $-2c_{12}$ & $2c_{13}$  &$-c_{20}(m_K^2-3m_{\pi}^2)$  & $-c_{20}(3m_K^2-m_{\pi}^2)$ \\
  \hline
  $(0, \frac{1}{2})$ & $\Xi_{cc}\pi\rightarrow \Xi_{cc}\pi$ 
  & $4c_{11}$  & $4c_{12}$ & $2(c_{13}+c_{14})$   & $-4c_{20}m_{\pi}^2$  & $4c_{20}m_{\pi}^2$\\
  $      $ & $\Xi_{cc}\eta\rightarrow \Xi_{cc}\eta$ 
  & $0$ & $0$ & $\frac{2}{3}c_{13}+2c_{14}$  & $0$ & $0$ \\
  $      $ & $\Omega_{cc}K\rightarrow \Omega_{cc}K$ 
  & $2c_{11}$  & $2c_{12}$  & $2(c_{13}+c_{14})$  & $-2c_{20}m_K^2$  & $2c_{20}m_K^2$ \\
  $      $ & $\Xi_{cc}\pi\rightarrow \Xi_{cc}\eta$ 
  & $0$  & $0$   & $2c_{13}$  & $0$  & $0$ \\
  $      $ & $\Xi_{cc}\pi\rightarrow \Omega_{cc}K$ 
  & $\sqrt{6}c_{11}$  & $\sqrt{6}c_{12}$ & $\sqrt{6}c_{13}$ & $\frac{\sqrt{6}}{2}c_{20}(m_K^2-3m_{\pi}^2)$ & $\frac{\sqrt{6}}{2}c_{20}(3m_{K}^2-m_{\pi}^2)$ \\
  $      $ & $\Xi_{cc}\eta\rightarrow \Omega_{cc}K$ 
  & $\sqrt{6}c_{11}$  & $\sqrt{6}c_{12}$ & $-\frac{\sqrt{6}}{3}c_{13}$ & $-\frac{\sqrt{6}}{2}c_{20}(3m_K^2-m_{\pi}^2)$ 
  & $\frac{1}{\sqrt{6}}c_{20}(5m_K^2+m_{\pi}^2)$
  \\
  \hline
  \hline
  \end{tabular}
  }
\end{table*}

\begin{sidewaystable*}[htbp]
  \centering
  \caption{\label{tab:op3coeff2} Coefficients in the tree amplitudes of $\mathcal{O}(p^3)$. The exchanged doubly charmed baryons are indicated in the square brackets. The combination $\mathcal{C}^{(3)}_{8-6}=\mathcal{C}^{(3)}_{8}-\mathcal{C}^{(3)}_{6}$ is used for brevity.
  }
  {
  \begin{tabular}{c c | c c}
  \hline
  \hline
   $(S, I)$ & Processes 
   &$\mathcal{C}^{(3)}_{6}$ &$\mathcal{C}^{(3)}_{8-6}$
   \\
  \hline
  $(-2, \frac{1}{2})$ & $\Omega_{cc}\bar{K}\rightarrow \Omega_{cc}\bar{K}$ & $0$   & $0$ \\
  \hline
  $(1, 1)$ & $\Xi_{cc}K\rightarrow \Xi_{cc}K$ & $0$  & $0$ \\
  \hline
  $(1, 0)$ & $\Xi_{cc}K\rightarrow \Xi_{cc}K$ & $0$  & $0$ \\
  \hline
  $(0, \frac{3}{2})$ & $\Xi_{cc}\pi \rightarrow \Xi_{cc}\pi$ & $0$   & $0$ \\
  \hline
  $(-1, 0)$ & $\Xi_{cc}\bar{K} \rightarrow \Xi_{cc}\bar{K}$  & $-\frac{8}{3}(2c_{15}+6c_{16}-3c_{18})m_{K}^2+(\frac{16}{3}c_{15}-8c_{16})m_{\pi}^2$  [$\Omega_{cc}$]
  & $0$ \\
  $       $ & $\Omega_{cc} \eta \rightarrow \Omega_{cc} \eta$ 
  & $\frac{8}{9}(8c_{15}-3c_{16}+6c_{17}-c_{18}-6c_{19})m_{\pi}^2-\frac{16}{9}(4c_{15}+3c_{16}+3c_{17}-2c_{18}-3c_{19})m_K^2$  [$\Omega_{cc}$]  & $0$ 
  \\
  $       $ & $\Xi_{cc}\bar{K}\rightarrow \Omega_{cc}\eta$ 
  & $\frac{8\sqrt{3}}{9}((2c_{15}+6c_{16}-3c_{18})m_K^2-(2c_{15}-3c_{16})m_{\pi}^2)$   [$\Omega_{cc}$]
  & $\Delta_1$
  \\%
  \hline
  $(-1, 1)$ & $\Omega_{cc} \pi \rightarrow \Omega_{cc} \pi$ & $0$     & $0$ \\
  $       $ & $\Xi_{cc}\bar{K} \rightarrow \Xi_{cc}\bar{K}$ & $0$   & $0$ \\
  $       $ & $\Omega_{cc} \pi \rightarrow \Xi_{cc}\bar{K}$ & $0$  & $0$ \\
  \hline
  $(0, \frac{1}{2})$ & $\Xi_{cc}\pi\rightarrow \Xi_{cc}\pi$ 
  & $4(2c_{15}-3c_{16})m_K^2-2(4c_{15}+3c_{16}-3c_{18})m_{\pi}^2$ [$\Xi_{cc}$]  & $0$ \\
  $      $ & $\Xi_{cc}\eta\rightarrow \Xi_{cc}\eta$ 
  & $\frac{4}{9}(2c_{15}-3c_{16}+6c_{17}+2c_{18}-6c_{19})m_K^2-\frac{2}{9}(4c_{15}+3c_{16}+12c_{17}+c_{18}-12c_{19})m_{\pi}^2$ [$\Xi_{cc}$] & 0  \\
  $      $ & $\Omega_{cc}K\rightarrow \Omega_{cc}K$ 
  & $(-\frac{8}{3}c_{15}-8c_{16}+4c_{18})m_K^2+(\frac{8}{3}c_{15}-4c_{16})m_{\pi}^2$ [$\Xi_{cc}$] & 0   \\
  $      $ & $\Xi_{cc}\pi\rightarrow \Xi_{cc}\eta$ 
  & $(\frac{8}{3}c_{15}-4c_{16})m_K^2-(\frac{8}{3}c_{15}+2c_{16}-2c_{18})m_{\pi}^2$ [$\Xi_{cc}$] & $\Delta_2$  \\
  $      $ & $\Xi_{cc}\pi\rightarrow \Omega_{cc}K$ 
  & $\frac{4\sqrt{6}}{3}(2c_{15}-3c_{16})m_K^2-\frac{2\sqrt{6}}{3}(4c_{15}+3c_{16}-3c_{18})m_{\pi}^2$ [$\Xi_{cc}$]  & $\Delta_3$ \\
  $      $ & $\Xi_{cc}\eta\rightarrow \Omega_{cc}K$ 
  & $\frac{4\sqrt{6}}{9}(2c_{15}-3c_{16}+6c_{17}+2c_{18}-6c_{19})m_K^2-\frac{2\sqrt{6}}{9}(4c_{15}+3c_{16}+12c_{17}+c_{18}-12c_{19})m_{\pi}^2$ [$\Xi_{cc}$]
  & $\Delta_4$ \\
  \hline
 \multicolumn{2}{c}{\multirow{4}*{Abbreviations:}} & \multicolumn{2}{l}{ $\Delta_1= \frac{8}{3\sqrt{3}}(6c_{15}+6c_{17}-c_{18}-6c_{19})(m_K^2-m_{\pi}^2)$} 
\\
\multicolumn{2}{c}{} &\multicolumn{2}{l}{ 
  $\Delta_2=\frac{8}{3}(3c_{17}+c_{18}-3c_{19})(m_K^2-m_{\pi}^2)$}\\
\multicolumn{2}{c}{} &\multicolumn{2}{l}{ 
  $\Delta_3=-2\sqrt{6}(2c_{15}-c_{18})(m_K^2-m_{\pi}^2)$}\\
\multicolumn{2}{c}{} &\multicolumn{2}{l}{ 
  $\Delta_4=-\frac{2\sqrt{6}}{9}(6c_{15}+12c_{17}+c_{18}-12c_{19})(m_K^2-m_{\pi}^2)$}\\
  \hline
  \end{tabular}
  }
\end{sidewaystable*}

\begin{sidewaystable*}[htbp]
  \centering
  \caption{\label{tab:op3coeff3} Coefficients in the tree amplitudes of $\mathcal{O}(p^3)$. The exchanged doubly charmed baryons are indicated in the square brackets. The combination $\mathcal{C}^{(3)}_{9-7}=\mathcal{C}^{(3)}_{9}-\mathcal{C}^{(3)}_{7}$ is used for brevity.}
  {
  \begin{tabular}{c c | c c}
  \hline
  \hline
   $(S, I)$ & Processes 
   &$\mathcal{C}^{(3)}_{7}$ & $\mathcal{C}^{(3)}_{9-7}$
   \\
  \hline
  $(-2, \frac{1}{2})$ & $\Omega_{cc}\bar{K}\rightarrow \Omega_{cc}\bar{K}$ 
  & $(-\frac{8}{3}c_{15}-8c_{16}+4c_{18})m_K^2+(\frac{8}{3}c_{15}-4c_{16})m_{\pi}^2$ [$\Xi_{cc}$]  &$0$   \\
  \hline
  $(1, 1)$ & $\Xi_{cc}K\rightarrow \Xi_{cc}K$ & $(-\frac{8}{3}c_{15}-8c_{16}+4c_{18})m_K^2+(\frac{8}{3}c_{15}-4c_{16})m_{\pi}^2$ [$\Omega_{cc}$]  &$0$  \\
  \hline
  $(1, 0)$ & $\Xi_{cc}K\rightarrow \Xi_{cc}K$ 
  & $(\frac{8}{3}c_{15}+8c_{16}-4c_{18})m_K^2-(\frac{8}{3}c_{15}-4c_{16})m_{\pi}^2$ [$\Omega_{cc}$]  &$0$ \\
  \hline
  $(0, \frac{3}{2})$ & $\Xi_{cc}\pi \rightarrow \Xi_{cc}\pi$ & $(\frac{16}{3}c_{15}-8c_{16})m_K^2-(\frac{16}{3}c_{15}+4c_{16}-4c_{18})m_{\pi}^2$  [$\Xi_{cc}$] 
  &$0$
  \\
  \hline
  $(-1, 0)$ & $\Xi_{cc}\bar{K} \rightarrow \Xi_{cc}\bar{K}$  & $0$   &$0$ \\
  $       $ & $\Omega_{cc} \eta \rightarrow \Omega_{cc} \eta$ 
  & $-\frac{16}{9}(4c_{15}+3c_{16}+3c_{17}-2c_{18}-3c_{19})m_K^2+\frac{8}{9}(8c_{15}-3c_{16}+6c_{17}-c_{18}-6c_{19})m_{\pi}^2$  [$\Omega_{cc}$] 
  &$0$
  \\
  $       $ & $\Xi_{cc}\bar{K}\rightarrow \Omega_{cc}\eta$ & $\frac{4}{3\sqrt{3}}(2(2c_{15}-3c_{16}+6c_{17}+2c_{18}-6c_{19})m_K^2-(4c_{15}+3c_{16}+12c_{17}+c_{18}-12c_{19})m_{\pi}^2)$ [$\Xi_{cc}$]
  &$\Delta_1^\prime$
  \\
  \hline
  $(-1, 1)$ & $\Omega_{cc} \pi \rightarrow \Omega_{cc} \pi$ & $0$   & $0$  \\
  $       $ & $\Xi_{cc}\bar{K} \rightarrow \Xi_{cc}\bar{K}$ & $0$   & $0$  \\
  $       $ & $\Omega_{cc} \pi \rightarrow \Xi_{cc}\bar{K}$ & $(-\frac{8}{3}c_{15}-8c_{16}+4c_{18})m_K^2+(\frac{8}{3}c_{15}-4c_{16})m_{\pi}^2$ [$\Xi_{cc}$]  & $\Delta_2^\prime$ \\
  \hline
  $(0, \frac{1}{2})$ & $\Xi_{cc}\pi\rightarrow \Xi_{cc}\pi$ & $(-\frac{8}{3}c_{15}+4c_{16})m_K^2+(\frac{8}{3}c_{15}+2c_{16}-2c_{18})m_{\pi}^2$ [$\Xi_{cc}$]  & $0$\\
  $      $ & $\Xi_{cc}\eta\rightarrow \Xi_{cc}\eta$ 
  & $\frac{4}{9}(2c_{15}-3c_{16}+6c_{17}+2c_{18}-6c_{19})m_K^2-\frac{2}{9}(4c_{15}+3c_{16}+12c_{17}+c_{18}-12c_{19})m_{\pi}^2$ [$\Xi_{cc}$] & $0$ \\
  $      $ & $\Omega_{cc}K\rightarrow \Omega_{cc}K$ & $0$  & $0$  \\
  $      $ & $\Xi_{cc}\pi\rightarrow \Xi_{cc}\eta$ & $\frac{4}{3}(2c_{15}-3c_{16}+6c_{17}+2c_{18}-6c_{19})m_K^2-\frac{2}{3}(4c_{15}+3c_{16}+12c_{17}+c_{18}-12c_{19})m_{\pi}^2$ [$\Xi_{cc}$] & $\Delta_3^\prime$
  \\
  $      $ & $\Xi_{cc}\pi\rightarrow \Omega_{cc}K$ & $0$  & $0$ \\
  $      $ & $\Xi_{cc}\eta\rightarrow \Omega_{cc}K$ 
  & $\frac{4\sqrt{6}}{9}((2c_{15}+6c_{16}-3c_{18})m_K^2-(2c_{15}-3c_{16})m_{\pi}^2)$ [$\Omega_{cc}$]  & $\Delta_4^\prime$ \\
 \hline
 \multicolumn{2}{c}{\multirow{4}*{Abbreviations:}} 
 & \multicolumn{2}{l}{ $\Delta_1^\prime=-\frac{4}{3\sqrt{3}}(6c_{15}+12c_{17}+c_{18}-12c_{19})(m_K^2-m_{\pi}^2)$} 
\\
 \multicolumn{2}{c}{} &\multicolumn{2}{l}{ 
  $\Delta_2^\prime=4(2c_{15}-c_{18})(m_K^2-m_{\pi}^2)$}\\
  \multicolumn{2}{c}{}&\multicolumn{2}{l}{ 
  $\Delta_3^\prime=-\frac{8}{3}(3c_{17}+c_{18}-3c_{19})(m_K^2-m_{\pi}^2)$}\\
 \multicolumn{2}{c}{} &\multicolumn{2}{l}{ 
  $\Delta_4^\prime=\frac{4\sqrt{6}}{9}(6c_{15}+6c_{17}-c_{18}-6c_{19})(m_K^2-m_{\pi}^2)$}
  \\
  \hline
  \hline
  \end{tabular}
  }
\end{sidewaystable*}


\subsection{Leading one-loop contributions\label{sec:3.4}}
\begin{figure*}[!htb] 
\centering
\includegraphics[width=0.9\textwidth]{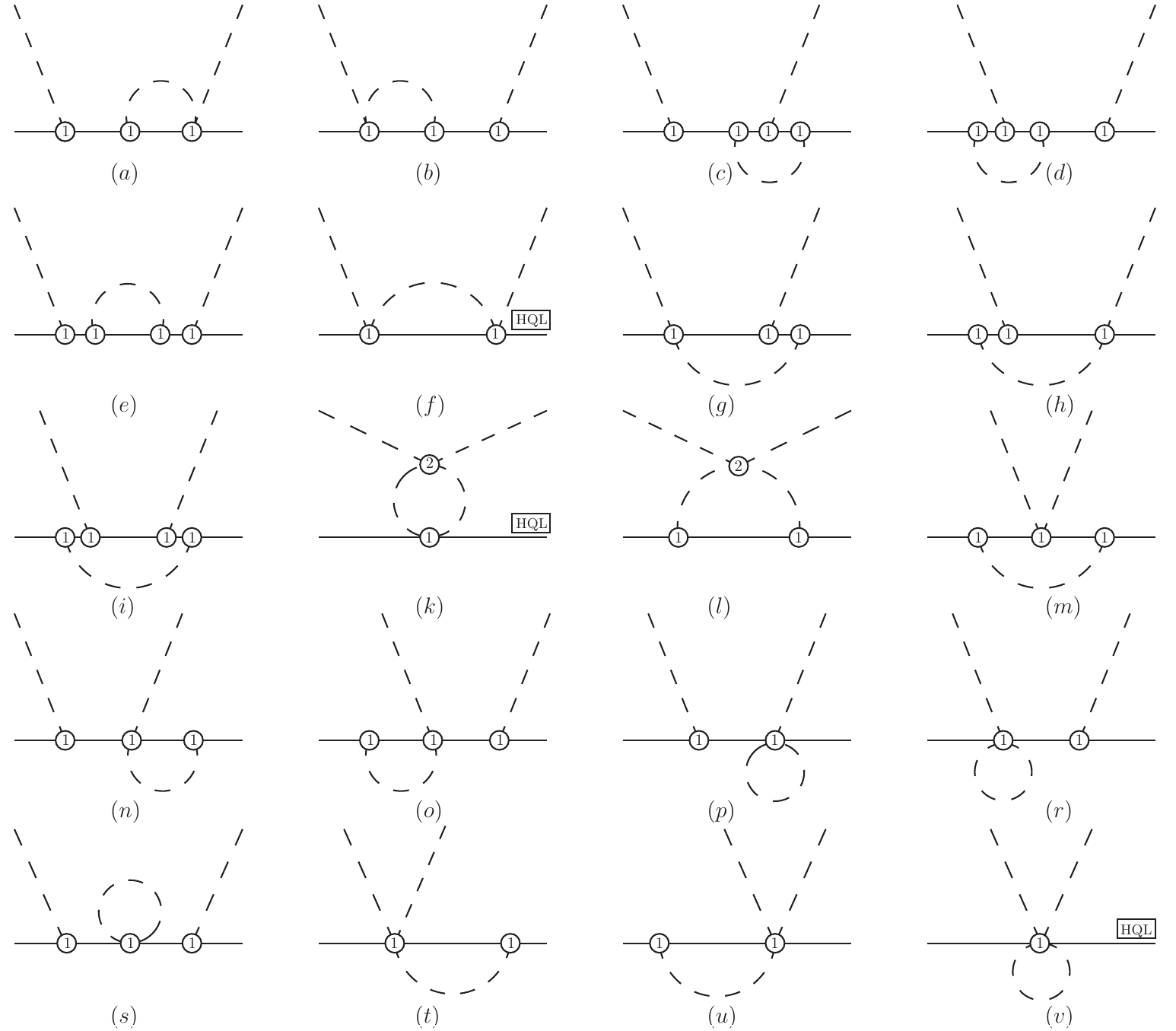} 
\caption{One-loop Feynman diagrams contributing to $\psi$$\phi$ scattering at $\mathcal{O}(p^3)$. Crossed diagrams are not shown. In addition, diagrams with loop corrections on external legs are taken into account via wave function renormalization. Diagrams surviving in the HQL are indicated by boxed \lq\lq HQL".}
\label{loopdiagram}
\end{figure*}
One-loop Feynman diagrams relevant to our calculation at $\mathcal{O}(p^3)$ are exhibited in Figure~\ref{loopdiagram}. Crossed diagrams are not shown. There are $34$ loop diagrams in total. We have calculated all of them. Explicit analytical expressions for the $16$ processes have been obtained. However, the expressions are too lengthy to be displayed here.\footnote{Explicit expressions of all the one-loop amplitudes are obtainable from the authors.} For the loop amplitudes, we do not need to distinguish the physical masses and the bare masses, since the caused difference is of higher order beyond our accuracy. Note that the contributions of diagrams corresponding to one-loop corrections on the external legs are incorporated via wave function renormalization, which will be discussed in the next section.  


\section{Renormalization\label{sec:4}}

\subsection{Masses and wave function renormalization constants\label{sec:4.1}}

\begin{figure}[!htb]
\centering
\includegraphics[width=0.6\textwidth]{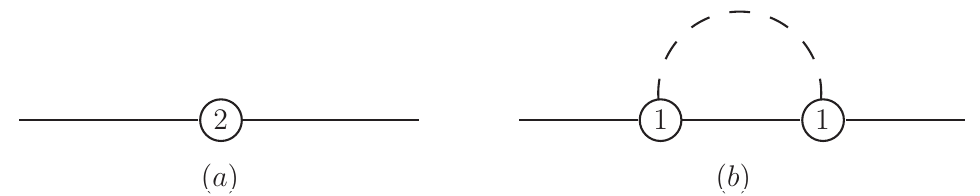}
\caption{Tree and one-loop self-energy diagrams up to $\mathcal{O}(p^3)$. }
\label{renoli-diagram}
\end{figure}

Let us begin with the baryonic sector. The dressed propagator of the doubly charmed baryons is defined as
\begin{align}\label{propa}
  iS_{\psi}(p) = \frac{i}{\slashed{p} - m - \Sigma_{\psi}(\slashed{p})}\ ,\quad \psi\in\{\Xi_{cc}^{++},\Xi_{cc}^{+},\Omega_{cc}^{+}\}\ ,
\end{align}
where $m$ denotes the bare baryon mass and $\Sigma_{\psi}(\slashed{p})$ refers to the baryon self-energy up to leading one-loop order. The sum of one-particle-irreducible diagrams contributing to the two-point function is denoted by $-i\Sigma_{\psi}(\slashed{p})$, which comprises contact and one-loop diagrams shown in Figure~\ref{renoli-diagram}. Namely, one has
\begin{align}\label{eq:SE.decom}
 -i\Sigma_{\psi}(\slashed{p})=-i\zeta_{\psi}^\dagger\big[\Sigma_{a}(\slashed{p})+\Sigma_{b}(\slashed{p})\big]\zeta_\psi \ ,
\end{align}
where $\zeta_{\psi}$ are unit vectors in the SU(3) flavor spaces, 
\begin{align}\label{eq:zeta}
\zeta_{\Xi^{++}_{cc}}=
\left(
\begin{array}{c}
    1  \\
    0  \\
    0 
\end{array}
\right) \ , 
\quad 
\zeta_{\Xi_{cc}^{+}}=
\left(
\begin{array}{c}
    0  \\
    1  \\
    0 
\end{array}
\right) \ , 
\quad 
\zeta_{\Omega^{+}_{cc}}=
\left(
\begin{array}{c}
    0  \\
    0  \\
    1 
\end{array}
\right) \ .
\end{align}
Furthermore, the chiral results for the self energies in Eq.~\eqref{eq:SE.decom} are given by
\begin{align}\label{selfenergy}
\Sigma_a^{ij}(\slashed{p}) &= -2\big[(b_1-\frac{1}{3}b_2)\langle \chi \rangle\delta^{ij}+b_2\chi^{ij}\big], \notag \\
\Sigma_b^{ij}(\slashed{p}) &= 
\frac{g^2}{8F^2 s}\lambda_c^{ik}\lambda_c^{kj}
\bigg\{
2s\, m_{\psi_k} \left[A_{0}(m_{\psi_k}^2)+ m_{\phi_c}^2 B_0(s,m_{\psi_k}^2,m_{\phi_c}^2) \right]
\notag \\
&
+\slashed{p}
\big[ (s-m_{\psi_k}^2)A_{0}(m_{\phi_c}^2)
+(s+m_{\psi_k}^2) A_{0}(m_{\psi_k}^2)
\notag \\
&
+[s(m_{\phi_c}^2-s)+m_{\psi_k}^2(2s+m_{\phi_c}^2)-m_{\psi_k}^4]B_{0}(s, m_{\psi_k}^2, m_{\phi_c}^2)
\big]
\bigg\} \ ,
\end{align}
with $i,j,k\in\{1,2,3\}$ and $c\in \{1,\cdots,8\}$. The definition of loop integrals $A_0$ and $B_0$ can be found in Appendix~\ref{app.wave}. The NGB mass matrix $\chi$ is defined in Eq.~\eqref{eq:GBmass} and $\lambda$'s are the standard Gell-mann matrices. Summation over repeated indices is implied. In addition, the masses of the intermediate states, showing up in the loop, are specified according to
\begin{align}
m_{\psi_k}=
\begin{cases}
    m_{\Xi_{cc}},    & k=1,2\\
    m_{\Omega_{cc}}, & k=3
\end{cases} \ ,
\quad 
m_{\phi_c}=
\begin{cases}
   m_{\pi} , & c=1,2,3 \\
   m_{K} ,   & c=4,\cdots,7 \\
   m_{\eta}, & c=8 
\end{cases}\ .
\end{align}
The pole position of the dressed propagator~\eqref{propa} defines the physical mass $m_\psi$ of the baryon. That is,
\begin{align}\label{eq:Bmass}
\big[\slashed{p}-m+\Sigma_{\psi}(\slashed{p})\big]_{\slashed{p}=m_\psi}=0\ .
\end{align}
With the help of Eq.~\eqref{eq:SE.decom} and Eq.~\eqref{selfenergy}, the physical masses of the doubly charmed baryons can be expressed as
\begin{align}
m_{\Xi_{cc}}&=m+\frac{4}{3}b_2(m_K^2-m_{\pi}^2)-2b_1(2m_K^2+m_{\pi}^2)+\delta m_{\Xi_{cc}}^{\rm loop}\ ,\label{eq:massXichpt}\\
m_{\Omega_{cc}}&=m-\frac{8}{3}b_2(m_K^2-m_{\pi}^2)-2b_1(2m_K^2+m_{\pi}^2)+\delta m_{\Omega_{cc}}^{\rm loop}\ ,\label{eq:massOmchpt}
\end{align}
where $\delta m_{\Xi_{cc}}^{\rm loop}$ and $\delta m_{\Omega_{cc}}^{\rm loop}$ are given in Eq.~\eqref{eq.mass.ren}.

The wave function renormalization constant is defined as the residue of the pole term of the dressed propagator,
\begin{align}
i S_{\psi} =\frac{i\mathcal{Z}_\psi}{\slashed{p}-m_{\psi}}+\text{non-pole pieces}\ ,
\end{align}
where $m_\psi$ is the physical baryon mass as specified in Eq.~\eqref{eq:Bmass}. In view of Eq.~\eqref{propa}, the wave function renormalization constant $\mathcal{Z}_\psi$ is given by 
\begin{align}
\mathcal{Z}_{\psi} = \frac{1}{1 - \Sigma_{\psi}^\prime(m_\psi)}\simeq 1+\Sigma_{\psi}^\prime(m_\psi)\ ,
\end{align}
where a prime means performing derivative with respect to $\slashed{p}$. Explicit expression of $\mathcal{Z}_\psi$ can be readily obtained by substituting Eq.~\eqref{selfenergy} into the above equation. The readers are referred to Appendix~\ref{app.wave} for the final results for $\mathcal{Z}_{\Xi_{cc}}$ and $\mathcal{Z}_{\Omega_{cc}}$.

Likewise, one can derive the wave function renormalization constants for the Goldstone bosons. Nevertheless, they have been extensively calculated elsewhere~\cite{Gasser:1984gg,GomezNicola:2001as,book:789407}. For completeness, we quote the results in the following
\begin{align}
\mathcal{Z}_{\pi}&=1-\frac{1}{3F^2}\bigg\{
24\left[2L_4 m_{K}^2+(L_4+L_5) m_{\pi}^2\right]
+A_{0}(m_{K}^2)
+2A_{0}(m_{\pi}^2)
\bigg\}
\ , \\
\mathcal{Z}_{K}&=1-\frac{1}{4F^2}
\bigg\{
32\left[(2L_4+L_5)m_{K}^2+L_4 m_{\pi}^2\right]
+A_{0}(m_{\eta}^2)+2A_{0}(m_{K}^2)+A_{0}(m_{\pi}^2)
\bigg\}
\ , \\
\mathcal{Z}_{\eta}&=1+\frac{1}{3F^2}
\bigg\{
8\left[
(L_5-3L_4)m_{\pi}^2
-(4L_5+6L_4)m_K^2
\right]
-3A_{0}(m_{K}^2)
\bigg\} \ .
\end{align}


\subsection{The full scattering amplitude within EOMS scheme\label{sec:4.2}}
\begin{figure}[!htb]
\centering
\includegraphics[width=0.95\textwidth]{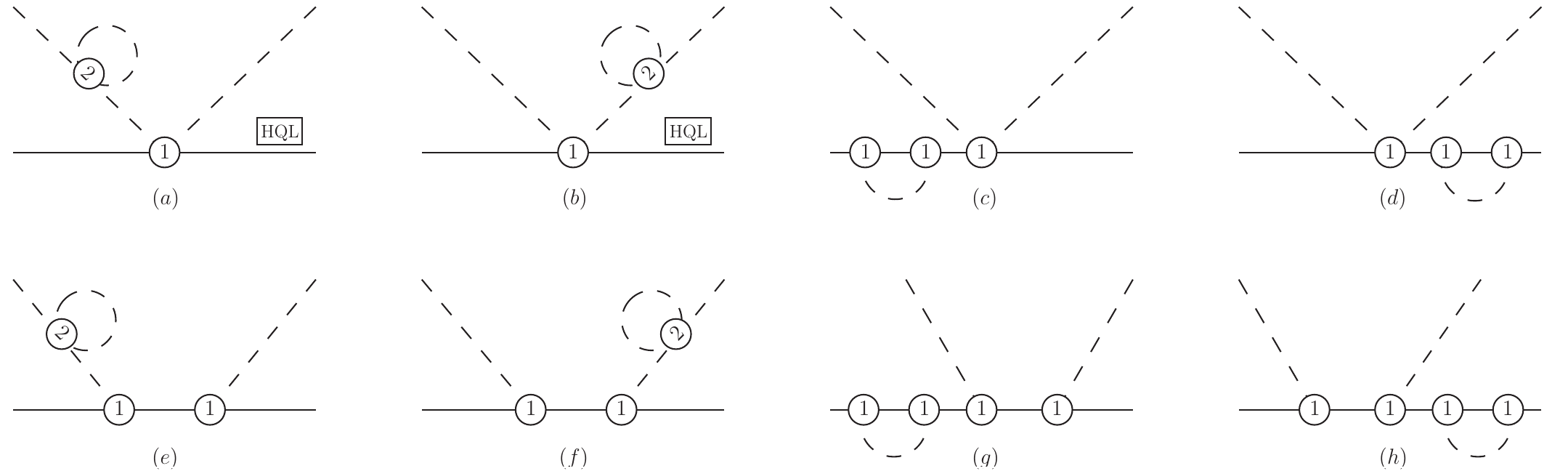} 
\caption{Loop diagrams of $\mathcal{O}(p^3)$ corresponding to wave function renormalization. Diagrams surviving in the HQL are indicated by boxed \lq\lq HQL".}
\label{wfdiagram}
\end{figure}

The Lehmann-Symanzik-Zimmermann (LSZ) reduction formula indicates that the full scattering amplitude, i.e. the on-shell transition amplitude, is related to the amputated Green function $\widehat{\mathcal{T}}$ in the momentum space through
\begin{align}\label{eq:LSZ}
\mathcal{T}_{\psi_1\phi_1\to\psi_2\phi_2}(s,t)=\mathcal{Z}_{\psi_1}^{\frac{1}{2}}\mathcal{Z}_{\phi_1}^{\frac{1}{2}}\mathcal{Z}_{\psi_2}^{\frac{1}{2}}\mathcal{Z}_{\phi_2}^{\frac{1}{2}} \bar{u}(p^\prime,\sigma^\prime)
\widehat{\mathcal{T}}_{\psi_1\phi_1\to\psi_2\phi_2} u(p,\sigma)\ ,
\end{align}
where $\widehat{\mathcal{T}}_{\psi_1\phi_1\to\psi_2\phi_2}$ has been calculated in Section~\ref{sec:ampcalc} and the wave function renormalization constants of the involved fields are presented in Subsection~\ref{sec:4.1}. Within our working accuracy, the full amplitude should be truncated at $\mathcal{O}(p^3)$. Therefore, the chiral expansion of the full amplitude can be written as
\begin{align}\label{T-SUM}
\mathcal{T}_{\psi_1\phi_1\to\psi_2\phi_2}(s,t) = \mathcal{T}^{(1)}_{\rm tree} + \mathcal{T}^{(2)}_{\rm tree} + \mathcal{T}^{(3)}_{\rm tree} + \mathcal{T}^{(3)}_{\rm loop} + 
\mathcal{T}^{(3)}_{\rm wf}\ ,
\end{align}
where the numbers in the superscripts denote the chiral orders. The last term is counted as $\mathcal{O}(p^3)$ and takes the form
\begin{align}
  \mathcal{T}_{\rm wf}^{(3)} = \frac{1}{2}\left(\delta \mathcal{Z}_{\phi_1} + \delta \mathcal{Z}_{\psi_1} +\delta \mathcal{Z}_{\phi_2}+\delta \mathcal{Z}_{\psi_2}\right) \mathcal{T}^{(1)}_{\rm tree}\ ,
\end{align}
with $\delta \mathcal{Z}_{\phi_i} = \mathcal{Z}_{\phi_i}-1$, $\delta \mathcal{Z}_{\psi_i}=\mathcal{Z}_{\psi_i} - 1$ ($i=1,2$). It is worth mentioning that this term perturbatively incorporates the effect of the multiplication of the wave function renormalization constants in Eq.~\eqref{eq:LSZ}, and the above procedure is usually called wave function renormalization. In another word, it actually takes into account the contribution of the diagrams with one-loop corrections on the external legs, as displayed in Figure~\ref{wfdiagram}.

The full scattering amplitude in Eq.~\eqref{T-SUM} has UV divergencies and  PCB terms. Here we adopt the EOMS scheme~\cite{Fuchs:2003qc} to address these issues. The EOMS scheme contains two steps: $\overline{{\rm MS}}$-1 subtraction and an extra finite renormalization. 

In $\overline{{\rm MS}}$-1 scheme, the bare baryon mass and axial-vector coupling constant, LECs are divided into renormalized and divergent parts. The divergent ones are used to cancel the UV divergence from loop amplitudes. To be specific, these quantities are written in the following form
\begin{align}\label{MS-UV}
  m &= m^{r}(\mu) + \beta_{m} \frac{R}{16 \pi^2 F^2}\ , \notag \\
  g &= g^{r}(\mu) + \beta_{g} \frac{R}{16 \pi^2 F^2}\ , \notag \\
  b_{i} &= b_{i}^{r}(\mu) + \beta_{b_{i}} \frac{R}{16 \pi^2 F^2} \quad \left(i = 1, 2, \dots, 7\right)\ , \notag \\
  c_{j} &= c_{j}^{r}(\mu) + \beta_{c_{j}} \frac{R}{16 \pi^2 F^2} \quad \left(j = 11, 12, \dots, 20\right)\ , 
\end{align}
where $\mu$ is the renormalization scale, and $R={2}/{(d-4)}-[\ln(4\pi)-\gamma_E+1]$ with $d$ being the number of space-time dimensions, $\gamma_E=0.577216$ being the Euler constant. The UV $\beta$-functions are gathered in Appendix~\ref{app:UV-divergent}. 

Then we utilize a finite renormalization to restore the correct power counting. Since the PCB terms are polynomials of chiral expansion parameters, they can be properly absorbed by the LECs in the tree amplitudes. The EOMS-renormalized parameters are defined by  
\begin{align}\label{EOMS-PCB}
  m^{r}(\mu) &= \widetilde{m} + \frac{\widetilde{\beta}_{m}}{16 \pi^2 F^2}\ , \notag \\
  g^{r}(\mu) &= \widetilde{g} +  \frac{\widetilde{\beta}_{g}}{16 \pi^2 F^2}\ , \notag \\
  b_{i}^{r}(\mu) &= \widetilde{b}_{i} + \frac{\widetilde{\beta}_{b_{i}} }{16 \pi^2 F^2} \quad \left(i=1, 2, \dots, 7\right)\ . 
\end{align}
Here, the EOMS $\widetilde{\beta}$-functions are collected in Appendix~\ref{app:EOMS-PCBterms}. Note that the $\mathcal{O}(p^3)$ LECs $c_{j}^{r}(\mu)$ maintain unchanged, i.e. $\widetilde{c}_{j}=c_{j}^{r}$. The renormalized amplitudes obtained in EOMS scheme possess of original analytic structure and respect the power counting rule. 

Finally, it should be mentioned that, in practical computations, the chiral limit decay constant $F$ is always replaced by the physical decay constants, $F_\pi$, $F_K$ and $F_\eta$. 
For the amplitude of the process $\psi_1\phi_1\to \psi_2\phi_2$, this is achieved by making the following substitution
\begin{align}
\frac{1}{F^{2n}} = \frac{1}{\left[\frac{F_{\phi_1}}{(1+\delta F_{\phi_1})}\right]^n\left[\frac{F_{\phi_2}}{(1+\delta F_{\phi_2})}\right]^n}=\frac{1}{F_{\phi_1}^nF_{\phi_2}^n}\bigg[1+n(\delta F_{\phi_1}+\delta F_{\phi_2})+\cdots\bigg]\ .\label{eq.Fsub}
\end{align}
It should be noted that such a substitution, when carried out for the $O(p^1)$ tree amplitude, generates $O(p^3)$ pieces that should be kept in our calculation. For the $O(p^2)$ and $O(p^3)$ amplitudes, one can merely make the replacement $F^{2n}\to (F_{\phi_1}F_{\phi_2})^n$, since the caused differences are of higher orders beyond the accuracy of our calculation. A merit of Eq.~\eqref{eq.Fsub} is that the physical decay constants are properly chosen according to the incoming and outgoing Goldstone bosons. For instance, the $F^2$ in the amplitude of $\Xi_{cc}\pi\to \Xi_{cc}\pi$ is changed to $F_\pi^2$ rather than e.g. $F_K^2$, while the one in the inelastic scattering amplitude of $\Xi_{cc}\pi\to \Xi_{cc}\eta$ is substituted by $F_\pi F_\eta$.

NLO expressions of the decay constants in ChPT can be found in Ref.~\cite{Gasser:1984gg}, which read
\begin{align}
    F_{\phi_i} \equiv F(1+\delta F_{\phi_i})\ ,\quad \phi_i\in \{\pi\,,K\,,\eta\}
\end{align}
with the chiral corrections being
\begin{align}
\delta F_{\pi}&=\frac{4m_{\pi}^2}{F^2}(L_4+L_5)+\frac{8m_K^2}{F^2}L_4+A_{0}(m_{\pi}^2)+\frac{1}{2}A_{0}(m_K^2) 
\ , \\
\delta F_{K}&=\frac{4m_{\pi}^2}{F^2} L_4+\frac{4m_{K}^2}{F^2}(2L_4+L_5)+\frac{3}{8}A_{0}(m_{\pi}^2)+\frac{3}{4}A_{0}(m_K^2)+\frac{3}{8}A_{0}(m_{\eta}^2) \ , \\
\delta F_{\eta}&=\frac{4(m_{\pi}^2+2m_{K}^2)}{F^2}L_4+\frac{4m_{\eta}^2}{F^2}L_5+\frac{3}{2}A_{0}(m_K^2) \ .
\end{align}
The bare parameters $L_i$'s are related to the corresponding UV-renormalized ones by
\begin{align}\label{eq:Gammas}
    L_i=L_i^r(\mu)+\frac{\Gamma_i}{32\pi^2}R\ .
\end{align}
The $\Gamma_i$ functions are given in Ref.~\cite{Gasser:1984gg} and are shown in Eq.~\eqref{eq:betaUV} for easy reference.


\section{Numerical results and discussions\label{sec:5}}
 

\subsection{Parameters\label{sec:5.1}}

Masses and decay constants used in our work are collected in Table~\ref{tab:phypa}. Since the $\Xi_{cc}^+$ and $\Xi_{cc}^{++}$ are treated as members of an isospin doublet, their masses are degenerate in the isospin limit and we take $m_{\Xi_{cc}}=3621.55$~MeV from the PDG review~\cite{ParticleDataGroup:2022pth}. Unlike the $\Xi_{cc}$ states, there is no experimental value for the $\Omega_{cc}$ baryon so far, therefore we employ $m_{\Omega_{cc}}=3738$~MeV determined by lattice QCD~\cite{Brown:2014ena}. 

For the decay constants of pion and kaon, we adopt their world-average values from Ref.~\cite{ParticleDataGroup:2022pth}. It is known that the physical $\eta$ and $\eta^\prime$ meson are superpositions of the singlet $\eta_0$ and octet $\eta_8$, that can be systematically formulated in large-$N_c$ ChPT, see e.g. Refs.~\cite{DiVecchia:1980yfw,Kaiser:2000gs}. Nevertheless, in the standard SU(3) ChPT we are using here, the singlet $\eta_0$ is absent and the octet member $\eta_8$ is approximately regarded as the physical $\eta$ meson. Based on the SU(3) ChPT calculation performed in Ref.~\cite{GomezNicola:2001as}, one has $F_\eta\simeq 1.27 F_\pi\simeq 117$ MeV. This value is more or less in agreement with the recent lattice QCD determination~\cite{Bali:2021qem}, if one assumes $F_\eta\simeq F_\eta^8$ with $F^8=115.0$~MeV.

\begin{table*}[htbp]
  \centering
  \caption{\label{tab:phypa} Masses and decay constants used in our work. All the quantities in this table are in units of MeV.}
  \renewcommand{\arraystretch}{1.1}
  \renewcommand{\tabcolsep}{1.1pc}
  \begin{tabular}{c c | c r  | c r}
  \hline
  \hline
  \multicolumn{2}{c}{Goldstone}   &\multicolumn{2}{c}{charm sector}   & \multicolumn{2}{c}{decay constant} \\
  \hline
  $m_{\pi}$   &$139.57$~\cite{ParticleDataGroup:2022pth}   &$m_{\Xi_{cc}}$     & $3621.55$~\cite{ParticleDataGroup:2022pth}   &$F_{\pi}$  &$92.2$~\cite{ParticleDataGroup:2022pth} \\
  $m_K$       &$493.68$~\cite{ParticleDataGroup:2022pth}  &$m_{\Omega_{cc}}$  & $3738$~\cite{Brown:2014ena} &$F_{K}$    &$112$~\cite{ParticleDataGroup:2022pth}   \\
  $m_{\eta}$ &$547.86$~\cite{ParticleDataGroup:2022pth} & $m_{D}$ & $1867$~\cite{ParticleDataGroup:2022pth} &$F_{\eta}$ &$117$
  \\
   & & $m_{D_s}$ & $1968$~\cite{ParticleDataGroup:2022pth}  & &
   \\
  \hline
  \hline
  \end{tabular} 
\end{table*}

Due to the lack of relevant experimental and lattice QCD data, a big challenge we encounter is the determination of the unknown LECs in the Lagrangians~\eqref{op1baryon}, \eqref{op2baryon} and \eqref{op3baryon}. Thanks to the HDA symmetry~\cite{Savage:1990di}, the doubly charmed baryon sector can be connected with the ones in the charmed meson sector. Derivation of the relations is detailed in Appendix~\ref{app.hda}.

Specifically, for the LO coupling constant, one has
\begin{align}
\tilde{g}=-\frac{1}{3\overline{m}_D}\tilde{g}_0 \ ,
\end{align}
where $\overline{m}_D=(m_D+m_{D_s})/2$ with $m_D$ and $m_{D_s}$ the physical masses of the $D$ and $D_s$ mesons, respectively. Here, $\tilde{g}_0$ denotes the axial coupling constant of the $D^\ast D\phi$ interaction, c.f. Eq.~\eqref{eq:dpi.lag.lo}. Its value can be fixed via the LO calculation of the decay width of $D^{\ast+}\to D^0\pi$~\cite{ParticleDataGroup:2022pth}, which leads to $\tilde{g}_0\simeq 1.095$~GeV.  
For the $O(p^2)$ LECs, one obtains
\begin{align}
\tilde{b}_1&=-\frac{1}{2\overline{m}_D}(\tilde{h}_0+\frac{1}{3} \tilde{h}_1) \ ,   \qquad 
\tilde{b}_2=-\frac{1}{2\overline{m}_D} \tilde{h}_1 \ , \qquad  
\tilde{b}_3=-\frac{1}{2\overline{m}_D} \tilde{h}_3 \ ,\notag \\
\tilde{b}_4&=\frac{1}{2\overline{m}_D} \tilde{h}_2 \ ,  \qquad 
\tilde{b}_5=\frac{\overline{m}_D}{8} \tilde{h}_5 \ , \qquad     
\tilde{b}_6=-\frac{\overline{m}_D}{8} \tilde{h}_4 \ . 
\label{eq:relation.p2}
\end{align}
At $O(p^3)$, the following pertinent relations are established,
\begin{align}
\tilde{c}_{11}&=\frac{\tilde{g}_2}{2} \ , \qquad 
\tilde{c}_{12}=\frac{\overline{m}_D^2}{4} \tilde{g}_3 \ ,\qquad 
\tilde{c}_{20}=-\frac{\tilde{g}_1}{2} \ .
\label{eq:relation.p3}
\end{align}
The parameters $\tilde{h}_{0,1,\cdots,5}$ in Eq.~\eqref{eq:relation.p2} and $\tilde{g}_{1,2,3}$ in Eq.~\eqref{eq:relation.p3} are the LECs involved in the NLO and NNLO effective Lagrangians, respectively, that describe the interactions between the NGBs and charmed mesons; see Eq.~\eqref{eq:dpi.lag.nlo} and Eq.~\eqref{eq:dpi.lag.nnlo} in Appendix~\ref{app.hda}. Their values have been determined by fitting to lattice QCD data in Ref.~\cite{Yao:2015qia}. Therein, four different fits were performed using the method of unitarized ChPT (UChPT). The fit denoted by UChPT-6($b^\prime$) was done by enforcing the naturalness requirements of the fitting parameters. Moreover, the validity range of UChPT is ensured by further excluding the lattice data with a pion mass larger than $600$~MeV. As it was expected, the obtained $D\phi$ LECs from the UChPT-6($b^\prime$) fit are more natural than the others in Ref.~\cite{Yao:2015qia}. Therefore we utilize the UChPT-6($b^\prime$) outputs, collected in the last column of Table~\ref{tab:reoflecs} for completeness, to estimate the values of $\tilde{b}_i$ ($i =1,\cdots, 6$) and $\tilde{c}_i$ ($i = 11, 12, 20$). The obtained results are shown in the 2nd column of Table~\ref{tab:reoflecs}. Importantly, those LECs determined by HDA symmetry are acceptable in the sense that they turn out be of natural size. Furthermore, the corresponding uncertainties are obtained by Monte Carlo propagating the errors of the $D\phi$ LECs of Ref.~\cite{Yao:2015qia}. \footnote{Specifically, sample groups of random values of the $D\phi$ LECs are generated by Monto Carlo method with normal distribution~\cite{Press:1992zz}. The \texttt{MINOS} algorithm in the package Minuit~\cite{James:1994vla} is followed to select good samples of normally-distributed parameters that satisfy the condition $\chi^2-\chi_{\rm min}^2\le 1$. Here, $\chi^2$ and $\chi^2_{\rm min}$ are calculated by using the randomly-generated parameters and the central values of the $D\phi$ LECs, respectively. A sufficient number of simulations have been performed, leading to 156 groups of good values of the $D\phi$ LECs. The 156 sets of parameter values are then utilized to estimate the uncertainties of the LECs in the doubly charmed baryon sector with the help of the HDA-symmetry relations~\eqref{eq:relation.p2} and \eqref{eq:relation.p3}. Moreover, the errors of scattering lengths and phase shifts to be discussed in the next subsections are propagated from the uncertainties of the $\psi\phi$ LECs. }

\begin{table*}[htbp]
\centering
\caption{\label{tab:reoflecs} Values of the LECs determined by making use of HDA symmetry. The LECs ${b}_i$'s and ${c}_j$'s are in units of $\rm{GeV}^{-1}$ and $\rm{GeV}^{-2}$, respectively.}
\renewcommand{\arraystretch}{1.1}
\renewcommand{\tabcolsep}{1.1pc}
\begin{tabular}{ c c |c c  }
\hline
\hline
\multicolumn{2}{c|}{$\psi\phi$ scattering} & \multicolumn{2}{c}{$D\phi$ scattering~\cite{Yao:2015qia}}\\
LECs     & Value      & LECs  & Value \\
\hline
$\tilde{g}$      & $-0.19$    & $\tilde{g}_0$ &{$1.095$} \\
$\tilde{b}_1$    & $-0.04$    & $\tilde{h}_0$    & $0.0172$\\
$\tilde{b}_2$    & $-0.11$    & $\tilde{h}_1$ &  $0.4266$\\
$\tilde{b}_3$    &$-1.46^{+0.43}_{-0.46}$  & $\tilde{h}_3$ &$5.59^{-2.07}_{-1.96}$ \\ 
$\tilde{b}_4$    &$0.66\pm 0.19$ & $\tilde{h}_2$ &$2.52^{+0.73}_{-0.74}$  \\ 
$\tilde{b}_5$    &$-0.17^{+0.05}_{-0.06}$ & $\tilde{h}_5$ &$-0.71^{+0.23}_{-0.24}$ \\
$\tilde{b}_6$    &$0.11\pm 0.04$ &  $\tilde{h}_4$ &$-0.47^{+0.17}_{-0.17}$ \\
\hline
$\tilde{c}_{11}$ &$-0.08^{+0.21}_{-0.14}$ & $\tilde{g}_2$ &$-0.16^{+0.52}_{-0.39}$  \\
$\tilde{c}_{12}$  &$0.08^{+0.03}_{-0.02}$ & $\tilde{g}_3$ &$0.08^{+0.03}_{-0.03}$  \\
$\tilde{c}_{20}$ &$0.49^{+0.09}_{-0.15}$ & $\tilde{g}_1$ &$-0.99^{+0.30}_{-0.18}$ \\
\hline
\hline
\end{tabular} 
\end{table*}

It is worth noting that $\tilde{h}_1$ in Table~\ref{tab:reoflecs} was fixed by the mass difference between $D$ and $D_s$ mesons in Ref.~\cite{Yao:2015qia}. Likewise, in our case the LEC $\tilde{b}_2$ can be estimated by the mass difference of the $\Xi_{cc}$ and $\Omega_{cc}$ baryons. With the help of Eq.~\eqref{eq:massXichpt} and Eq.~\eqref{eq:massOmchpt}, one obtains
\begin{align}\label{eq:b2.massdiff}
\tilde{b}_2\simeq \frac{ m_{\Xi_{cc}}-m_{\Omega_{cc}}}{4(m_K^2-m_\pi^2)} \simeq - 0.13~{\rm GeV}^{-1}\ .
\end{align}
Mention that the loop corrections have been neglected and the masses in Table~\ref{tab:phypa} have been used. One can see that the $\tilde{b}_2$ value in Eq.~\eqref{eq:b2.massdiff} is comparable with the one in Table~\ref{tab:reoflecs}, justifying the validity of HDA symmetry to a certain extent.

In fact, a more reliable determination of $\tilde{b}_1$ and $\tilde{b}_2$ has been conducted in Ref.~\cite{Yao:2018ifh}. The two LECs are pinned down by fitting to the lattice QCD data of the baryon masses, leading to\footnote{The LECs ${b}_1$ and ${b}_2$ are related to the ones used in Ref.~\cite{Yao:2018ifh} via $b_1=\hat{c}_1+\frac{1}{3}c_7$ and $b_2=c_7$.  }
\begin{align} \label{eq:b1b2lattice}
    \tilde{b}_1=-0.09\pm0.08~{\rm GeV}^{-1},\quad \tilde{b}_2=-0.09\pm0.09~{\rm GeV}^{-1}\ .
\end{align}
The $\tilde{b}_1$-$\tilde{b}_2$ parameter space allowed by $1$-$\sigma$ uncertainties cover the values of $\tilde{b}_1$ and $\tilde{b}_2$ in Table~\ref{tab:reoflecs} and Eq.~\eqref{eq:b2.massdiff}. Therefore, we use Eq.~\eqref{eq:b1b2lattice} for $\tilde{b}_1$ and $\tilde{b}_2$ throughout this paper. It should be also stressed that the numbers in Eq.~\eqref{eq:b1b2lattice} are obtained at the renormalization scale $\mu=1$~GeV. For consistency, the same renormalization scale is chosen, during our numerical computation of one-loop corrections. 

Apart from the fixed parameters discussed above, there are still eight unknown LECs. Since they can not be estimated via HDA symmetry,  we assume them to be zero, $\tilde{b}_7=0.0~{\rm GeV}^{-1}$ and $
\tilde{c}_{k}=0.0~{\rm GeV}^{-2},\quad k\in\{13, \cdots, 19\}
$. Such an assumption is more or less reasonable in view of the smallness for most of the LECs in Table~\ref{tab:reoflecs}. A solid determination of these parameters is expected to be done only when lattice QCD data of e.g. scattering lengths are available in the future.

Finally, the values of the LECs $L_4^r$ and $L_5^r$ can be found in Ref~\cite{Gasser:1984gg}, 
$L_4^r=-0.3\times 10^{-3}$ and $L_5^r=1.4\times 10^{-3}$, which are obtained at the scale $\mu=M_\rho$. An recent update of the mesonic LECs is summarized in Ref.~\cite{Bijnens:2014lea}, and comparable results of $L_4^r$ and $L_5^r$ are achieved. Those values at $\mu=M_\rho$ can be translated to the ones at $\mu=1$~GeV with the help of the following relations
\begin{align}
    L_i^r(\mu)=L_i^r(M_\rho)+\frac{\Gamma_i}{16\pi^2}\ln\bigg(\frac{M_\rho
    }{\mu}\bigg)\ .
\end{align}

\subsection{Prediction of scattering lengths\label{sec:5.2}}

Once all the involved LECs are pinned down, we are now in the position to calculate the $S$- and $P$-wave scattering lengths numerically. By definition, scattering lengths can be calculated via
\begin{align}
    a_{\ell\pm} = \lim_{{\bf |q|}\to 0} \frac{f_{\ell\pm}(s)}{{\bf q}^{2\ell}}\ .
\end{align}
However, the fraction on right hand side can not be computed numerically exact at threshold for $\ell\geq 1$. That is, the fraction value becomes undefined when one has zeros in the denominator, as the modulus of the three momentum is vanishing. To avoid this issue, by making use of Eq.~\eqref{eq:sca.len.ana}, we have derived analytical expressions for the $S$- and $P$-wave scattering lengths, with which numerical computations can be carried out precisely. For calculations with accuracy up to NNLO, derivatives of one-loop integrals are involved, which are handled by adopting the techniques proposed in Ref.~\cite{Devaraj:1997es}.

\begin{table*}[tbp]
\centering
\caption{\label{tab:swaveSL} $S$-wave scattering lengths $a_{0+}^{(S,I)}$ ($J^{P}=\frac{1}{2}^{-}$) in units of $\rm{fm}$.} 
\renewcommand{\tabcolsep}{0.3pc}
{\small
\begin{tabular}{c c | c c c c | c | c}
\hline \hline
\multirow{2}{1 cm}{$(S, I)$} &\multirow{2}{2 cm}{Processes}  &\multirow{2}*{$\mathcal{O}(p^1)$}
&\multirow{2}*{$\mathcal{O}(p^2)$}
&\multicolumn{2}{c|}{$\mathcal{O}(p^3)$}
&\multicolumn{1}{c|}{\multirow{2}*{Total}} &\multicolumn{1}{c}{\multirow{2}*{Ref.~\cite{Meng:2018zbl}}}
\\
\cline{5-6}
& & & & Tree & Loop  &  &
\\
\hline
$(-2, \frac{1}{2})$  & $\Omega_{cc}\bar{K}\rightarrow \Omega_{cc}\bar{K}$ & $-0.27$ & $0.29$ & $-0.11$  &$-0.001$ & $-0.09^{+0.12}_{-0.13}$  & $-0.20(1)$\\
$(1, 1)$             & $\Xi_{cc}K\rightarrow \Xi_{cc}K$ & $-0.27$       & $0.27$ & $-0.13$  & $-0.47$ & $-0.60\pm0.13$  & $-0.25(1)$\\
$(1, 0)$             & $\Xi_{cc}K\rightarrow \Xi_{cc}K$ & $0.27$       & $0.34$ & $0.13$  & $0.30$ & $1.03\pm0.19$    &$0.92(2)$\\
$(0, \frac{3}{2})$   & $\Xi_{cc}\pi \rightarrow \Xi_{cc}\pi$           & $-0.12$ &  $0.04$ & $-0.01$  & $-0.06$ & $-0.16\pm0.02$    & $-0.10(2)$
\\
\hline
$(-1, 0)$            & $\Xi_{cc}\bar{K} \rightarrow \Xi_{cc}\bar{K}$   & $0.54$ &  $0.24$ & $0.25$
& $0.16$ & $1.19^{+0.22}_{-0.21}$  & $2.15(11)$ \\
$       $            & $\Omega_{cc} \eta \rightarrow \Omega_{cc} \eta$ & $-0.001$  &  $0.37$ & $0.0$  & $0.05+0.55i$ & $0.42^{+0.18}_{-0.19}+0.55 i$  
&$0.57(3)+0.21i$
\\
\hline
$(-1, 1)$            & $\Omega_{cc} \pi \rightarrow \Omega_{cc} \pi$ & $0.0$  &  $0.04$ & $0.0$ & $-0.04$ & $-0.01\pm0.02$   & $-0.002(1)$\\
$       $            & $\Xi_{cc}\bar{K} \rightarrow \Xi_{cc}\bar{K}$ & $0.0$  &  $0.31$ & $0.0$ & $-0.04+0.10i$ & $0.27^{+0.13}_{-0.13}+0.10i$   
&$0.26(1)+0.19i$
\\
\hline
$(0, \frac{1}{2})$   & $\Xi_{cc}\pi\rightarrow \Xi_{cc}\pi$          &  $0.25$    &  $0.04$ & $0.01$  & $0.04$        &  $0.34\pm0.02$    &$0.36(1)$ \\
$      $             & $\Xi_{cc}\eta\rightarrow \Xi_{cc}\eta$        &  $-0.001$  & $0.32$  & $0.0$   & $-0.26$       & $0.06^{+0.14}_{-0.15}$    
&$0.34(1)+0.10i$ \\
$      $             & $\Omega_{cc}K\rightarrow \Omega_{cc}K$        &  $0.27$    & $0.29$  & $0.11$  & $-0.01+0.55i$ &  $0.66^{+0.13}_{-0.13}+0.55i$   
&$1.18(6)+0.29i$ \\
\hline \hline
\end{tabular}
}
\end{table*}

Results of the $S$-wave scattering lengths with quantum numbers $J^{P}=\frac{1}{2}^{-}$ are collected in Table~\ref{tab:swaveSL}. The uncertainties are propagated from the errors of the LECs by means of Monte Carlo methodology. We are concerning ourselves only with the $11$ elastic scattering processes that are indicated by the first and second columns of Table~\ref{tab:swaveSL}. Contributions from $O(p^1)$, $O(p^2)$, $O(p^3)$ trees, $O(p^3)$ loops and their sum are displayed separately. The $\mathcal{O}(p^1)$ $S$-wave scattering lengths are dominated by the contributions from the WT term, without any unknown LECs. The contributions of baryon-exchanging diagrams are suppressed. The explicit expressions of the NLO $S$-wave scattering lengths do not contain $b_7$, while the other NLO LECs $b_i$ ($i=1,\cdots,6$) are well fixed by HDA symmetry. On the contrary, the $O(p^3)$ tree contributions are roughly estimated due to the fact that, except for $c_{11}$, $c_{12}$ and $c_{20}$, most of the NNLO LECs are simply set to zero.    

The convergence of SU(3) ChPT has remained under debate for over several decades, see e.g. Refs.~\cite{Geng:2013xn}. Generally speaking, the convergence in the three-flavor ChPT calculations is usually worse than the two-flavor case, due to the relatively large strange quark mass.\footnote{{It should be pointed out that a one-loop calculation is not really sufficient to make a solid statement about convergence. It is well-known from both mesonic and baryonic sectors of ChPT that one needs at least two loops to make a significant statement about the chiral expansion. For example, an estimate for the convergence radius of the chiral expansion of the nucleon mass at the two-loop level was performed in Refs.~\cite{Schindler:2006ha,Schindler:2007dr}, indicating a breakdown of convergence already below $m_\pi\approx 360~{\rm MeV}$. The convergence might be even worse in the SU(3) case, since the kaons and the eta are substantially heavier. }} Here it would be also interesting to have a look at the convergence properties of the chiral expansion of the $\psi\phi$ scattering lengths. We first discuss the processes involving pion mesons. It can be seen from Table~\ref{tab:swaveSL} that the chiral series for the elastic $\Xi_{cc}\pi$ scatterings, with $(S,I)=(0,3/2)$ or $(0,1/2)$, converge well if one only concerns the first two orders. Namely, the $O(p^1)$ contributions are significantly larger than the $O(p^2)$ ones. Nevertheless, although the sums of $O(p^3)$ trees and loops in the two channels are small, there are comparable to the $O(p^2)$ trees. We ascribe the failure of convergence to the underestimation of the $O(p^3)$ trees for poor information on the LECs. As for $\Omega_{cc}\pi\to\Omega_{cc}\pi$, scattering length starts to contribute at $O(p^2)$, since the LO term for this channel (see Eq.~\eqref{o1tofT} and Table~\ref{tab:op1coeff}) is identical to zero exactly.

On the other hand, the scatterings of kaon and eta mesons off doubly charmed baryons have bad convergence properties, due to the emergence of large masses of kaon and eta as chiral expansion parameters. Such a non-convergent behaviour usually implies resummation of diagrams is required so that higher-order contributions can be implemented non-perturbatively. Similar situations happened for SU(3) meson-meson scatterings~\cite{GomezNicola:2001as}, and kaon-nucleon scatterings~\cite{Mai:2009ce,Lu:2018zof}. In general, the resummation procedure restores unitarity and extends the applicability range of ChPT, see Refs.~\cite{Yao:2020bxx} for a recent review. It should be emphasized that, the so-called UChPT amplitude plays a crucial role in investigating the spectrum of doubly charmed baryons. For instance, possible candidates of negative-parity doubly charmed baryons have been found on the basis of the LO~\cite{Guo:2017vcf} and NLO~\cite{Yan:2018zdt} ChPT amplitudes. It is also worth noting that the $O(p^3)$ amplitudes turn out to be significant and an inclusion of their effects may improve the studies made in Refs.~\cite{Guo:2017vcf,Yan:2018zdt}. Especially for the two channels, $\Xi_{cc}K\to\Xi_{cc}K$ with $(S,I)=(1,1)$ and $\Omega_{cc}\bar{K}\to\Omega_{cc}\bar{K}$ with $(S,I)=(-2,1/2)$, since there exist large cancellations between the LO and NLO contributions, the NNLO corrections become dominating. In a word, the ChPT amplitudes up to NNLO, we obtain in this work, can be applied to systematically scrutinize the doubly-charmed-baryon spectroscopy in the future, once more experimental or lattice QCD data are available.

For comparison, the HB results of Ref.~\cite{Meng:2018zbl} are shown in the last column of Table~\ref{tab:swaveSL}. For all the channels, our relativistic results are qualitatively consistent with the ones obtained in the HB formalism. Relativistic corrections are mainly responsible for the differences. Another source might be owing to the fact that the HQL-vanishing diagrams are not taken into account in the calculation of Ref.~\cite{Meng:2018zbl}. Nevertheless, their contributions are negligible, which will be illustrated below. Lastly, the spin-$3/2$ doubly charmed baryons are incorporated as explicit degrees of freedom in Ref.~\cite{Meng:2018zbl}, which would lead to discrepancies as well. As we can see, good agreements are observed within one-sigma uncertainties for the five channels: $\Omega_{cc}\bar{K}$ with $(S,I)=(-2,1/2)$, $\Xi_{cc}K$ with $(S,I)=(1,0)$, $\Omega_{cc}\eta$ with $(S,I)=(-1,0)$, $\Omega_{cc}\pi$ with $(S,I)=(-1,1)$ and $\Xi_{cc}\pi$ with $(S,I)=(0,1/2)$. Such an observation implies that the net effects of relativistic corrections, contributions of HQL-vanishing diagrams and resonance-exchanging diagrams are slight for those channels in $S$ wave.

$P$-wave scattering lengths with $J^P=\frac{3}{2}^+$ and $J^P=\frac{1}{2}^+$ are complied in Table~\ref{tab:ppwaveSL} and Table~\ref{tab:pmwaveSL}, respectively. It is found that, at $\mathcal{O}(p^1)$, the $P$-wave scattering lengths are entirely saturated by the crossing partner of diagram $(a)$. For some channels like $\Xi_{cc}\pi$ scatterings, contributions from $\mathcal{O}(p^3)$ loops to the $P$-wave scattering lengths turn out to be sizeable.

\begin{table*}[tbp]
\centering
\caption{\label{tab:ppwaveSL} $P$-wave scattering lengths $a_{1+}^{(S,I)}$ ($J^P=\frac{3}{2}^+$) in units of $10^{-2}~{\rm fm}^3$.}
\renewcommand{\arraystretch}{1.1}
\renewcommand{\tabcolsep}{0.4pc}
{
\begin{tabular}{c c | c c c c |c }
\hline \hline
\multirow{2}{1 cm}{$(S, I)$} &\multirow{2}{2 cm}{Processes}  &\multirow{2}*{$\mathcal{O}(p^1)$}
&\multirow{2}*{$\mathcal{O}(p^2)$}
&\multicolumn{2}{c|}{$\mathcal{O}(p^3)$}
&\multicolumn{1}{c}{\multirow{2}*{Total}}
\\
\cline{5-6}
& & & & Tree & Loop  &  
\\
\hline
$(-2, \frac{1}{2})$ & $\Omega_{cc}\bar{K}\rightarrow \Omega_{cc}\bar{K}$ & $0.16$ & $0.60$  & $-0.22$   &$-3.00$  & $-2.47^{+3.04}_{-2.64}$  \\
\hline
$(1, 1)$ & $\Xi_{cc}K\rightarrow \Xi_{cc}K$ & $0.10$   & $0.59$ & $-0.22$   & $-1.19$  & $-0.73^{+3.02}_{-2.64}$ \\
\hline
$(1, 0)$ & $\Xi_{cc}K\rightarrow \Xi_{cc}K$ & $-0.10$  & $-8.77$ & $0.22$   & $1.71$ & $-6.93^{+2.83}_{-3.21}$ \\
\hline
$(0, \frac{3}{2})$ & $\Xi_{cc}\pi \rightarrow \Xi_{cc}\pi$ &  $0.62$ & $0.75$  & $-0.18$  & $-41.8$   & $-40.6^{+3.20}_{-2.97}$ \\
\hline
$(-1, 0)$ & $\Xi_{cc}\bar{K} \rightarrow \Xi_{cc}\bar{K}$ & $0.0$ & $5.27$   & $0.45$     & $0.48$    & $6.19^{+4.78}_{-5.40}$  \\
$       $ & $\Omega_{cc} \eta \rightarrow \Omega_{cc} \eta$ & $0.07$  & $2.0$   & $0.0$   & $-1.13+0.01 i$  & $0.93^{+2.04}_{-1.96}+0.01i$     \\
\hline
$(-1, 1)$ & $\Omega_{cc} \pi \rightarrow \Omega_{cc} \pi$ & $0.0$  & $-6.23$  & $0.001$  & $-0.10$  & $-6.32^{+1.85}_{-1.82}$ \\
$       $ & $\Xi_{cc}\bar{K} \rightarrow \Xi_{cc}\bar{K}$ & $0.0$  & $-4.09$  & $0.0$  & $-0.11+0.01i$  & $-4.2^{+1.23}_{-1.21}+0.01i$ \\
\hline
$(0, \frac{1}{2})$ & $\Xi_{cc}\pi\rightarrow \Xi_{cc}\pi$ &  $-0.31$ & $0.75$  & $0.35$  &$21.2$   & $21.9^{+3.39}_{-3.70}$  \\
$      $ & $\Xi_{cc}\eta\rightarrow \Xi_{cc}\eta$ &  $0.02$  & $-2.31$  & $0.0$  & $-0.01+0.01i$  & $-2.30^{+1.13}_{-1.13}+0.01i$ \\
$      $ & $\Omega_{cc}K\rightarrow \Omega_{cc}K$ &  $0.0$   & $0.6$  & $0.22$  & $0.19+0.01i$  & $1.0^{+2.67}_{-3.01}+0.01i$  \\
\hline \hline
\end{tabular}
}
\end{table*}

\begin{table*}[tbp]
\centering
\caption{\label{tab:pmwaveSL} $P$-wave scattering lengths $a_{1-}^{(S,I)}$ ($J^P=\frac{1}{2}^+$) in units of $10^{-2}~{\rm fm}^3$.}
\renewcommand{\arraystretch}{1.1}
\renewcommand{\tabcolsep}{0.4pc}
{
\begin{tabular}{c c | c c c c | c }
\hline \hline
\multirow{2}{1 cm}{$(S, I)$} &\multirow{2}{2 cm}{Processes}  &\multirow{2}*{$\mathcal{O}(p^1)$}
&\multirow{2}*{$\mathcal{O}(p^2)$}
&\multicolumn{2}{c|}{$\mathcal{O}(p^3)$}
&\multicolumn{1}{c}{\multirow{2}*{Total}}
\\
\cline{5-6}
& & & & Tree & Loop  &  
\\
\hline
$(-2, \frac{1}{2})$ & $\Omega_{cc}\bar{K}\rightarrow \Omega_{cc}\bar{K}$ & $-0.38$ & $0.58$   & $-0.34$  &$0.02$  & $-0.13^{+3.03}_{-2.64}$   \\
\hline
$(1, 1)$ & $\Xi_{cc}K\rightarrow \Xi_{cc}K$ &$-0.36$    & $0.57$   & $-0.37$   &$-1.74$  & $-1.90^{+3.01}_{-2.61}$  \\
\hline
$(1, 0)$ & $\Xi_{cc}K\rightarrow \Xi_{cc}K$ & $0.36$  & $-8.80$ & $0.37$   &$0.46$  & $-7.59^{+2.82}_{-3.20}$ \\
\hline
$(0, \frac{3}{2})$ & $\Xi_{cc}\pi \rightarrow \Xi_{cc}\pi$ &  $-0.80$ & $0.75$ & $-0.2$   & $19.5$  & $19.3^{+3.19}_{-2.97}$ \\
\hline
$(-1, 0)$ & $\Xi_{cc}\bar{K} \rightarrow \Xi_{cc}\bar{K}$ & $0.16$ & $5.25$   & $0.74$  & $-9.77$ & $-3.61^{+4.77}_{-5.37}$  \\
$       $ & $\Omega_{cc} \eta \rightarrow \Omega_{cc} \eta$ & $-0.13$  & $1.97$   & $0.0$   & $-2.16+0.01i$   & $-0.32^{+2.03}_{-1.95}+0.01i$\\
\hline
$(-1, 1)$ & $\Omega_{cc} \pi \rightarrow \Omega_{cc} \pi$ & $0.0$  & $-6.23$   & $0.0$  & $-0.53$  & $-6.75^{+1.85}_{-1.82}$  \\
$       $ & $\Xi_{cc}\bar{K} \rightarrow \Xi_{cc}\bar{K}$ & $0.0$  & $-4.11$   & $0.0$  & $-0.60+0.01i$  & $-4.72^{+1.24}_{-1.22}+0.01i$    \\
\hline
$(0, \frac{1}{2})$ & $\Xi_{cc}\pi\rightarrow \Xi_{cc}\pi$ &  $-0.27$ & $0.75$   & $0.39$ & $-104.9$  & $-104.1^{+3.38}_{-3.70}$ \\
$      $ & $\Xi_{cc}\eta\rightarrow \Xi_{cc}\eta$ &  $-0.03$  & $-2.33$ & $0.0$  & $-1.43+0.01i$  & $-3.79^{+1.13}_{-1.14}+0.01i$ \\
$      $ & $\Omega_{cc}K\rightarrow \Omega_{cc}K$ &  $0.16$   & $0.58$  & $0.34$   & $-3.77+0.01i$ & $-2.69^{+2.67}_{-3.00}+0.01i$\\
\hline \hline
\end{tabular}
}
\end{table*}

Before ending this subsection, we intend to discuss the contributions of HQL-surviving diagrams in more detail. 
In the HQL, most of the diagrams in Figures~\ref{treediagram}, \ref{loopdiagram} and \ref{wfdiagram} do not contribute at threshold. This can be illustrated by performing an expansion in terms of the inverse of the baryon mass, i.e. HB projection. For the axial term in Eq.~\eqref{op1baryon}, the HB projection yields
\begin{align}
\bar{\psi}\bigg\{\frac{g}{2}\slashed{u}\gamma_5 \bigg\}\psi \longrightarrow\bar{\cal N}_v\big\{
g S_\nu\cdot u\big\}
{\cal N}_v+ \alpha_1\, m^{-1}+\alpha_2\, m^{-2}+\cdots
\end{align}
where the spin matrix is $S_v^\mu \equiv \frac{i}{2}\gamma_5\sigma^{\mu\nu}v_\nu$ with the four vector $v^\mu=(1,0,0,0)$. In the HQL, $m\to \infty$, all the inverse mass terms approach to zero. Only the first term that is independent of mass survives. However, it vanishes at threshold due to the feature of derivative coupling of the NGBs. Specifically, one has 
\begin{align}
\bar{\cal N}_v\big\{
g S_\nu\cdot u\big\}
{\cal N}_v \propto \bar{\cal N}_v\big\{
g S_v\cdot \partial\phi \big\}
{\cal N}_v 
\end{align}
which corresponds to $S_v\cdot q_\phi$ in the momentum space, with $q_\phi$ being the momentum of Goldstone boson. At threshold, $q_\phi=(m_\phi,\Vec{0})$, leading to $S_v\cdot q_\phi=0$. Subsequently, all the diagrams containing axial vertices disappear in the HQL. Therefore, only the diagrams that are irrelevant to the axial coupling $g$ survive. For easy reference, they have been labelled by boxed \lq\lq HQL" in Figures~\ref{treediagram}, \ref{loopdiagram} and \ref{wfdiagram}. 

Results of scattering lengths obtained by including only the HQL-surviving diagrams are shown in Table~\ref{tab.HQL.SL}. As expected, for the $S$-wave scattering lengths the deviation of the HQL results in Table~\ref{tab.HQL.SL} from the ones in Table~\ref{tab:swaveSL} is negligible. However, the results of $P$-wave scattering lengths change dramatically by switching off the HQL-vanishing diagrams, especially for $a_{1+}$ with $J^P=\frac{3}{2}^+$. One may probably ascribe such large variations to the absence of the HQS partners like spin-$3/2$ doubly charmed baryons. They are expected to contribute equally as the spin-$3/2$ baryons, since the HQS symmetry is exact in HQL. In the next subsection, we will evaluate the effect of the spin-$3/2$ doubly charmed baryons.

\begin{table*}[htbp]
\centering
\caption{\label{tab.HQL.SL}Results of scattering lengths obtained by taking only the HQL-surviving diagrams into consideration. The $S$- and $P$-wave scattering lengths are in units of $\rm{fm}$ and $10^{-2}~{\rm fm}^3$, respectively.}
\renewcommand{\arraystretch}{1.1}
\renewcommand{\tabcolsep}{0.4pc}
\begin{tabular}{cc|ccc}
\hline \hline 
{$(S, I)$} &{Processes}  
&$a_{0+}$~[$J^P=\frac{1}{2}^-$]
&$a_{1+}$~[$J^P=\frac{3}{2}^+$]
&$a_{1-}$~[$J^P=\frac{1}{2}^+$]
\\
\hline
$(-2, \frac{1}{2})$ & $\Omega_{cc}\bar{K}\rightarrow \Omega_{cc}\bar{K}$     &$-0.08^{+0.12}_{-0.13}$  &$0.34^{+3.04}_{-2.64}$    &$-1.42^{+3.03}_{-2.64}$    \\
\hline
$(1, 1)$ & $\Xi_{cc}K\rightarrow \Xi_{cc}K$  &$-0.62\pm{0.13}$   &$0.39^{+3.02}_{-2.62}$ &$-2.51^{+3.01}_{-2.61}$   \\
\hline
$(1, 0)$ & $\Xi_{cc}K\rightarrow \Xi_{cc}K$  &$1.03\pm{0.19}$    &$-8.19^{+2.83}_{-3.21}$ &$-7.41^{+2.82}_{-3.20}$  \\
\hline
$(0, \frac{3}{2})$ & $\Xi_{cc}\pi \rightarrow \Xi_{cc}\pi$ &$-0.15\pm0.02$   &$0.63^{+3.20}_{-2.97}$    &$-1.26^{+3.19}_{-2.97}$  \\
\hline
$(-1, 0)$ & $\Xi_{cc}\bar{K} \rightarrow \Xi_{cc}\bar{K}$  &$1.19^{+0.22}_{-0.21}$ &$6.36^{+4.78}_{-5.40}$  &$4.66^{+4.77}_{-5.37}$   \\
$       $ & $\Omega_{cc} \eta \rightarrow \Omega_{cc} \eta$ &$0.42^{+0.18}_{-0.19}+0.56i$   &$2.03^{+2.04}_{-1.96}$  &$-0.12^{+2.03}_{-1.95}+0.02i$  \\
\hline
$(-1, 1)$ & $\Omega_{cc} \pi \rightarrow \Omega_{cc} \pi$  &$0.0\pm0.02$      &$-6.23^{+1.85}_{-1.82}$   &$-6.72^{+1.85}_{-1.82}$     \\
$       $ & $\Xi_{cc}\bar{K} \rightarrow \Xi_{cc}\bar{K}$  &$0.28_{-0.13}^{+0.13}+0.10i$   &$-4.14^{+1.23}_{-1.21}$    &$-4.71^{+1.24}_{-1.22}$  \\
\hline
$(0, \frac{1}{2})$ & $\Xi_{cc}\pi\rightarrow \Xi_{cc}\pi$ &$0.33\pm0.02$   &$1.41^{+3.39}_{-3.70}$     &$1.51^{+3.38}_{-3.70}$ \\
$      $ & $\Xi_{cc}\eta\rightarrow \Xi_{cc}\eta$   &$0.05^{+0.14}_{-0.15}$
  &$-2.16\pm1.13$    &$-3.32^{+1.13}_{-1.14}$   \\
$      $ & $\Omega_{cc}K\rightarrow \Omega_{cc}K$   &$0.64_{-0.13}^{+0.13}+0.55i$     &$1.14^{+2.67}_{-3.01}$    &$-1.01^{+2.67}_{-3.00}+0.03i$  \\
\hline \hline
\end{tabular}
\end{table*}


\subsection{Effect of spin-$3/2$ doubly charmed baryons\label{sec:5.3}}

According to the naive quark model with flavour SU(4) ($u,d,s,c$ quarks) symmetry, there are three ground $20$-plets, i.e. ${\bf 4}\otimes{\bf 4}\otimes{\bf 4}={\bf 20}_S\oplus {\bf 20}_M\oplus {\bf 20}_M\oplus {\bf 4}_A$. The spin-$1/2$ doubly charmed baryons, $\Xi_{cc}^{++}$, $\Xi_{cc}^{+}$ and $\Omega_{cc}^{+}$, belong to one of the mixed-symmetric $20$-plets, while the spin-$3/2$ doubly charmed baryons, $\Xi_{cc}^{\prime ++}$, $\Xi_{cc}^{\prime +}$ and $\Omega_{cc}^{\prime +}$, to the symmetric $20$-plet ${\bf 20}_S$; See e.g. Ref.~\cite{Crede:2013kia} for a detailed review. In the HQL, the two doubly-charmed-baryon triplets degenerate and should be treated on equal footing, c.f. Eq.~\eqref{eq:super.super.field}, in line with HQS symmetry~\cite{Falk:1991nq}. In this sense, the effects of spin-$3/2$ baryons are as important as those of the spin-$1/2$ ones. Therefore, in this subsection, we aim to assess the impact of spin-$3/2$ doubly charmed baryons on the scattering lengths within the framework of covariant BChPT.

\begin{figure}
\centering
\includegraphics[width=0.65\textwidth]{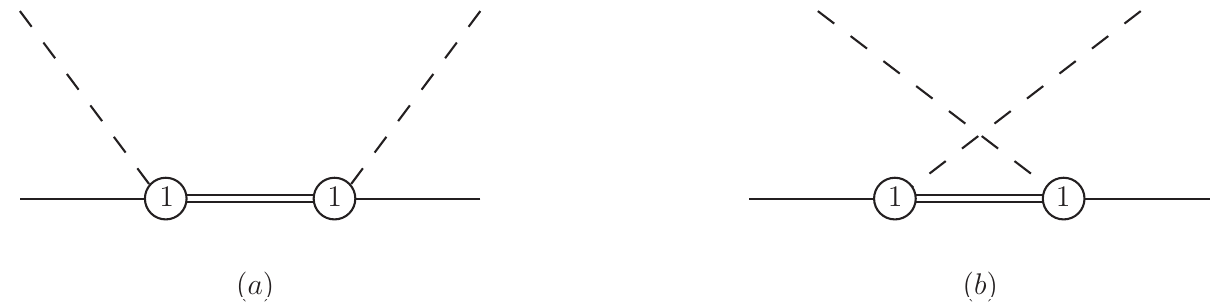}
\caption{Tree diagrams contributing to $\psi$$\phi$ scattering at LO. The solid, dashed and double lines represent the spin-$1/2$ baryons, the pions and the spin-$3/2$ doubly baryons, in order.}
\label{fig:dcSL_tree_EB}
\end{figure}

{The inclusion of spin-$3/2$ baryons in covariant BChPT is complicated. An appropriate power counting rule should be assigned to the new parameter, i.e. the mass difference between the spin-$1/2$ and spin-$3/2$ baryons, denoted by $\Delta$ hereafter. Like the treatment of the $\Delta(1232)$ resonances in the traditional BChPT, here we adopt the so-called $\delta$-counting rule that was proposed in Ref.~\cite{Pascalutsa:2002pi}. Specifically, in the energy region near threshold the mass splitting $\Delta$ is counted as $O(p^{1/2})$ in addition to Eq.~\eqref{eq:3.2}. According to this power counting rule, the spin-$3/2$ baryon propagator is $O(p^{-1/2})$. Loops with internal spin-$3/2$ baryon lines are at least of order $O(p^{7/2})$, and therefore their contributions are beyond the accuracy of our calculation. For the calculation up to $O(p^3)$, we only need to take into account the LO and NLO tree-level contributions from explicit spin-$3/2$ baryons, which are $O(p^{3/2})$ and $O(p^{5/2})$, respectively. In order to avoid the emergence of too many new unknown LECs, the NLO Born-term contribution is omitted here. In fact, as pointed out in Ref.~\cite{Yao:2016vbz}, the NLO Born term is redundant in the sense that its contribution can be absorbed by redefining the coupling $h_A$ in the LO Born term (see Eq.~\eqref{eq.exBprime}) and the LECs in the contact terms~\eqref{o2tofT} and~\eqref{o3tofT}.   } 

For the purpose of our calculation the following LO Lagrangian are needed,
\begin{align}
\mathscr{L}_{\frac{3}{2}}
&=
-\bar{\psi}^{\prime\mu}[g_{\mu\nu}(i\slashed{D}-\mt)
+iA(\gamma_{\mu}D_{\nu}+\gamma_{\nu}D_{\mu})+\frac{i}{2}(3A^2+2A+1)\gamma_{\mu}\slashed{D}\gamma_{\nu}
\notag \\
&
+\mt(3A^2+3A+1)\gamma_{\mu}\gamma_{\nu}
]\psi^{\prime\nu}+
\frac{h_A}{2}(\bar{\psi}u^{\mu}\psi^{\prime}_{\mu}+H.c.) \ ,
\end{align}
where $\mt$ and $h_A$ are the mass of the spin-$3/2$ baryons and the coupling constant of the $\psi\psi^\prime\phi$ interactions in the chiral limit, respectively. $A$ is an arbitrary real parameter but $A\neq -1/2$. The spin-$3/2$ doubly charmed baryons are collected in the triplet
\begin{align}
\psi^{\prime}_\mu=
\left(
\begin{array}{c}
 \Xi_{cc}^{\prime ++} \\
 \Xi_{cc}^{\prime +}   \\
 \Omega_{cc}^{\prime +} \\
\end{array} 
\right)_\mu \ .
\end{align}
Feynman diagrams relevant to our calculation are displayed in Figure~\ref{fig:dcSL_tree_EB}. Tree-level amplitudes are derived, which can be written in the form as
\begin{align}
\mathcal{A}_{\rm tree}^{(1)}&=-\frac{h_A^2}{48F^2}
[\mathcal{C}_S^{(1)}\mathcal{F}^\prime(s)
+\mathcal{C}_{U}^{(1)} \mathcal{F}^\prime(u)]
\ , \notag\\
\mathcal{B}_{\rm tree}^{(1)}
&=\frac{h_A^2}{24F^2}
[\mathcal{C}_{S}^{(1)} \mathcal{G}^\prime(s)
-\mathcal{C}_{U}^{(1)} \mathcal{G}^\prime(u)]
\ ,\label{eq.exBprime}
\end{align}
where the coefficients $\mathcal{C}_{S}^{(1)}$ and $\mathcal{C}_{U}^{(1)}$ are the same as those in Eq.~\eqref{o1tofT}, which are shown in Table~\ref{tab:op1coeff}. Nevertheless, the exchanged spin-$1/2$ baryons indicated in Table~\ref{tab:op1coeff} should be substituted by their spin-$3/2$ partners.
Explicit expressions of the functions $\mathcal{F}^\prime$ and $\mathcal{G}^\prime$ are given by
\begin{align}
\mathcal{F}^\prime(s)&=\frac{1}{M^2(s-M^2)}
\bigg\{
(m_{\psi_1}+m_{\psi_2})(m_{\psi_1}^2-m_{\phi_1}^2-s)(m_{\psi_2}^2-m_{\phi_2}^2-s)
\notag \\
&
-M[
(m_{\psi_1}-m_{\psi_2})^2
(m_{\psi_1}m_{\psi_2}+s)
-m_{\psi_1}m_{\psi_2}(m_{\phi_1}^2+m_{\phi_2}^2)
-2s(m_{\phi_1}^2+m_{\phi_2}^2)
\notag \\
&
+3(m_{\psi_2}^2m_{\phi_1}^2+m_{\psi_1}^2m_{\phi_2}^2)
-4m_{\phi_1}^2m_{\phi_2}^2)
]
-M^2[
(m_{\psi_1}^2+m_{\psi_2}^2-2s-3t)(m_{\psi_1}+m_{\psi_2})
\notag \\
&
+2m_{\psi_2}(2m_{\phi_1}^2+m_{\phi_2}^2)
+2m_{\psi_1}(m_{\phi_1}^2+2m_{\phi_2}^2)
]
\notag \\
&
-2M^3
[
m_{\psi_1}^2+m_{\psi_2}^2+3(m_{\phi_1}^2+m_{\phi_2}^2-t)-2s
]
\bigg\}
\ , \notag \\
\mathcal{G}^\prime(s)&=
\frac{1}{M^2(s-M^2)}
\bigg\{
2M^3(m_{\psi_1}+m_{\psi_2})
-(m_{\psi_1}^2-m_{\phi_1}^2-s)(m_{\psi_2}^2-m_{\phi_2}^2-s)
\notag \\
&+M[m_{\psi_2}(m_{\psi_1}^2-m_{\phi_1}^2-s)
+m_{\psi_1}(m_{\psi_2}^2-m_{\phi_2}^2-s)
]
\notag \\
&
+M^2[(m_{\psi_1}+m_{\psi_2})^2
+2(m_{\phi_1}^2+m_{\phi_2}^2)
-3t
]
\bigg\}
\ .
\end{align}
As argued in Ref.~\cite{Nath:1971wp}, physical quantities are independent of the parameter $A$. Therefore, we have set $A=-1$ for convenience as done in Ref.~\cite{Yao:2016vbz}.

{\footnotesize
\begin{table*}[htbp]
\centering
\renewcommand{\tabcolsep}{0.22pc}
\caption{LO contributions of the spin-$3/2$ $\psi^\prime$ baryons to the scattering lengths. $\Delta$ denotes the mass difference between the spin-$3/2$ and spin-$1/2$ baryons. The $S$- and $P$-wave scattering lengths are in units of ${\rm fm}$ and $10^{-2}~{\rm fm}^3$, respectively. \label{tab:spin3o2}}
\begin{tabular}{cc|ccc|ccc|ccc}
\hline \hline
\multirow{2}*{$(S, I)$} 
&\multirow{2}*{Processes} 
&\multicolumn{3}{|c|}{$\Delta=50~{\rm MeV}$}
&\multicolumn{3}{c|}{$\Delta=100~{\rm MeV}$}
&\multicolumn{3}{c}{$\Delta=150~{\rm MeV}$}
\\
\cline{3-11}
&
&$a_{0+}$
&$a_{1+}$
&$a_{1-}$
&$a_{0+}$
&$a_{1+}$
&$a_{1-}$
&$a_{0+}$
&$a_{1+}$
&$a_{1-}$
\\
\hline
$(-2, \frac{1}{2})$ & $\Omega_{cc}\bar{K}\rightarrow \Omega_{cc}\bar{K}$   &$-0.02$   &$0.31$   &$1.25$ &$-0.02$   &$0.28$   &$1.11$  &$-0.02$   &$0.25$   &$1.00$   \\
\hline
$(1, 1)$ & $\Xi_{cc}K\rightarrow \Xi_{cc}K$  &$-0.02$  &$0.19$   &$0.80$  &$-0.02$  &$0.18$   &$0.74$ &$-0.02$  &$0.17$   &$0.69$  \\
\hline
 $(1, 0)$ & $\Xi_{cc}K\rightarrow \Xi_{cc}K$   &$0.02$    &$-0.19$    &$-0.80$  &$0.02$    &$-0.18$    &$-0.74$   &$0.02$  &$-0.17$     &$-0.69$     
\\
\hline
$(0, \frac{3}{2})$ & $\Xi_{cc}\pi \rightarrow \Xi_{cc}\pi$  &$-0.003$    &$0.94$   &$3.79$ &$-0.003$    &$0.75$   &$2.99$  &$-0.003$   &$0.61$    &$2.47$   \\
\hline
$(-1, 0)$ & $\Xi_{cc}\bar{K} \rightarrow \Xi_{cc}\bar{K}$   &$-0.04$  &$-1.89$   &$-0.02$ &$-0.04$  &$-2.23$   &$-0.02$ & $-0.04$    &$-2.73$   &$-0.02$   \\
$       $ & $\Omega_{cc} \eta \rightarrow \Omega_{cc} \eta$  &$-0.03$ &$-0.24$ &$0.54$  &$-0.03$ &$-0.30$ &$0.50$  &$-0.03$  &$-0.36$   &$0.46$   \\
\hline
$(-1, 1)$ & $\Omega_{cc} \pi \rightarrow \Omega_{cc} \pi$   &$0.0$    &$0.0$  &$0.0$  &$0.0$    &$0.0$  &$0.0$   &$0.0$    &$0.0$    &$0.0$    \\
$       $ & $\Xi_{cc}\bar{K} \rightarrow \Xi_{cc}\bar{K}$  &$0.0$ &$0.0$  &$0.0$   &$0.0$    &$0.0$  &$0.0$   &$0.0$    &$0.0$    &$0.0$  \\
\hline
$(0, \frac{1}{2})$ & $\Xi_{cc}\pi\rightarrow \Xi_{cc}\pi$ &$-0.003$ &$-8.85$ &$-1.90$  &$-0.003$ &$-19.3$ &$-1.51$  &$-0.003$   &$71.6$   &$-1.25$ \\
$      $ & $\Xi_{cc}\eta\rightarrow \Xi_{cc}\eta$  &$-0.01$ &$-0.06$  &$0.14$  &$-0.01$ &$-0.07$  &$0.12$   &$-0.01$    &$-0.09$  &$0.12$   \\
$      $ & $\Omega_{cc}K\rightarrow \Omega_{cc}K$   &$-0.02$ &$-0.55$  &$-0.01$ &$-0.02$ &$-0.61$  &$-0.01$  &$-0.02$    &$-0.68$  &$-0.01$   \\
\hline \hline
\end{tabular}
\end{table*}
}

There are two unknown parameters: $h_A$ and $M$. We need to assign appropriate values to them, so that scattering lengths can be evaluated numerically. Thanks to HQS, the LO coupling constant $h_A$ can be related to the $g$ in the Lagrangian~\eqref{op1baryon} by $h_A=2\sqrt{3}g$. The chiral limit mass $M$ is replaced by the physical mass $M_{\psi^\prime}$. The physical mass of spin-$3/2$ baryon is assumed to be $M_{\psi^\prime}=m_{\psi}+\Delta$, where $\Delta$ is the mass splitting. Since the spin-$1/2$ and spin-$3/2$ baryons differ only in the relative orientation of the quark spins, the resultant difference in their masses is attributed to the spin-spin interaction in the viewpoint of potential model. Here, following Ref.~\cite{Meng:2018zbl}, we roughly take the difference to be around $100$~MeV. Three values of $\Delta$ are adopted: $\Delta=50$~MeV, $\Delta=100$~MeV and $\Delta=150$~MeV. Contributions of spin-$3/2$ baryons to the $S$- and $P$-wave scattering lengths are compiled in Table~\ref{tab:spin3o2}.

It can be seen from Table~\ref{tab:spin3o2} that for $S$ wave the size of the contribution of spin-$3/2$ baryons is nearly negligible. The $s$- and $u$-channel exchanges of $\psi^\prime$, corresponding to diagram $(a)$ and $(b)$ in Figure~\ref{fig:dcSL_tree_EB} respectively, mainly affect the $P$-wave scattering lengths. 
For a given strong interaction, strangeness and isospin are conserved definitely. Therefore, the $s$-channel exchange of spin-$3/2$ $\Xi_{cc}^\prime$ and $\Omega_{cc}^\prime$ states contribute only to the scattering processes with $(S,I)=(0,1/2)$ and $(-1,0)$, respectively. This can also be justified by seeing the coefficients $C_S^{(1)}$ in Table~\ref{tab:op1coeff}. Furthermore, we have checked that it dominates the $a_{1+}$ scattering lengths in the two coupled channels with $(S,I)=(0,1/2)$ and $(-1,0)$, which is in accordance with the conservation law of total angular momentum. The remaining processes get contributions entirely from the crossed diagram, i.e. diagram $(b)$ in Figure~\ref{fig:dcSL_tree_EB}. 

In parallel, one may also study the resonance contribution to the LECs with Eq.~\eqref{eq.exBprime}. In the viewpoint of effective field theory, a low-energy effective Lagrangian contains only low-lying degrees of freedom, while resonances at hard scale have been integrated out. The information of resonance contributions is regarded to be encoded in the LECs, i.e, the coefficients of the operators in the chiral effective Lagrangian without resonances. The pertinent contributions of resonances to the LECs can be efficiently achieved by a matching procedure carried out at the level of effective Lagrangian. Such a procedure has already been used, e.g., in the analyses of the LECs in pion-nucleon scattering~\cite{Bernard:1996gq,Chen:2012nx} and $D$-$\phi$ interactions~\cite{Du:2016tgp}. 
In our current situation, we intend to utilize the above-mentioned technique to evaluate the influence of the $\psi^\prime$ states on the LECs. Specifically, the $\psi^\prime$-exchange amplitudes in Eq.~\eqref{eq.exBprime} are expanded in terms of $\nu=(s-u)/(4m)$, $t$ and $m_\phi$, and then compared to the contact-term contribution in Eqs.~\eqref{o2tofT} and~\eqref{o3tofT}. As a result, the $\psi^\prime$-exchange contributions to the ${\cal O}(p^2)$ LECs read
\begin{align}
  b_{1,2,4,6}^{\psi^\prime} =0\ ,\quad
  b_3^{\psi^\prime} = -\frac{h_A^2}{6\Delta}\ ,\quad
  b_5^{\psi^\prime} = -\frac{h_A^2m^2}{48\Delta(m+\Delta)^2}\ ,\quad
  b_7^{\psi^\prime} = -\frac{h_A^2m^2}{24\Delta}\ ,\label{eq.5.14}
\end{align}
where $m$ is the spin-$1/2$ baryon mass in the chiral limit. For the ${\cal O}(p^3)$ LECs, the $\psi^\prime$ states contribute as
\begin{align}
c_{11}^{\psi^\prime}&=-\frac{h_A^2(2m^2+3m\,\Delta+3\Delta^2)}{48\Delta^2(m+\Delta)^2}\ ,\quad 
c_{12}^{\psi^\prime}=-\frac{h_A^2m^2}{96\Delta^2(m+\Delta)^2}\ ,\quad \label{eq.5.15}\\
c_{13}^{\psi^\prime}&=\frac{h_A^2\,m}{96\Delta^2(m+\Delta)}\ ,\quad 
c_{14}^{\psi^\prime}=0\ ,\quad 
c_{20}^{\psi^\prime}=\frac{h_A^2(3m+\Delta)}{48\Delta(m+\Delta)^2}\ . \label{eq.5.16}
\end{align}
For the sake of numerical estimation, we take the chiral limit mass $m$ equal to the average of the physical masses of the $\Xi_{cc}$ and $\Omega_{cc}$, i.e., $m=({m_{\Xi_{cc}}+m_{\Omega_{cc}}})/{2}=3679.8$~MeV. Numerical results, obtained with three different mass splitting values $\Delta=(50,100,150)$~MeV, are displayed in Table~\ref{tab.res.con.lec}.

\begin{table}[htbp]
\centering
\caption{Results of Born-term $\psi^\prime$-exchange contributions to the LECs. $\Delta$ denotes the mass splitting between the spin-$1/2$ and spin-$3/2$ doubly charmed baryons. \label{tab.res.con.lec}}
\renewcommand{\arraystretch}{1.2}
\renewcommand{\tabcolsep}{1pc}
\begin{tabular}{cc|cc}
\hline\hline
NLO LECs                   &$\Delta=(50,100,150)~{\rm MeV}$      & NNLO LECs                 &$\Delta=(50,100,150)~{\rm MeV}$  \\     
\hline 
$b_1^{\psi^\prime}$        &$0$                                  &$c_{11}^{\psi^\prime}$     &$(-7.17, -1.78, -0.79)$           \\
$b_2^{\psi^\prime}$        &$0$                                  &$c_{12}^{\psi^\prime}$     &$(-1.76,-0.43,-0.19)$             \\
$b_3^{\psi^\prime}$        &$(-1.44,-0.72,-0.48)$                &$c_{13}^{\psi^\prime}$     &$(1.78, 0.44, 0.19)$               \\
$b_4^{\psi^\prime}$        &$0$                                  &$c_{14}^{\psi^\prime}$     &$0$                               \\ 
$b_5^{\psi^\prime}$        &$(-0.18, -0.09,-0.06)$               &$c_{20}^{\psi^\prime}$     &$(0.14, 0.07, 0.05)$ \\ 
$b_6^{\psi^\prime}$        &$0$                                  &                           &     \\
$b_7^{\psi^\prime}$        &$(-4.89, -2.44,-1.63)$               &                           &     \\
\hline\hline
\end{tabular}
\end{table}

{
In Table~\ref{tab.res.con.lec}, the magnitudes of the LECs become smaller as the mass splitting $\Delta$ gets larger, which can be inferred from Eqs.~\eqref{eq.5.14}, \eqref{eq.5.15} and \eqref{eq.5.16}. Consequently, the spin-$3/2$ baryon contribution to the scattering lengths is expected to decrease as $\Delta$ increases. In Table~\ref{tab:spin3o2}, such a behavior can be clearly observed e.g. for the $P$-wave scattering lengths $a_{1-}$. However, anomaly happens for the scattering lengths $a_{1+}$ with $(S,I)=(-1,0)$ and $(S,I)=(0,1/2)$. This anomaly actually indicates that the spin-$3/2$ baryon fields can not be integrated out in the two channels. In other words, explicit inclusion of spin-$3/2$ doubly charmed baryons is necessary if one intends to well determine the $P$-wave scattering length with $J^P=(3/2)^+$ and $(S,I)\in \{(-1,0),(0,1/2)\}$.
}

\subsection{$S$-wave Phase shifts\label{sec:5.4}}

With the chiral amplitudes, one can compute partial-wave phase shifts straightforwardly, which are functions of the CM energy $\sqrt{s}$. Although it is undoable to extract phase shifts experimentally, they can be related to energy levels according to L{\"u}scher formula~\cite{Luscher:1986pf} and its extensions~\cite{Rummukainen:1995vs,Kim:2005gf,Gockeler:2012yj,Briceno:2014oea}, which can be computed by lattice simulations in future.

Usually, the partial-wave amplitude $f^{(S,I)}_{\ell\pm}(s)$ from BChPT do not obey partial wave unitarity exactly, since they are derived perturbatively up to a certain order. The method of extracting phase shifts from perturbative amplitudes has been discussed, e.g, in Ref.~\cite{Yao:2016vbz}. Namely, one can calculate the phase shifts in the elastic scattering region by using
\begin{align}
\delta^{(S,I)}_{\ell\pm}(s)= \arctan\bigg\{|{\bf q}|\, {\rm Re} \left[f^{(S,I)}_{\ell\pm}(s)\right]\bigg\} \ ,
\end{align}
where $|{\bf q}|$ is the modulus of the CM momentum. In the present work, we are only interested in the $S$-wave interactions, whose strength is expected to be stronger than that of higher partial waves. Moreover, one is allowed to ignore the effects of the spin-$3/2$ HQS cousins of the spin-$1/2$ doubly charmed baryons, since their impact on the $S$-wave phase shifts are negligible, as discussed in the preceding subsection.

\begin{figure}
\centering
\includegraphics[width=0.95\textwidth]{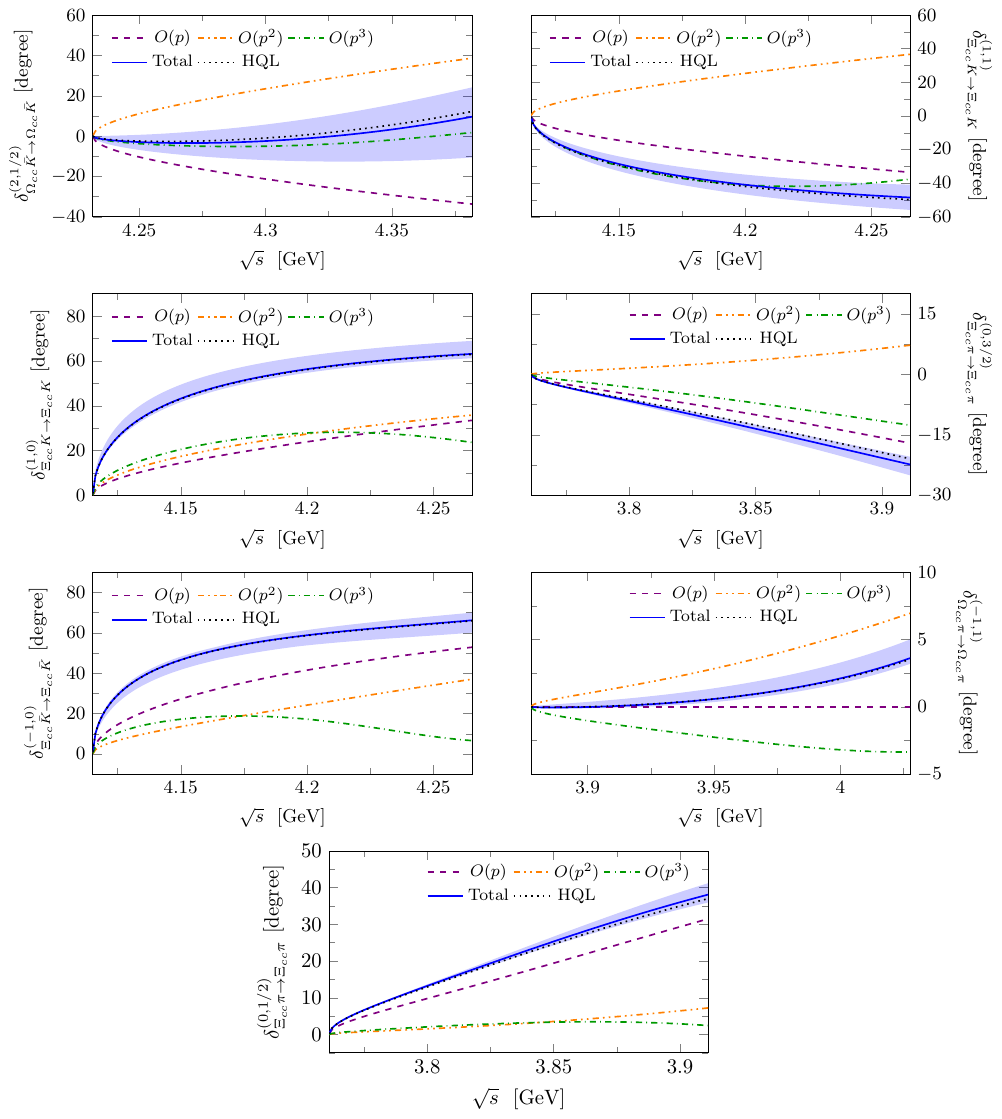}
\caption{\label{fig.sps} Results of $S$-wave phase shifts. The $\mathcal{O}(p)$, $\mathcal{O}(p^2)$ and $\mathcal{O}(p^3)$ contributions are represented by violet dashed, orange dash-dot-dotted, green dash-dotted lines, in order. The blue solid lines with bands stand for the total contribution. For comparison, the HQL results (black dotted lines) are also shown.}
\label{fig:dcSL_ps}
\end{figure}

In Figure~\ref{fig.sps}, the $S$-wave phase shifts for the processes of elastic scattering are plotted for the energy region from threshold $\sqrt{s_{\rm th}}$ to the point $\sqrt{s_{\rm th }}+150~{\rm  MeV}$. The blue solid lines stand for our results up to the order of $\mathcal{O}(p^3)$. The light-blue bands corresponds to the uncertainties propagated from the errors of LECs via the method of Monte Carlo technique. In the figure, we also show the phase shifts order by order. The $\mathcal{O}(p)$, $\mathcal{O}(p^2)$ and $\mathcal{O}(p^3)$ contributions are represented by violet dashed, orange dash-dot-dotted, green dash-dotted lines, in order. It can be seen that the convergence properties of the chiral expansion in the two elastic channels, $\Xi_{cc}\pi$ with $(S,I)=(0,1/2)$ and $(0,3/2)$ are relatively better, compared to the other channels. In fact, most of the LECs we used here are estimated by imposing HDA symmetry, while some of the LECs in the $\mathcal{O}(p^3)$ tree amplitudes are simply assumed to be zero under the requirement of naturalness. Therefore, a solid conclusion on the convergence properties can be drawn only when relevant lattice QCD data are available. For comparison, the HQL results of phase shifts are shown as well, which are obtained by setting $g=0$. They are represented by the black dotted lines in the figure. It is found that the HQL-vanishing diagrams contribute slightly to the phase shifts, which is similar to the case of $S$-wave scattering lengths.

\section{Summary and outlook\label{sec:6}}
In this work, we have performed a NNLO calculation of the scattering amplitudes for the interactions between NGBs and doubly charmed baryons within the framework of relativistic BChPT. The EOMS scheme is employed to handle with the UV divergences and the PCB terms originating from the loops.
We find that most of the unknown LECs in the chiral effective Lagrangians can be pinned down by making use of the HDA symmetry. The obtained LEC values enable us to make predictions of the $S$- and $P$-wave scattering lengths. We show that the HQL-surviving diagrams dominate the contribution to the $S$-wave scattering lengths, while the HQL-vanishing ones contribute marginally. The influence of spin-$3/2$ double-charm baryons on the scattering lengths is also evaluated in detail. Their contributions to the LECs are estimated as well. For future reference, the energy-dependent $S$-wave phase shifts are plotted for the elastic scattering channels in the energy regions near the lowest thresholds. Our chiral results can be applied to perform chiral extrapolations of future lattice QCD data, and can also be used to investigate the spectroscopy of doubly heavy baryons systematically.


\appendix
\section{$\Gamma$-functions and $\beta$-functions\label{app:UV-divergent}}
In this appendix, the $\Gamma$-functions in Eq.~\eqref{eq:Gammas} and the UV- renormalization $\beta$-functions in Eq.~\eqref{MS-UV} are listed. Their specific expressions read
\begin{align}\label{eq:betaUV}
  \Gamma_{4} &= \frac{1}{8}, \notag \\
  \Gamma_{5} &= \frac{3}{8}, \notag \\
  \Gamma_{6} &= \frac{11}{144}, \notag \\
  \Gamma_{8} &= \frac{5}{48}, \notag \\
  \beta_{m} &= \frac{8g^2m^{3}}{3}, \notag \\
  \beta_{g} &= \frac{g(-9 + 2g^2)m^{2}}{3}, \notag \\
  \beta_{b_{1}} &= -\frac{4g^2m}{9}, \notag \\
  \beta_{b_{2}} &= -\frac{5g^2m}{12}, \notag \\
  \beta_{b_{3}} &= -\frac{3(-1 + 2g^2 + 3g^4)m}{16}, \notag \\
  \beta_{b_{4}} &= \frac{(1 - 10g^2 + 5g^4)m}{16}, \notag \\
  \beta_{b_{5}} &= -\frac{3(-1 + g^2)^2}{64m}, \notag \\
  \beta_{b_{6}} &= -\frac{(-1 + g^2)^2}{32m}, \notag \\
  \beta_{b_{7}} &= \frac{3(-1 - 2g^2 + 3g^4)m}{32}, \notag \\
  \beta_{c_{11}} &= \frac{1 - 4g^2 + 3g^4}{64}, \notag \\
  \beta_{c_{12}} &= 0, \notag \\
  \beta_{c_{13}} &= -\frac{3(-1 + g^2)^2}{128m}, \notag \\
  \beta_{c_{14}} &= -\frac{(-1 + g^2)^2}{64m}, \notag \\
  \beta_{c_{15}} &=-\frac{3 \left(g^3-g\right)}{32}, \notag \\
  \beta_{c_{16}} &= \frac{g(1 -  g^2)}{8}, \notag \\
  \beta_{c_{17}} &= -\frac{g-g^3}{16}, \notag \\
  \beta_{c_{18}} &= 0, \notag \\
  \beta_{c_{19}} &= 0, \notag \\
  \beta_{c_{20}} &= \frac{1 - g^2}{32}.
\end{align}

\section{$\widetilde{\beta}$-functions\label{app:EOMS-PCBterms}}
The $\widetilde{\beta}$-functions in Eq.~\eqref{EOMS-PCB} are given as follows.
\begin{align}
  \widetilde{\beta}_{m} &= -\frac{8g^2m}{3} A_{0}(m^{2}), \notag \\
  \widetilde{\beta}_{g} &= \frac{2 g^3 m^2}{3}-\frac{g(-9 + 2g^2)}{3} A_{0}(m^{2}), \notag \\
  \widetilde{\beta}_{b_{1}} &=\frac{4g^2 m}{9} + \frac{4g^2}{9m} A_{0}(m^{2}), \notag \\
  \widetilde{\beta}_{b_{2}} &=\frac{5  g^2 m }{12}+ \frac{5 g^2 }{12 m} A_{0}(m^2), \notag \\
  \widetilde{\beta}_{b_{3}} &= \frac{g^2 \left(g^2+3 \right) m}{4 } + \frac{3 \left(3 g^4 + 2 g^2 - 1\right) }{16 m} A_{0}(m^2), \notag \\
  \widetilde{\beta}_{b_{4}} &= \frac{  g^2 \left(5 g^2-1\right) m}{4 } -\frac{5 g^4 - 10 g^2 + 1 }{16  m} A_{0}(m^2), \notag \\
\widetilde{\beta}_{b_{5}} &= \frac{5 g^4+9}{96  m}+\frac{3 \left(g^2 - 1\right)^2}{64 m^3} A_{0}(m^2),\notag \\
\widetilde{\beta}_{b_{6}} &= \frac{g^4+1}{16m} + \frac{\left(g^2 - 1\right)^2}{32  m^3} A_{0}(m^2), \notag \\
\widetilde{\beta}_{b_{7}} &=-\frac{ g^2 \left(3 g^2+5\right)m}{12 } -\frac{3 \left(3g^4 - 2 g^2 - 1\right)}{32 m} A_{0}(m^2)\ ,
\end{align}
where $A_0$ is defined in Appendix~\ref{app.wave}.

\section{Loop corrections to the baryon masses and wave function renormalization constants\label{app.wave}}
In our calculation, the $N$-point ($N\le 4$) one-loop scalar integrals~\cite{Denner:2005nn} are defined by 
\begin{align}
T^{N}=\frac{(2\pi\mu)^{4-d}}{i\pi^2}\int_{}^{}  \, \frac{{\rm d}^{d}k}{\left[k^2-m_0^2+i\epsilon\right]\left[\left(k+p_1\right)^2-m_1^2+i\epsilon\right]\cdots\left[\left(k+p_{N-1}\right)^2-m_{N-1}^2+i\epsilon\right]}\ , \notag 
\end{align}
with $\epsilon$ being an infinitesimal positive number. Traditionally, the one-, two-, three- and four-point one-loop scalar integrals are denoted by $A_0$, $B_0$, $C_0$ and $D_0$, in order. To be specific, one has
\begin{align}
&T^{1}=A_0(m_0^2)\ , \\&T^{2}=B_0(p_1^2,m_0^2,m_1^2)\ , 
\\& T^{3}=C_0(p_1^2,(p_2-p_1)^2,p_2^2,m_0^2,m_1^2,m_2^2)\ ,  \\
&T^{4}=D_0(p_1^2,(p_2-p_1)^2,(p_3-p_2)^2,p_3^2,p_2^2,(p_3-p_1)^2,m_0^2,m_1^2,m_2^2,m_3^2)\ . 
\end{align}

With these definitions, the chiral corrections concerning the baryon masses and wave function renormalization constants are clearly shown in what follows. The one-loop corrections to the $\Xi_{cc}$ and $\Omega_{cc}$ baryons read
\begin{align}
\delta m_{\Xi_{cc}}^{\rm loop}
&=\frac{g^2}{12 m_{\Xi_{cc}}F^2}
\bigg\{
3(m_{\Xi_{cc}}^2-m_{\Omega_{cc}}^2) A_{0}(m_K^2) 
+3(m_{\Xi_{cc}}+m_{\Omega_{cc}})^2 A_{0}(m_{\Omega_{cc}}^2)
\notag \\
&+20 m_{\Xi_{cc}}^2 A_{0}(m_{\Xi_{cc}}^2)
+2m_{\eta}^2 m_{\Xi_{cc}}^2 B_{0}(m_{\Xi_{cc}}^2, m_{\eta}^2, m_{\Xi_{cc}}^2) 
\notag \\
&
-3[(m_{\Omega_{cc}}-m_{\Xi_{cc}})^2-m_K^2](m_{\Omega_{cc}}+m_{\Xi_{cc}})^2 B_{0}(m_{\Xi_{cc}}^2, m_{K}^2, m_{\Omega_{cc}}^2) 
\notag \\
&
+18 m_{\pi}^2 m_{\Xi_{cc}}^2 B_{0}(m_{\Xi_{cc}}^2, m_{\pi}^2, m_{\Xi_{cc}}^2) 
\bigg\} \ ,
\notag \\
\delta m_{\Omega_{cc}}^{\rm loop}
&=\frac{g^2}{6 m_{\Omega_{cc}}F^2}
\bigg\{
3(m_{\Omega_{cc}}^2-m_{\Xi_{cc}}^2)A_{0}(m_K^2)
+4m_{\Omega_{cc}}^2 A_{0}(m_{\Omega_{cc}}^2)
\notag \\
&
+3(m_{\Omega_{cc}}+m_{\Xi_{cc}})^2 A_{0}(m_{\Xi_{cc}}^2)
+
4m_{\eta}^2 m_{\Omega_{cc}}^2 B_{0}(m_{\Omega_{cc}}^2, m_{\eta}^2, m_{\Omega_{cc}}^2)
\notag \\
&
-3[(m_{\Omega_{cc}}-m_{\Xi_{cc}})^2-m_K^2]
(m_{\Omega_{cc}}+m_{\Xi_{cc}})^2
B_{0}(m_{\Omega_{cc}}^2, m_K^2, m_{\Xi_{cc}}^2)
\bigg\}
\ . \label{eq.mass.ren}
\end{align}
For the $\Xi_{cc}^{++}$ and $\Xi_{cc}^+$ states, the wave function renormalization constants are defined by
\begin{align}
\mathcal{Z}_{\Xi_{cc}}=1+\delta \mathcal{Z}_{\Xi_{cc}}\ ,
\end{align}
with
\begin{align}
\delta Z_{\Xi_{cc}}&=\frac{g^2}{12F^2}
\bigg\{
\frac{1}{m_{\eta}^2-4m_{\Xi_{cc}}^2}
\big[
(2d-3)m_{\eta}^2-4(d-1)m_{\Xi_{cc}}^2
\big]
A_{0}(m_{\eta}^2)
\notag \\
&
+\frac{3}
{m_{\Xi_{cc}}^2\Delta_{\psi K}^2} 
\big\{
(d-1)(m_{\Omega_{cc}}+m_{\Xi_{cc}})^2(m_{\Omega_{cc}}^2+m_{\Xi_{cc}}^2-m_K^2)
+2m_K^2m_{\Omega_{cc}} m_{\Xi_{cc}}
\big\}
A_{0}(m_K^2)
\notag \\
&
-\frac{3(m_{\Omega_{cc}}+m_{\Xi_{cc}})
}{m_{\Xi_{cc}}^2\Delta_{\psi K}^2}
\big\{
(d-1)(m_{\Omega_{cc}}+m_{\Xi_{cc}})(m_{\Omega_{cc}}^2-m_{\Xi_{cc}}^2-m_K^2)
+2m_K^2m_{\Xi_{cc}}
\big\}
A_{0}(m_{\Omega_{cc}}^2)
\notag \\
&
+\frac{9}{m_{\pi}^2-4m_{\Xi_{cc}}^2}
[(2d-3)m_{\pi}^2-4(d-1)m_{\Xi_{cc}}^2]
A_{0}(m_{\pi}^2)
\notag \\
&
+
\frac{4(d-2)}{(m_{\eta}^2-4m_{\Xi_{cc}}^2)(m_{\pi}^2-4m_{\Xi_{cc}}^2)} 
[18m_{\pi}^2 m_{\Xi_{cc}}^2-m_{\eta}^2(5m_{\pi}^2-2m_{\Xi_{cc}}^2)]
A_{0} (m_{\Xi_{cc}}^2)
\notag \\
& 
-\frac{2m_{\eta}^2}{m_{\eta}^2-4m_{\Xi_{cc}^2}}
[(d-2)m_{\eta}^2-2(d-1)m_{\Xi_{cc}}^2]
B_{0}(m_{\Xi_{cc}}^2, m_{\eta}^2, m_{\Xi_{cc}}^2)
\notag \\
&
+\frac{3(m_{\Omega_{cc}}+m_{\Xi_{cc}})}{m_{\Xi_{cc}}^2\Delta_{\psi K}^2}
\big\{
(d-1)(m_{\Omega_{cc}}+m_{\Xi_{cc}})
(m_{\Omega_{cc}}^4-m_{\Xi_{cc}}^4+m_K^4-
2m_K^2 m_{\Omega_{cc}}^2)
\notag \\
&
+2m_K^2m_{\Xi_{cc}}
[
(m_{\Omega_{cc}}-m_{\Xi_{cc}})^2
-m_K^2
]
\big\}
B_{0}(m_{\Xi_{cc}}^2, m_K^2, m_{\Omega_{cc}}^2)
\notag \\
&
-\frac{18m_{\pi}^2}{(m_{\pi}^2-4m_{\Xi_{cc}}^2)}
\big\{
(d-2)m_{\pi}^2-2(d-1)m_{\Xi_{cc}}^2
\big\}
B_{0}(m_{\Xi_{cc}}^2, m_{\pi}^2, m_{\Xi_{cc}}^2)
\bigg\} 
\ , 
\end{align}
and
\begin{align}
\delta Z_{\Omega_{cc}}&=
\frac{g^2}{6F^2}
\bigg\{  
\frac{2}{m_{\eta}^2-4m_{\Omega_{cc}}^2} 
\big\{
[(2d-3)m_{\eta}^2-4(d-1)m_{\Omega_{cc}}^2]
\big\}
A_{0}(m_{\eta}^2)
-\frac{4(d-2)m_{\eta}^2}{m_{\eta}^2-4m_{\Omega_{cc}}^2}
A_{0}(m_{\Omega_{cc}}^2)
\notag \\
&
+\frac{3}{m_{\Omega_{cc}}^2
\Delta_{\psi K}^2} 
\big\{
(d-1)(m_{\Omega_{cc}}+m_{\Xi_{cc}})^2(m_{\Omega_{cc}}^2+m_{\Xi_{cc}}^2-m_K^2)
+2m_K^2
m_{\Omega_{cc}} m_{\Xi_{cc}}
\big\}
A_{0}(m_K^2)
\notag \\
&
+\frac{3(m_{\Omega_{cc}}+m_{\Xi_{cc}})
}{m_{\Omega_{cc}}^2\Delta_{\psi K}^2}
\big\{
(d-1)(m_{\Omega_{cc}}+m_{\Xi_{cc}})(m_{\Omega_{cc}}^2-m_{\Xi_{cc}}^2+m_K^2)
-2m_K^2m_{\Omega_{cc}}
\big\}
A_{0}(m_{\Xi_{cc}}^2)
\notag \\
&
-\frac{4m_{\eta}^2}{m_{\eta}^2-4m_{\Omega_{cc}}^2}
[(d-2)m_{\eta}^2-2(d-1)m_{\Omega_{cc}}^2]
B_{0}(m_{\Omega_{cc}}^2, m_{\eta}^2, m_{\Omega_{cc}}^2)
\notag \\
&
+\frac{3(m_{\Omega_{cc}}+m_{\Xi_{cc}})}{m_{\Omega_{cc}}^2\Delta_{\psi K}^2}
\big\{
(d-1)(m_{\Xi_{cc}}+m_{\Omega_{cc}})
(m_K^4-(m_{\Omega_{cc}}^4-m_{\Xi_{cc}}^4)-2m_K^2m_{\Xi_{cc}}^2)
\notag \\
&
+2m_K^2
m_{\Omega_{cc}}
[
(m_{\Xi_{cc}}-m_{\Omega_{cc}})^2-m_K^2
]
\big\}
B_{0}(m_{\Omega_{cc}}^2, m_K^2, m_{\Xi_{cc}}^2)
\bigg\} 
\ ,
\end{align}
where the abbreviation $\Delta^2_{\psi K}\equiv[(m_{\Omega_{cc}}+m_{\Xi_{cc}})^2-m_K^2]$ has been used.

\section{Heavy diquark-antiquark symmetry\label{app.hda}}
In HQL, charmed mesons and doubly charmed baryons can form a super multiplet according to the HDA symmetry. A uniform Lagrangian can be constructed with common LECs. Following Refs.~\cite{Hu:2005gf,Meng:2018zbl,Liu:2011mi,Jiang:2019hgs}, we define a super field $\mathcal{S}$ in the form as 
\begin{align}
\mathcal{S}\equiv 
\left(\begin{array}{cc}
\bar{\tilde{H}} & 0 \\
  0                 & T
\end{array}
\right) \ , \quad 
\mathcal{\bar{S}}\equiv 
\left(\begin{array}{cc}
\tilde{H}    & 0              \\
0                & \overline{T}
\end{array}
\right) \ ,
\end{align}
which comprises both the charmed mesons and doubly charmed baryons. The pseudoscalar and vector charmed mesons are collected in the $\tilde{H}$ and $\bar{\tilde{H}}$ fields as 
\begin{align}
\tilde{H}&=\sqrt{\overline{m}_D}(\slashed{\tilde{P}}^{\ast}+i\tilde{P}\gamma_{5})\frac{1-\slashed{\upsilon}}{2}, 
&\bar{\tilde{H}}&=\frac{1-\slashed{\upsilon}}{2}\sqrt{\overline{m}_D}(\slashed{\tilde{P}}^{\ast\dagger} + i\tilde{P}^{\dagger}\gamma_{5}), 
\notag \\
\tilde{P}&=( \bar{D}^{0}, D^{-},D_{s}^{-} ), 
&\tilde{P}^{\ast}_{\mu}&=(\bar{D}^{\ast0}, D^{\ast-},D_{s}^{\ast-})_{\mu}, 
\notag 
\end{align}
where $\overline{m}_D$ is the charmed meson mass in the HQL.
Likewise, the spin-$1/2$ and spin-$3/2$ doubly charmed baryons are contained in $T^{\mu}$ and $\overline{T}^{\mu}$ fields as
\begin{align}
T^{\mu}&=\psi^{\prime \mu}+\sqrt{\frac{1}{3}} (\gamma^{\mu} + \upsilon^{\mu}) \gamma^{5}\psi, 
&\overline{T}^{\mu}&=\bar{\psi}^{\prime \mu}-\sqrt{\frac{1}{3}} \bar{\psi}\gamma^{5} (\gamma^{\mu} + \upsilon^{\mu}), 
\notag \\
\psi&=(\Xi_{cc}^{++}, \Xi_{cc}^{+}, \Omega_{cc}^{+})^{T},
&\psi^{\prime}_{\mu}&=(\Xi_{cc}^{\prime ++}, \Xi_{cc}^{\prime +}, \Omega_{cc}^{\prime +})_{\mu}^{T}.
\label{eq:super.super.field}
\end{align}
Accordingly, auxiliary chiral blocks for the NGBs should be introduced. We need
\begin{align}
\widehat{\chi}_{\pm}&=E_2\otimes\chi_{\pm}=
\left(\begin{array}{cc}
\chi_{\pm}   &  0 \\
0            &  \chi_{\pm}
\end{array}
\right) \ ,  \quad 
\widehat{\tilde{\chi}}_{\pm}=E_2\otimes \tilde{\chi}_{\pm} \ , 
\notag\\
\widehat{u}&=E_2\otimes u \ , \quad
\widehat{D}=E_2\otimes D \ , \quad 
\widehat{h}=E_2\otimes h \ , \quad  
\end{align}
with $E_2$ the two-dimensional identity matrix. With the superfield $S$ and the super chiral blocks, chiral effective Lagrangians in HQL up to $\mathcal{O}(p^3)$ can be constructed straightforwardly, which read
\begin{align}
\mathscr{L}_{\mathcal{S}\phi}^{(1)}
&=i {\rm Tr} (\mathcal{S} \upsilon \cdot D \bar{\mathcal{S}})
-\frac{g_h}{2}{\rm Tr} (\mathcal{S} \widehat{u}^{\mu} \gamma_{\mu}\gamma_5 \bar{\mathcal{S}}) \ , 
\notag \\
\mathscr{L}_{\mathcal{S}\phi}^{(2)}
 &= d_1 {\rm Tr}(\mathcal{S} \bar{\mathcal{S}})  \left \langle \chi_{+} \right \rangle + d_2 {\rm Tr}( \mathcal{S} \widehat{\chi}_{+} \bar{\mathcal{S}}) - d_3 {\rm Tr}(\mathcal{S}\bar{\mathcal{S}}) \left \langle (\upsilon \cdot u)^2 \right \rangle - d_4 {\rm Tr}(\mathcal{S} (\upsilon \cdot \widehat{u})^2 \bar{\mathcal{S}}) \notag \\
  &- d_5 {\rm Tr}(\mathcal{S} \bar{\mathcal{S}})  \left \langle u^2 \right \rangle - d_6 {\rm Tr}(\mathcal{S} \widehat{u}^2 \bar{\mathcal{S}}) - id_7 {\rm Tr}(\mathcal{S} [\widehat{u}^{\mu}, \widehat{u}^{\nu}]\sigma_{\mu\nu}\bar{\mathcal{S}})\ , \notag \\
%
\mathscr{L}^{(3)}_{\mathcal{S}\phi} 
&=ie_{1} {\rm Tr} (\mathcal{S} \left[\widehat{u}_{\mu},\widehat{h}^{\mu \nu} \right]\upsilon_{\nu} \bar{\mathcal{S}})
+ie_2 {\rm Tr} (\mathcal{S}  \left[\upsilon \cdot \widehat{u}, \widehat{h}^{\mu\nu}\right] \upsilon_{\mu} \upsilon_{\nu} \bar{\mathcal{S}})
+e_3 {\rm Tr} (\mathcal{S} \left\{\widehat{u}^{\mu},\widehat{h}^{\nu\rho }\right\}\sigma_{\mu\nu}\upsilon_{\rho}
\bar{\mathcal{S}})
\notag \\
&
+e_4 {\rm Tr} (\mathcal{S}
\sigma_{\mu\nu} \left \langle u^{\mu}h^{\nu \rho} \right \rangle \upsilon_{\rho}
\bar{\mathcal{S}})
+e_{5} {\rm Tr} (\mathcal{S}
\left\{\widehat{u}^{\nu},\widehat{\chi}_{+}\right\}\gamma_{5}\gamma_{\nu}
\bar{\mathcal{S}})
+e_{6}{\rm Tr} (\mathcal{S}
\widehat{u}^{\mu}\gamma_{5}\gamma_{\mu} \left \langle 
\chi_{+} \right \rangle 
\bar{\mathcal{S}})
\notag \\
&+e_{7}{\rm Tr} (\mathcal{S}
\gamma_{5}\gamma_{\mu} \left \langle u^{\mu}\widetilde{\chi}_{+}\right \rangle
\bar{\mathcal{S}})
+
ie_{8} {\rm Tr}(\mathcal{S} \gamma_{5}\gamma_{\nu}\left[\widehat{D}^{\nu},\widehat{\widetilde{\chi}}_{-}\right] \bar{\mathcal{S}})
+
ie_9 {\rm Tr}(\mathcal{S} \gamma_{5}\gamma_{\nu}\left \langle \left[D^{\nu}, \chi_{-} \right]\right \rangle \bar{\mathcal{S}}) 
\notag \\
&
+e_{10} ({\rm Tr}(\mathcal{S} \left[\widehat{\widetilde{\chi}}_{-}, \upsilon \cdot \widehat{u} \right]
\bar{\mathcal{S}})  \ .
\label{eq.Lag.sp}
\end{align}
Here, $d_i$ ($i = 1, 2, \cdots, 7$) and $e_i$ ($i= 1, 2, \cdots, 10$) are LECs common to $D\phi$, $D^\ast\phi$, $\psi\phi$ and $\psi^\prime\phi$ interactions. The symbol ${\rm Tr}(\cdots)$ denotes the trace in the Dirac space, and $\upsilon_{\mu}$ is the baryon four-velocity satisfying $\upsilon^2 = 1$.

On the other hand, relativistic $D\phi$ Lagrangians can be found in Ref.~\cite{Yao:2015qia}, which are 
\begin{align}
\label{eq:dpi.lag.lo} \mathscr{L}_{P\phi}^{(1)} &= D_{\mu} P D^{\mu} P^\dagger - m_{P}^2 P P^\dagger
+ i g_{0} (P^{\ast}_{\mu} u^{\mu} P^\dagger
- P u^\mu P^{\ast\dagger}_{\mu})
\ ,
\\
\label{eq:dpi.lag.nlo} \mathscr{L}_{P\phi}^{(2)} &=-h_0 P \left \langle \chi_{+} \right \rangle P^{\dagger} -h_1 P \chi_{+} P^{\dagger} + h_2 P \left \langle u^2 \right \rangle P^{\dagger} -h_3 P u^2 P^{\dagger} 
+ h_4 D_{\mu} P \left \langle u^{\mu} u^{\nu} \right \rangle D_{\nu} P^{\dagger} 
\notag \\
&
- h_5 D_{\mu} P \{u^{\mu}, u^{\nu}\} D_{\nu} P^{\dagger}\ , 
\\
\label{eq:dpi.lag.nnlo} \mathscr{L}_{P\phi}^{(3)} &= ig_1 P [\chi_{-}, u^{\mu}] D_{\mu} P^{\dagger} + g_2 P [u_{\mu},h^{\mu \nu} ] D_{\nu} P^{\dagger} + \frac{g_3}{2} P [u^{\mu}, h^{\nu \rho}] \{D_{\mu}, \{D_{\nu}, D_{\rho}\}\} P^{\dagger}\ . 
\end{align}
Here $h_{i}$ ($i = 0, 1, \cdots, 5$) and $g_i$ ($i = 1, 2, 3$) are LECs, whose values have already been determined in Ref.~\cite{Yao:2015qia}. 

Our aim is to derive relations between the $\psi\phi$ LECs in the Lagrangians~\eqref{op1baryon},~\eqref{op2baryon} and~\eqref{op3baryon} and the $D\phi$ LECs in Lagrangians ~\eqref{eq:dpi.lag.lo},~\eqref{eq:dpi.lag.nlo} and \eqref{eq:dpi.lag.nnlo}. This can be established in three steps. 
First, one performs non-relativistic projection of the relativistic $D\phi$ Lagrangians to get their HQL counterparts. By comparing with the HQL Lagrangians in Eq.~\eqref{eq.Lag.sp}, one finds  
\begin{align}
g_0 &=g_h \overline{m}_D,  
& h_0 &= 2 d_1 \overline{m}_D, 
& h_1 &= 2 d_2 \overline{m}_D, 
& h_2 &= 2 d_5 \overline{m}_D, 
& h_3 &= -2 d_6 \overline{m}_D, \notag \\
h_4 &= \frac{2 d_3}{\overline{m}_{D}}, 
&h_5 &= -\frac{d_4}{\overline{m}_{D}}; 
& g_1 &=2 e_{10}, 
& g_2 &= -2 e_{1}, 
& g_3 &= \frac{e_{2}}{\overline{m}_{D}^{2}}\ .\label{eq:dphi.relation}
\end{align}
Second, one repeats the same procedure for the relativistic $\psi\phi$ Lagrangians and gets
\begin{align}\label{eq:psiphi.relations}
g&=-\frac{1}{3}g_h , 
& b_1 &= -(d_1 + \frac{1}{3} d_2), 
& b_2 &= -d_2, 
& b_3 &= d_6, 
& b_4 &= d_5, 
\notag \\
b_5 &= -\frac{d_4}{8}, 
& b_6 &= -\frac{d_3}{4}, 
& b_7 &= -\frac{d_7}{3}; 
& c_{11} &= -e_{1}, 
& c_{12} &= \frac{e_2}{4}, \notag \\
c_{13} &= \frac{e_3}{6},
& c_{14} &= \frac{e_4}{6}, 
& c_{15} &= \frac{e_5}{3}, 
& c_{16} &= \frac{e_6}{3}, 
& c_{17} &= \frac{e_7}{3},  \notag \\
c_{18} &= \frac{e_8}{3}, 
& c_{19} &= \frac{e_9}{3}, 
& c_{20} &= -e_{10}. 
\end{align} 
Eventually, the identities in Eq.~\eqref{eq:dphi.relation} and Eq.~\eqref{eq:psiphi.relations} lead to the following relations between the relativistic $\psi\phi$ and $D\phi$ LECs:
\begin{align}
{g}&=-\frac{1}{3\overline{m}_D}{g}_0 \ ,\qquad  
{b}_1=-\frac{1}{2\overline{m}_D}({h}_0+\frac{1}{3} {h}_1) \ ,   \qquad 
{b}_2=-\frac{1}{2\overline{m}_D} {h}_1 \ , \notag \\
{b}_3&=-\frac{1}{2\overline{m}_D} {h}_3 \ ,\qquad 
{b}_4=\frac{1}{2\overline{m}_D} {h}_2 \ ,  \qquad 
{b}_5=\frac{\overline{m}_D}{8} {h}_5 \ ,  \notag \\  
{b}_6&=-\frac{\overline{m}_D}{8} {h}_4 \ , \qquad 
{c}_{11}=\frac{{g}_2}{2} \ , \qquad 
{c}_{12}=\frac{\overline{m}_D^2}{4} {g}_3 \ ,\qquad 
{c}_{20}=-\frac{{g}_1}{2} \ .
\end{align}

\acknowledgments

We would like to thank Shao-Zhou~Jiang and Zhan-Wei~Liu for helpful discussions. This work is supported by National Nature Science Foundations of China (NSFC) under Contract Nos. 12275076, 11905258 and by the Fundamental Research Funds for the Central Universities under Contract No. 531118010379.



\end{document}